\documentclass[twocolumn,tighten]{aastex63}
\usepackage{amsmath}

\newcommand{\genv}{$g_\text{Ne,env}$}
\newcommand{\gfit}{$g_\text{Ne,fit}$}
\newcommand{\gN}{$g_\text{Ne}$}
\newcommand{\eenv}{$e_\text{N,env}$}

\newcommand{\eN}{$e_\text{N}$}
\newcommand{\et}{$\tilde{e}$}
\newcommand{\etenv}{$\tilde{e}_\text{env}$}
\newcommand{\etfit}{$\tilde{e}_\text{fit}$}
\newcommand{\pap}{Paper~I}

\shorttitle{The External Field Effect and Large-Scale Structure}
\shortauthors{Chae et al.}

\begin{document}

\title{Testing the Strong Equivalence Principle. II. Relating the External Field Effect in Galaxy Rotation Curves to the Large-Scale Structure of the Universe}
\correspondingauthor{Kyu-Hyun Chae, Harry Desmond}

\author{Kyu-Hyun Chae}
\affil{Department of Physics and Astronomy, Sejong University, 209 Neungdong-ro Gwangjin-gu, Seoul 05006, Republic of Korea}
\email{KHC: chae@sejong.ac.kr, kyuhyunchae@gmail.com}

\author{Harry Desmond}
\affil{Astrophysics, University of Oxford, Denys Wilkinson Building, Keble Road, Oxford, OX1 3RH, UK}
\email{HD: harry.desmond@physics.ac.uk}

\author{Federico Lelli}
\affil{INAF, Arcetri Astrophysical Observatory, Largo Enrico Fermi 5, I-50125 Florence, Italy}
\email{FL: federico.lelli@inaf.it}

\author{Stacy S. McGaugh}
\affil{Department of Astronomy, Case Western Reserve University, Cleveland, OH 44106, USA}
\email{SSM: ssm69@case.edu}

\author{James M. Schombert}
\affil{Department of Physics, University of Oregon, Eugene, OR 97403, USA}
\email{JMS: schombe@gmail.com}
\begin{abstract}
Theories of modified gravity generically violate the strong equivalence principle, so that the internal dynamics of a self-gravitating system in free fall depends on the strength of the external gravitational field (the external field effect). We fit rotation curves (RCs) from the SPARC database with a model inspired by Milgromian dynamics (MOND), which relates the outer shape of a RC to the external Newtonian field from the large-scale baryonic matter distribution through a dimensionless parameter $e_{\rm N}$. We obtain a $>4\sigma$ statistical detection of the external field effect (i.e.\ $e_{\rm N}>0$ on average), confirming previous results. We then locate the SPARC galaxies in the cosmic web of the nearby Universe and find a striking contrast in the fitted $e_{\rm N}$  {values} for galaxies in underdense versus overdense regions. Galaxies in an underdense region between 22 and 45 Mpc from the celestial axis in the northern sky have RC fits consistent with $e_{\rm N}\simeq0$, while those in overdense regions adjacent to the CfA2 great wall and the Perseus-Pisces supercluster return $e_{\rm N}$ that are a factor of two larger than the median for SPARC galaxies. We also calculate independent estimates of $e_{\rm N}$ from galaxy survey data and find that they agree with the $e_{\rm N}$ inferred from the RCs within the uncertainties, the chief uncertainty being the spatial distribution of baryons not contained in galaxies or clusters.
\end{abstract}


\section{Introduction} \label{sec:intro}

General relativity and its Newtonian limit are unique in that the internal gravitational dynamics of a self-gravitating system is unaffected by a constant external gravitational field \citep{Will2014}. This strong equivalence principle (SEP) is generally violated by Milgromian dynamics (MOND; \citealt{Mil1983}), which alters Newton's laws at low accelerations to explain the dynamics of galaxies without resorting to dark matter (DM). Specifically, MOND violates local position invariance for gravitational experiments because it is non-linear even in the non-relativistic regime, so an external acceleration is expected to affect the internal dynamics of a system. This is known as the external field effect (EFE). The most prominent observable consequence of the MOND EFE is that, whereas isolated galaxies should have asymptotically flat rotation curves (RCs), galaxies in dense environments should have declining RCs at large radius. Thus, the EFE produces a smoking gun signature of MOND as contrasted with Newtonian dynamics plus DM, as already emphasized in the first paper proposing MOND \citep{Mil1983}. 

Non-relativistic field theories of gravity embodying the MOND EFE include the AQUAdratic Lagrangian (AQUAL) theory \citep{BM1984} and the quasi-linear MOND (QUMOND) theory \citep{Mil2010}. Relativistic theories, which reduce to either AQUAL or QUMOND in the non-relativistic limit, are under active development (see \citealt{FM2012,Mil2014} for reviews). In particular, the proposal of \citet{Skor2020} appears particularly promising as it can reproduce the angular power spectrum of the cosmic microwave background as well as the linear matter power spectrum.

There have been several attempts to test the EFE empirically \citep{McGaugh_Milgrom, McGaugh_Milgrom_2, Hees, Haghi2016, Wu_Kroupa, Famaey_DF2}, all of which find better agreement with galaxy dynamics when the EFE is included. Recently, \cite[][hereafter Paper~I]{Chae2020b} studied the EFE in 153 rotationally supported galaxies from the Spitzer Photometry and Accurate Rotation Curves (SPARC) database \citep{Lel2016}. They presented both individual and statistical EFE detections from fitting RCs with an AQUAL-inspired function, which contains the MOND external field strength $g_{\rm e}$ as a free parameter (see also \citealt{Chae2021}). They also showed that the median $g_{\rm e}$ agrees with an independent estimate of the external field strength from the galaxies' large-scale environments. This is a surprising result from the DM perspective because the observed decline in the RCs occurs well within the tidal radius. \pap, however, estimated the environmental external fields under the $\Lambda$ cold dark matter (CDM) paradigm using DM halos as surrogates of the MOND effects; this was expedient because $g_{\rm e}$ from the large-scale environment cannot be easily computed in AQUAL due to its non-linearity. 

Here we use a formalism that quantitatively relates the decline of a galaxy's RC to the strength of the \emph{Newtonian} external field $g_{\rm Ne}$ at its position. As $g_{\rm Ne}$ is sourced linearly, it can be calculated by summing the Newtonian fields from the large-scale distribution of baryons around the galaxy. Thus, the EFE can be tested directly with the observable Newtonian external field, and the baryonic content of the nearby Universe can be studied as a byproduct. The detailed calculation of the Newtonian external gravitational field produced by the baryonic large scale structure, for direct comparison with the fitted values, is the key advance of this work over \pap. 

In Section~\ref{sec:method} we describe our formalism and RC-fitting method, with technical details given in Appendix~\ref{sec:form}. In Section~\ref{sec:env} we describe our method for estimating the environmental Newtonian fields, first with an approximate all-sky calculation (Section~\ref{sec:method_allsky}) and then with a more detailed calculation within the SDSS footprint (Section~\ref{sec:method_sdss}). We present our results in Section~\ref{sec:res} and discuss their implications in Section~\ref{sec:disc}. In Appendix~\ref{sec:app}, we provide tables of the fitted parameters and the estimated Newtonian environmental field strengths.

\section{Formalism and Fitting Method} \label{sec:method}

Our goal is to infer the external Newtonian field directly from the RCs  {of galaxies to} compare  {it} with the Newtonian field estimated from their cosmic environment, so we need an EFE model that is parameterized by  $e_{\rm N}\equiv g_{\rm Ne}/a_0$, where $a_0= 1.2\times 10^{-10}$ m~s$^{-2}$ is the MOND acceleration scale \citep{MLS2016,Lel2017,Li2018,Chae2020a,Li2021}. In \pap\ we used an AQUAL-based model that was parameterized by $e \equiv g_\text{e}/a_0$ where $g_e$ is the MOND external field. In this model, the expected acceleration of a test particle ($V^2/R$ for circular motion) is given by:
\begin{equation}
    g = \nu_e (g_\text{N}/a_0) g_\text{N},
    \label{eq:rare}
\end{equation}
where $g_\text{N}$ is the Newtonian internal gravitational field (sourced by the galaxy itself) and $\nu_e (y)$ is given by Equation~(6) of \pap\ with $y \equiv g_\text{N}/a_0$ (see also Appendix~\ref{sec:form}). This equation is based on the simple interpolating function \citep{FB2005,Chae2019} and the analytic solution of the AQUAL equations in one dimension \citep{FM2012}.

 In the QUMOND formalism, one can analytically derive the ratio $g/g_\text{N}$ as a function of \eN\ in one dimension. Incidentally, as shown in Appendix~\ref{sec:form}, this ratio is equivalent to Equation~(6) of \pap\ with $e^2/(1+e) = e_\text{N}$ for non-negative values of $e_\text{N}$. Thus, Equation~(\ref{eq:rare}) holds with the replacement of $\nu_e(y)$ with the following
\begin{equation}
  \nu_{e_{\rm N}}(y) = \frac{1}{2} + \sqrt{\frac{1}{4}D_{e_{\rm N}}^2(y)+ \frac{D_{e_{\rm N}}(y)}{y}} - \frac{e_{\rm N}}{\sqrt{|e_{\rm N}|}} \frac{C_{e_{\rm N}}}{y},
  \label{eq:nueN}
\end{equation}  
where $C_{e_{\rm N}}\equiv \sqrt{1+|e_{\rm N}|/4}$ and $D_{e_{\rm N}}(y)\equiv 1+ {|e_{\rm N}|}/{y}$. Note that $e_{\rm N}<0$ is unphysical (similarly to $e<0$), but we allow it because the measured RCs may prefer negative values empirically. For typical $e_\text{N}<0.01$ and the acceleration range probed by galactic RCs, the last term of Equation~(\ref{eq:nueN}) mainly controls the EFE  {and scales almost linearly with $e_\text{N}/\sqrt{|e_\text{N}|}$}. Thus, we define the fitting parameter $\tilde{e}\equiv e_\text{N}/\sqrt{|e_\text{N}|}$. Our notations are summarized in Table~\ref{tab:notation}.

 \begin{table}
\caption{Summary of notations}\label{tab:notation}
\begin{center}
  \begin{tabular}{cc}
  \hline
 Notation & Meaning \\
 \hline
 $g_{\rm Ne}$ & Newtonian external field strength \\
 $g_{\rm Ne,fit}$ & $g_{\rm Ne}$ from fitting the RC \\
 $g_{\rm Ne,env}$ & $g_{\rm Ne}$ from the environment \\
  $e_{\rm N}$ & $g_{\rm Ne}/a_0$ \\
  $e_{\rm N,fit}$ & $g_{\rm Ne,fit}/a_0$ \\
  $e_{\rm N,env}$ & $g_{\rm Ne,env}/a_0$ \\
  $\tilde{e}$ & $e_{\rm N}/\sqrt{|e_{\rm N}|}$ \\
  $\tilde{e}_{\rm fit}$ & $e_{\rm N,fit}/\sqrt{|e_{\rm N,fit}|}$ \\
  $\tilde{e}_{\rm env}$ & $\sqrt{e_{\rm N,env}}$ \\
 \hline
\end{tabular}
  Note. (1) The subscript `fit' or `env' may be dropped when it is obvious from the context. 
 \newline (2) The RC-fitted quantities allow negative values. 
 
\end{center}
\end{table}

 Values of \et\ for SPARC galaxies may be obtained simply from $e$ of \pap\ through an analytically extended transformation such as $\tilde{e}=e/\sqrt{1+e}$ to include negative values. We prefer, however, to obtain values of \et\ directly from fitting Equation~(\ref{eq:nueN}) to the RCs. Our procedure of obtaining probability density functions (PDFs) of \et\ and galactic parameters is the same as that of \pap. The galactic parameters are mass-to-light ratio(s) of the disk (and the bulge if present), total gas-to-hydrogen mass ratio, distance to the galaxy normalized by the SPARC reported value, and inclination of the disk.

We also revise our selection to maximise the sample size. From the SPARC database of 175 rotationally-supported galaxies within distance $D \lesssim 120$~Mpc, the RCs of 163 galaxies have good or acceptable quality ($Q=1$ or $2$). Out of these, 10 galaxies have low measured inclination angles of $< 30^\circ$. \pap\ excluded these low-inclination galaxies for their analysis following \cite{MLS2016} and thus used only 153 galaxies. Here we use all 163 galaxies including the low-inclination ones for the purpose of enlarging the sample and as our MCMC procedure allows the inclination angle to vary freely. However, including or excluding low-inclination galaxies does not affect our general results.

Examining all posterior PDFs, we find that for one galaxy (UGC~06787) the PDFs are double-peaked and the posterior distance is an order of magnitude larger than the SPARC reported value ($D=21\pm5$ Mpc). Its RC (from \citealt{Noordermeer2007}) has an unusual sinusoidal shape that is hard to fit in any context, either with MOND \citep{Li2018} or with DM halos \citep{Li2020}. The galaxy has a strongly warped HI disk, so the RC shape depends on the details of the warp modeling (the adopted radial variation of position angle and inclination). Considering the unreliable value of the best-fit distance (and probably of \et), we drop this galaxy because we aim to compare the external Newtonian field with the location of galaxies on the large-scale structure of the Universe. Thus, our analysis is based on  {a starting sample of} 162 SPARC galaxies.  As illustrated in Figure~4 of \pap\ and discussed further in Appendix~\ref{sec:app}, only RCs that reach very low accelerations can probe the EFE. We select 143 galaxies as described in Appendix~\ref{sec:app}. The fitted \et\ values of these galaxies are compared with the cosmic large-scale structure.

\begin{figure}
  \centering
  \includegraphics[width=1.03\linewidth, trim=0mm 20mm 0mm 15mm, clip]{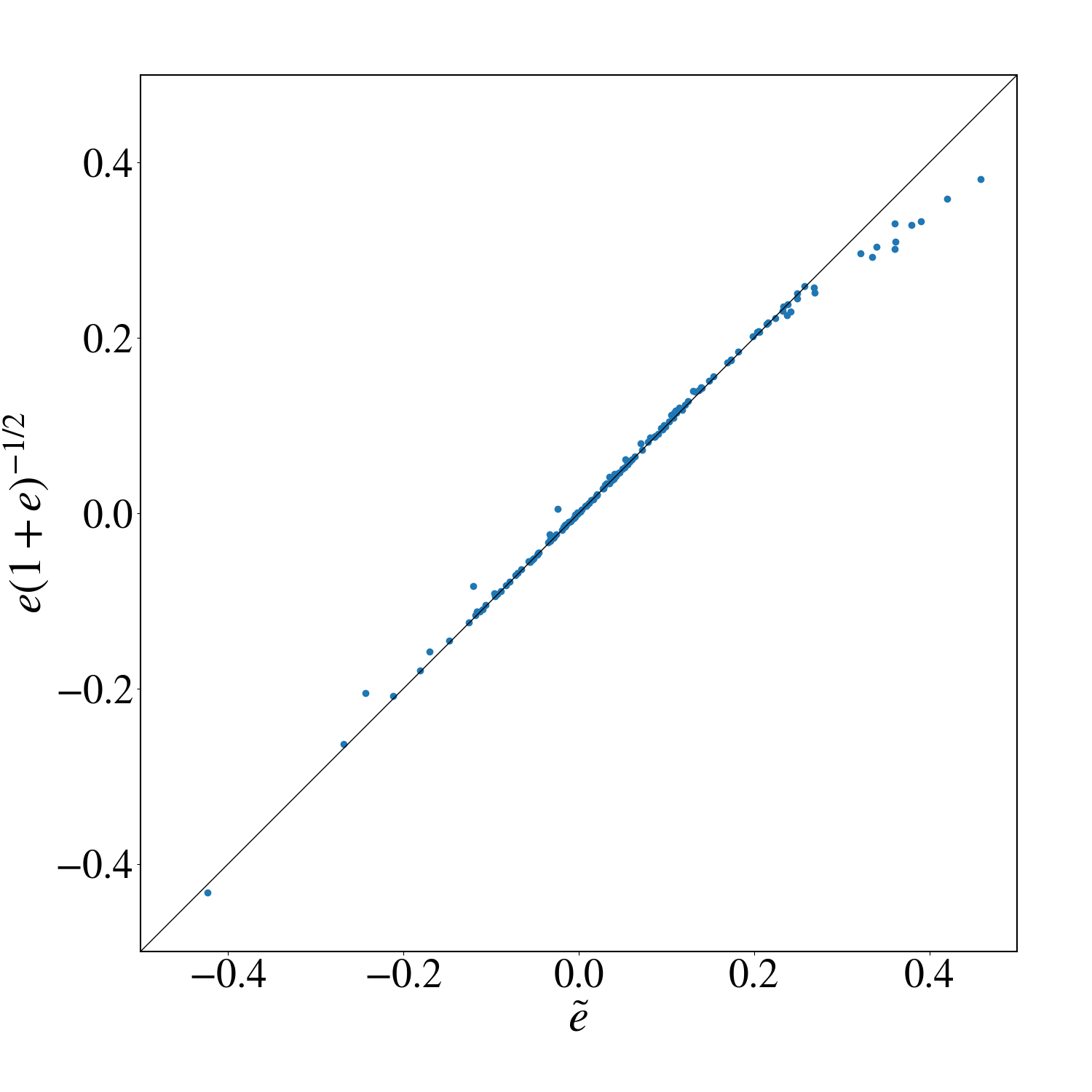}
    \vspace{-0.5truecm}  
        \caption{\small  Fitted \et\ values based on Equation~(\ref{eq:nueN}) from this work are compared with $e$ values based on Equation~(\ref{eq:nue}) from \pap\ for 152 overlapping galaxies.
    } 
  \label{ete}
\end{figure}

Considering the equivalence of Equation~(\ref{eq:nueN}) and Equation~(\ref{eq:nue}) with $e/\sqrt{1+e}=\tilde{e}$, we expect the fitted value of \et\ from this work to agree with $e/\sqrt{1+e}$ from \pap. Figure~\ref{ete} shows that there is a good match between the two for all galaxies except some at large \et.

We stress that our work is not specific to a particular MOND theory because Equation~(\ref{eq:nueN}) is valid in both AQUAL and QUMOND for the idealized one-dimensional case. In Appendix~\ref{sec:form}, we show that Equation~(\ref{eq:nueN}) is a good approximation to numerical calculations in the AQUAL context. More generally, Equation~(\ref{eq:nueN}) can be viewed simply as a fitting function to characterise non-flat outer RCs; then the extra free parameter $e_{\rm N}$ is expected to be equal to the external Newtonian gravitational field (in units of $a_0$) only if the MOND paradigm is essentially correct.

\section{The Newtonian field from baryonic large scale structure} \label{sec:env}
 
We estimate the environmental Newtonian field \genv\ by summing the contributions from baryonic mass in various forms in the nearby Universe. In MOND, gravitational collapse is expected to become strongly non-linear at high $z$ (e.g.\ \citealt{McGaugh2015}), so the matter power spectrum at $z\simeq0$ cannot be estimated analytically; one would need large-scale hydrodynamical simulations in a fully-fledged MOND cosmology. We therefore compile a census of observed baryonic structures to estimate their Newtonian accelerations directly. We use four main catalogues: the 2M++ galaxy catalogue \citep{LH2011}, the MCXC galaxy clusters catalogue \citep{MCXC}, the Karachentsev galaxy catalogue \citep{Kar1, Kar2} and the NASA-Sloan Atlas (NSA)\footnote{\url{https://www.sdss.org/dr13/manga/manga-target-selection/nsa/}} of galaxies. 2M++ and MCXC are used for the approximate all-sky calculation of \genv\ (Section~\ref{sec:method_allsky}) while MCXC, Karachentsev and the NSA are used for the more detailed calculation within the SDSS footprint (Section~\ref{sec:method_sdss}). 

\subsection{An all-sky estimate of $g_\text{Ne,env}$}
\label{sec:method_allsky}

We begin with an approximate all-sky calculation of \genv\ to give a sense of its magnitude and investigate its angular variation. This is based primarily on 2M++ \citep{LH2011}, an all-sky catalogue comprising the Two-Micron-All-Sky-Survey (2MASS) Redshift Survey (2MRS), the Six-Degree-Field Galaxy Redshift Survey Data Release 3 (6dFGRS-DR3), and the Sloan Digital Sky Survey Data Release 7 (SDSS-DR7). The catalogue reaches a depth of $K_S \le 11.5$ over the full sky, and $K_S \le 12.5$ in the 6dF and SDSS regions. The Galactic plane Zone of Avoidance is populated with mock galaxies by cloning large scale structure just outside the Zone to reproduce both 1- and 2-point statistics of the density field. In view of the approximate nature of this preliminary calculation, we estimate distances from the galaxy's recession velocity in the rest frame of the Cosmic Microwave Background assuming a Hubble constant of $H_0=73$ km~s$^{-1}$~Mpc$^{-1}$ (as assumed in SPARC, \citealt{Lel2016}). We estimate stellar masses assuming a mass-to-light ratio of 0.64 in the $K_S$ band (consistent with 0.5 in the Spitzer [3.6] band as assumed in SPARC; \citealt{McGaugh_Schombert}). We follow the prescription of \citet{LH2011} for relating apparent and absolute magnitude. We remove sources with $K_S>11.5$ which are preferentially seen in the deeper 6dF and SDSS fields and cut the catalogue at 200 Mpc. This leaves 54,483 galaxies out of 69,160 in total.

\subsubsection{Adding the intra-cluster medium}\label{sec:ICM}

We also consider the hot intra-cluster medium of galaxy clusters. MCXC is a meta-catalogue of 1743 X-ray detected clusters from 12 separate catalogues, many of which derive from the ROSAT satellite \citep{ROSAT}. The catalogue provides redshift, coordinates, X-ray luminosity $L_{500}$ in the range 0.1-2.4 keV standardised between component catalogues at a critical overdensity of 500, and the total mass $M_{500}$. This $\Lambda$CDM-based mass is not suitable for our purposes here. In MOND, the mass that sources \genv\ is simply the total baryonic mass. The observed baryonic masses of galaxy clusters, however, systematically disagree with the MOND-predicted total masses (from X-ray hydrostatic equilibrium) by a factor $\sim$2 \citep{Sanders1999}. Recent X-ray data show a smaller discrepancy \citep{Ettori2019} but it remains clear that MOND has a missing mass problem in cluster centres. This discrepancy may be driven by undetected baryons (such as very cold gas, e.g. \citealt{Mil2008}), massive sterile neutrinos \citep{Angus}  {or a gravitating scalar field \citep{Skor2020}}. In the following, we will estimate the MOND dynamical mass and use it to calculate $g_{\rm Ne, env}$ regardless of the nature of the missing cluster mass.

We begin with a scaling relation between $L_{500}$ and the observed mass of X-ray-emitting gas ($M_X$) from an X-ray flux-limited sample of 62 clusters (HIFLUGCS; \citealt{HIFLUGCS}):
\begin{equation}\label{eq:L500-Mx}
M_X = \frac{10^{14} M_\odot}{E(z)} 10^{(\log(L_{500}/E(z))-A)/B},
\end{equation}
where $A=44.44$, $B=1.11$ and $E(z) = 0.3(1+z)^3+0.7$. We then build a scaling relation between the observed gas mass and the MOND dynamical mass ($M_{\rm MOND}$) using the results of the hydrostatic equilibrium analysis of \citet{Angus} for 26 X-ray emitting clusters,
\begin{equation}
\log(M_\text{\rm MOND}/M_\odot) = C + D \log(M_X/M_\odot),
\end{equation}
finding $C=3.814$ and $D=0.728$ to provide a good fit to their data.

The results for the clusters' masses and hence \genv\ are not significantly altered if we use the $L_{500}-M_\text{X}$ relation of \citet{WtG} instead of Equation~(\ref{eq:L500-Mx}), or if we use the results of \citet{Angus} to convert $M_{500}$ from MCXC directly to $M_\text{MOND}$. The latter approach has two disadvantages: first it is more sensitive to the $\Lambda$CDM-driven choice to measure the mass within a critical overdensity of 500, and second it involves an implicit $L_X-M_{500}$ scaling relation assumed by the MCXC team. $M_\text{MOND}$ is a factor of several lower than $M_{500}$, so it is important not to use $M_{500}$ directly.

\subsubsection{Total gravitational field}

We calculate \genv\ by linearly summing the \gN\ values of each of the objects in the combined catalogue. For galactic sources -- and clusters beyond $R_{500}$ -- we treat the sources as point objects because the distances between the source and test points typically greatly exceed the dimensions of the source objects themselves. In case a test point is a distance $d<R_{500}$ from the centre of a cluster, we scale the \genv\ contribution of the cluster by a factor $(d/R_{500})^3$, i.e. approximating the cluster density as uniform. This does not appreciably affect the results. As our primary interest here is in the angular variation of \genv, we calculate it on a \texttt{healpix} \citep{hp} grid of \texttt{nside}=54 ($\Delta$RA, $\Delta$Dec $\simeq1\deg$) every 1.5 Mpc in line-of-sight distance out to 150 Mpc. To remove numerical artifacts we exclude source objects within 10 kpc of a test point when evaluating \genv\ at that point.

As we discuss in detail below, many sources are neglected in this method so that \genv\ is underestimated on average. This underestimation is not however expected to be a strong function of sky position, so the calculation should be sufficient to estimate the relative variation of \genv\ with \{RA, Dec\}.

\subsection{A more precise estimate within the SDSS footprint}
\label{sec:method_sdss}

We can calculate \genv\ in greater detail within the SDSS footprint, where deeper photometry is available. For this purpose we use the NSA, a value added catalogue based primarily on SDSS data. We calculate \genv\ within two regions defined by $\{128 < \text{RA/deg} < 230, 0 < \text{Dec/deg} < 60\}$ and $\{-25 < \text{RA/deg} < 25, -9 < \text{Dec/deg} < 30\}$. We use \textsc{elpetro\_mass} to estimate stellar mass and \textsc{zdist} with $H_0 = 73$ km~s$^{-1}$~Mpc$^{-1}$ to estimate distances. We only cut sources with $D>500$ Mpc, to ensure that we include sufficient long-range contributions to \genv. The intra-cluster medium is added in the same way as in Section~\ref{sec:ICM}. In addition, we develop a probabilistic Monte Carlo framework to propagate uncertainties in various input quantities into \genv, as described in more detail below.

In the Local Volume ($D<11$ Mpc), we replace the NSA with the catalogue of \citet{Kar1} which is complete above a $B$-band absolute magnitude of about $-$12 mag and therefore contains fainter galaxies across the sky. It also provides both $K$-band luminosities and H{\small I} masses. We estimate stellar masses from $K$-band luminosities, again adopting $M/L=0.64$, and cold gas masses from H{\small I} masses using the molecular gas correction of \citet{McGaugh_MolecularGas}:
\begin{equation}
\log(M_\text{gas}/M_\odot) = \log(M_\text{HI}/M_\odot + 0.07 M_*/M_\odot) + \log(1.38).
\end{equation}
If only an upper limit on $M_\text{HI}$ is recorded in the catalogue, then we estimate it from a scaling relation with $M_\star$ (see Section~\ref{sec:gas}), with a cap at $0.8 \: M_\text{HI,max}$ (this choice does not affect \genv\ appreciably). The final Karachentsev catalogue contains 1109 galaxies to a sufficiently low mass that a correction for galaxies below the completeness limit is unnecessary.

\subsubsection{Adding galaxies below the SDSS detection limit}
\label{sec:mocks}

The SDSS is limited to $m_r < 17.77$ ($m_r < 17.6$ conservatively), so galaxies of increasing stellar mass are progressively absent from the NSA at larger distances. If not corrected for, this would cause the average value of \genv\ to fall with distance. To mitigate this selection bias, we add mock galaxies too faint to be measured by the survey. We use the triple-Schechter fit to the \citet{Li_White} stellar mass function (SMF)\footnote{This agrees well with the elliptical Petrosian stellar mass function of the NSA \citep{Stiskalek}.} to model stellar masses for mock galaxies. In particular, we take 80 logarithmically uniform bins in stellar mass $M_\star$ between $10^5$ and $10^{13} M_\odot$, and 50 linear bins in distance $D$ between 11 and 500 Mpc. In each $\{M_\star, D\}$ bin we use the SMF to calculate the expected number of galaxies. We then assign absolute $r$-band magnitudes to these mock galaxies using the scaling relation
\begin{equation}
M_r = E \log(M_\star/M_\odot) + F,
\end{equation}
with $E=-1.991$, $F=-0.265$ and a Gaussian scatter in $M_r$ of 0.385 dex, fitted directly from the NSA data, and convert to apparent magnitude in the same manner as for 2M++. We conservatively consider the galaxies visible to SDSS to be those with $m_r<17.6$. This lets us calculate the average fraction of the total stellar mass that is included in each radial bin when using the NSA catalogue directly; the remaining mass must now be added back in so that \genv\ is not biased low at large $D$. Ideally a similar completeness correction would be applied to the galaxy cluster catalogue. We do not attempt this here because the cluster mass function is not well characterised, but note that the higher intrinsic brightness of galaxy clusters makes them visible to much larger distances than individual galaxies, so we expect the problem to be less severe.

While the above calculates the 1-point function of the missing mass, its 2-point function (i.e. clustering) is also relevant to the \genv\ that it produces. To model this we consider the two most extreme scenarios: we assume either that the mock galaxies are randomly distributed in space within each distance annulus (i.e. unclustered), or that they are satellites of the galaxies included in the NSA so that they may be considered coincident with them (i.e.\ maximum clustering). These models bracket the full range of possible clustering strengths; we will refer to them as the ``no clustering'' and ``max clustering'' models respectively.

For the ``max clustering'' method we multiply the mass of each NSA galaxy by the reciprocal of the mass fraction calculated above for the corresponding distance bin, so that the correct total amount of mass is modelled in each bin. For the ``no clustering'' method we explicitly add mock galaxies with $m_r>17.6$ to our source catalogue before calculating \genv; this is done separately for each mass and distance bin described above, and the galaxies are given random positions within the bin. This primarily increases the uncertainty in \genv\ when the random positions of the mock galaxies are marginalised over by Monte Carlo sampling, while the max clustering method primarily boosts \genv. We find this boost to \genv\ to be $\sim$0.1 dex, and hence not negligible. On the other hand, altering the widths of the mass or distance bins, the functional form of the SMF (within reasonable limits), or the minimum stellar mass down to which mock galaxies are included has a negligible effect on the results. Note that these methods for adding undetected galaxies preserve any over- or under-densities in galaxy number counts within the NSA as a function of distance, allowing them to manifest as a distance or other environment-dependence of $g_\text{Ne,env}$.

\subsubsection{Adding gas to galaxies}\label{sec:gas}

Galaxies contain appreciable amount of gas, either ``cold'' ($T\lesssim10^{4}$ K) gas, residing in star-forming atomic or molecular disks, or ``hot'' ($T\gtrsim10^5$ K) gas, residing in X-ray emitting ionised halos. The gas content depends on both stellar mass and morphology, with low-mass late-type galaxies having a larger fraction of cold gas and high-mass early-type galaxies having a larger fraction of hot gas.

To model the gas content of our source galaxies, we begin by estimating the galaxy type as a function of its stellar mass. From \cite{Henriques}, the early-type fraction is given by $f_\text{early} \simeq 0.28\log(M_\star/M_\odot) - 2.12$. We use this as a probability to assign an early-type or late-type flag to each source galaxy, separately within each Monte Carlo realisation of the model so that the uncertainty it induces is naturally propagated into $g_\text{Ne,env}$.  For late-type galaxies, the gas mass is given by
\begin{eqnarray}\label{eq:a_nfw}
M_\text{g,cold} = 11500 \: (M_\star/M_\odot)^{0.54} + \: 0.07 \: M_\star
\end{eqnarray}
where the first and second terms in the sum consider, respectively, atomic \citep{Lel2016} and molecular \citep{McGaugh_MolecularGas} gas. For early-type galaxies, the gas mass is instead given by
\begin{equation}
\log(M_\text{g,hot}/M_\odot) = 1.47 \log(M_\star/M_\odot) - 5.414.
\end{equation}
We derived this scaling relation considering 94 early-type galaxies with hot gas masses from the \textit{Chandra} X-ray observatory \citep{Babyk} and stellar masses from our own WISE photometry (following the same procedures as in \citealt{Lel2016}), adopting $M_\star/L_{W1}=0.7$ as appropriate for early-type galaxies \citep{Schom2019}. We take a scatter in both $\log(M_\text{g,cold}/M_\odot)$ and $\log(M_\text{g,hot}/M_\odot)$ of 0.2 dex, and add gas masses in the same way to the faint mock galaxies of Section~\ref{sec:mocks}.

\subsubsection{Adding baryonic mass outside the SDSS footprint}

Thus far, our catalogue of \genv\ sources is all-sky within 11 Mpc but restricted to the SDSS footprint beyond that. Since gravity is a long-range force, this will lead to an underestimation of \genv\ on average at $D>11$ Mpc and a spurious dependence of \genv\ and its direction on sky position within the NSA footprint. To mitigate this effect, we add a uniform grid of mass outside the SDSS footprint out to 500 Mpc. We calculate the density of this grid by integrating the SMF over all masses, including the corrections for cold and hot gas as a function of stellar mass as described above, as well as adding in the average density of cluster mass. This gives $\rho_\text{tot} = 10^{8.549} \: M_\odot \: \text{Mpc}^{-3}$. We use a Cartesian grid of spacing 5 Mpc in each direction, so that each grid cell has a mass of $4.4\times10^{10} M_\odot$. To soften its \gN\ contribution, we assume this mass to have a spherical Gaussian density profile with width $\sigma=1.25$ Mpc, so that the gravitational field it sources at a displacement $\vec{r}$ from its centre is
\begin{equation}
\vec{g}_\text{N}(\vec{r}) = - \frac{GM}{r^2} \left(\text{erf}\left(\frac{r}{\sqrt{2} \sigma}\right) - \sqrt{\frac{2}{\pi}} \frac{r}{\sigma} \exp\left(-\frac{r^2}{2 \sigma^2}\right)\right) \: \hat{r}
\end{equation}
This can have a $\sim$0.2 dex effect on $g_\text{Ne,env}$ and tends to reduce variation in \genv\ with distance due to the homogeneity of the grid mass. The result is not significantly altered if the grid spacing is reduced to 2.5 Mpc, the SMF of \citet{Li_White} is replaced by that of \citet{GAMA_SMF} or \citet{Bernardi_SMF}, or if the grid cells are uniformly spaced in $D^2$ rather than $D$ so that finer structure is resolved at smaller $D$. This indicates that the manner in which mass is distributed outside the SDSS footprint is not critical for the gravitational field within it, although it is important that the correct mass on average is modelled.

\subsubsection{Adding cosmological missing baryons}

The above method accounts for baryons in each of the forms in which they are readily visible: stars, cold and hot gas in galaxies, and an X-ray-emitting intra-cluster medium. However, there may be ``missing baryons'' not found in one of these forms. This is a necessity in $\Lambda$CDM where $\Omega_b$ is known precisely from the Cosmic Microwave Background, but found to be much larger than that implied by a census of visible mass locally (e.g. \citealt{Shull2012}). The result for $\Omega_b$ is likely to hold in a MOND cosmology as well, where at early times (e.g.\ during Big Bang Nucleosynthesis) the high average density of the Universe makes the MOND modification to gravity ineffective. This is however subject to the considerable uncertainty in MOND cosmology, stemming largely from the unknown relativistic parent theory; see, e.g., \citet{FM2012}. In the case of the relativistic MOND theory of \citet{Skor2020}, we expect Big Bang Nucleosynthesis to work in MOND in nearly the same way as in $\Lambda$CDM, so the missing baryons problem should be analogous in both cosmologies.

We adopt the census of baryons from \citet{Shull2012}, who find that baryons in the forms we explicitly model amount to $\sim \Omega_b/8$ (with $\Omega_b \simeq 0.046$; \citealt{baryoncensus}), meaning that there are potentially $\sim$8$\times$ more baryons than we have included so far. As an additional consistency check on our mass model, we calculate that it produces total densities of all baryonic component within the SDSS footprint that are approximately consistent with the results of \citet{Shull2012}. Similarly to our method for adding faint galaxies, we consider two extreme clustering models for these missing baryons: 1) they are completely unclustered (or equivalently non-existent) so do not alter $g_\text{Ne,env}$, or 2) they are maximally clustered with the structures we model and equally distributed between them, causing a uniform increase in the magnitude of the \genv\ field by a factor of 8. 

Our assumption is that the cosmological missing baryons are too far from the centres of galaxies and clusters to appreciably alter their measured kinematics. This is reasonable even for the ``max clustering'' model because the scales involved in calculating \genv\ ($\gtrsim10$ Mpc) are much larger than those over which kinematics can be probed ($<$Mpc). These missing baryons are therefore fully independent of those potentially invoked in Sec.~\ref{sec:ICM} to explain the dynamical masses of clusters. Nevertheless it is possible that the potential missing baryons inside clusters
may account for some fraction of the cosmological missing baryons, leading us to overestimate \genv\ when multiplying by a factor of 8. This effect is however small ($\sim10\%$), and the max clustering model is intended to give an upper bound on plausible \genv\ values, so that it brackets the possible range along with the ``no clustering'' model. In fact, the difference between $M_X$ and $M_\text{MOND}$ in clusters may be due to neutrinos or a massive scalar field rather than dark baryons.

\subsubsection{Modelling uncertainties}

Finally, it is important to estimate the uncertainty in \genv\ due to uncertainties in our baryonic mass model of the Universe. This includes uncertainties in the scaling relations that we use to determine gas and cluster masses, the locations of mock galaxies above the magnitude limit of SDSS, and the basic properties of the source objects themselves (e.g. distances and stellar masses). Our Monte Carlo framework makes it easy to propagate these uncertainties: at each test point we evaluate \genv\ 2000 times, in each case randomly sampling from the distributions describing the input variables. This is a sufficient number for convergence of the \genv\ PDFs.

We use a 30\% uncertainty in $M_\star$ and $M_\text{\rm MOND}$, describing uncertainties in stellar mass-to-light ratios and the calculation of the MOND dynamical mass of galaxy clusters. We use a 0.2 dex scatter in cold and hot gas masses of galaxies, and 0.4 dex in cluster masses to account for the additional uncertainty in the conversion from X-ray luminosity to hot gas mass. Following \citet{Lel2016}, distance uncertainties are based on the following scheme (suitable for Hubble-flow distances): 30\% for $D<20$ Mpc, 25\% for $20 \le D/\text{Mpc} < 40$, 20\% for $40 \le D/\text{Mpc} < 60$, 15\% for $60 \le D/\text{Mpc} < 80$, and 10\% for $D > 80$ Mpc. 

The remaining uncertainties derive from the assignment of an early-type/late-type flag to each galaxy, and the distribution of faint mock galaxies within each distance bin in the ``no clustering'' model for filling in faint galaxies. We note that the ``no clustering'' model for cosmic missing baryons is different to that for faint mock galaxies in that it does not discretize the missing mass and assign it a random position. The uncertainty induced by cosmic missing baryons is therefore entirely in the systematic difference between the no clustering and max clustering models. We summarise the resulting \genv\ distribution over all Monte Carlo realisations at each test point by its mean and standard deviation.

We evaluate \genv\ at the positions of the Karachentsev and NSA galaxies themselves out to 150 Mpc to assess its trend with environment and distance, and at the positions of the SPARC galaxies for a direct comparison with $g_\text{Ne,fit}$. We exclude from the source catalogue the test galaxy under consideration, either by simply masking out the corresponding galaxy from the source array (for a Karachentsev or NSA galaxy), or, for a SPARC galaxy, by using NED to determine if it is in the NSA and an RA/Dec match with tolerance $0.1\deg$ to determine if it is in the Karachentsev catalogue.

\section{Results} \label{sec:res}

\subsection{Qualitative consistency with large-scale structure}
 
\begin{figure*}
  \centering
  \includegraphics[width=1.\linewidth, trim=0mm 0mm 0mm 0mm, clip]{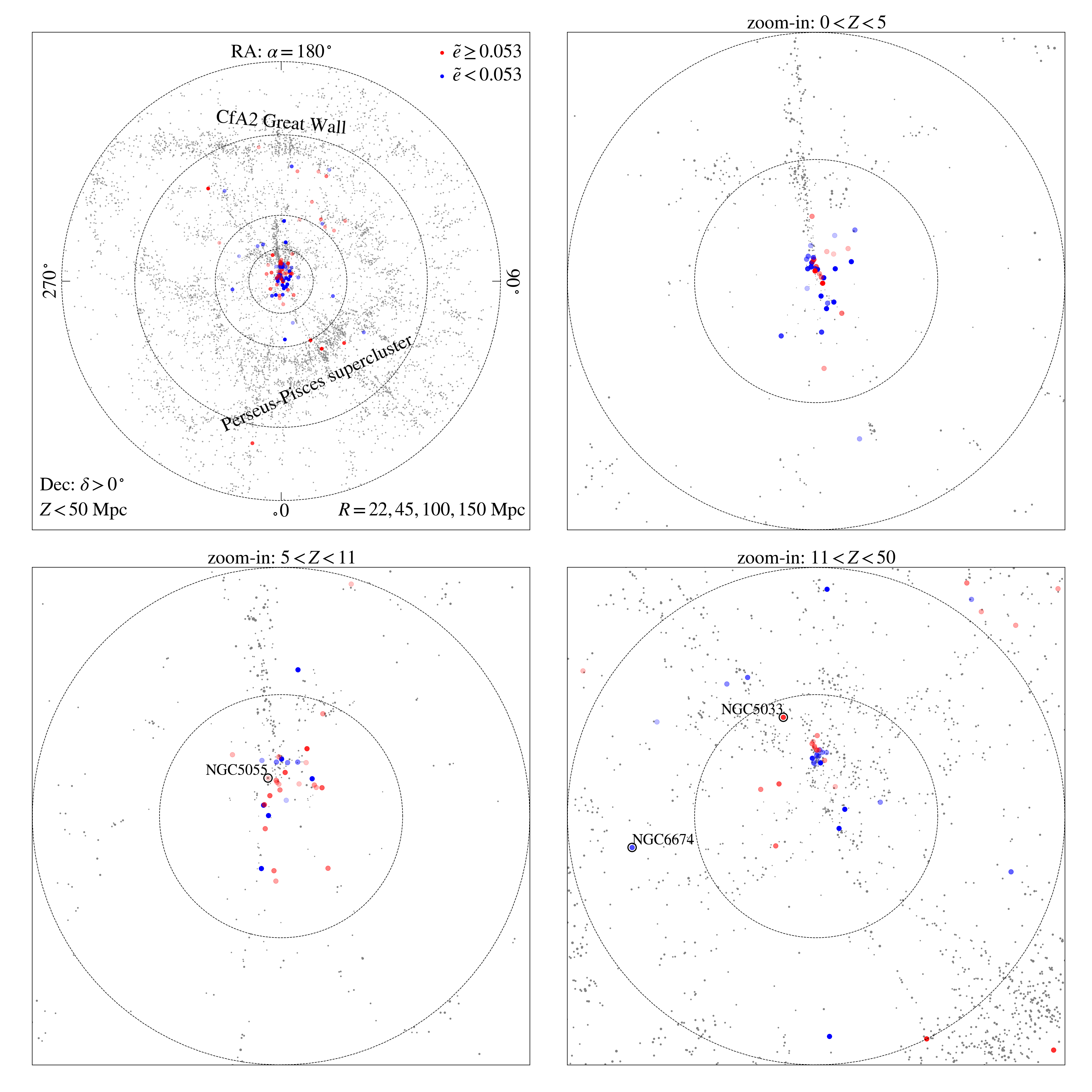}
    \vspace{-0.3truecm}  
    \caption{\small The top left panel shows a map of galaxies in a cylindrical space of the northern sky with a radius of 150~Mpc and a height of 50~Mpc. Galaxies are taken from the 2M++ catalog  (accepting only those brighter than 11.5 in $K$-magnitude to prevent any direction bias) and the area of each point is proportional to the stellar mass of the galaxy. SPARC galaxies with external Newtonian fields from RC fits are indicated by coloured points. External Newtonian fields higher and lower than the median ($\tilde{e}=0.053$) are shown in red and blue respectively, and the opacity of the point is proportional to $|\tilde{e}-0.053|$/(uncertainty of $\tilde{e}$). The central $R<45$Mpc region is sliced and zoomed-in in the other panels. These panels show an overdensity-underdensity contrast in the two opposite directions of RA $\alpha=180^\circ$ and $0^\circ$. Two ``golden'' galaxies and one control galaxy from \pap\ are indicated. The first ring-like region with $R=(22,45)$~Mpc is relatively underdense and the fitted $\tilde{e}$ values are mostly lower in this region. The second ring-like region of $R=(45,100)$~Mpc includes the CfA2 great wall and the Perseus-Pisces supercluster. The fitted $\tilde{e}$ values are mostly higher in this region.
   }
   \label{map}
\end{figure*}

\begin{figure*}
  \centering
  \includegraphics[width=\textwidth]{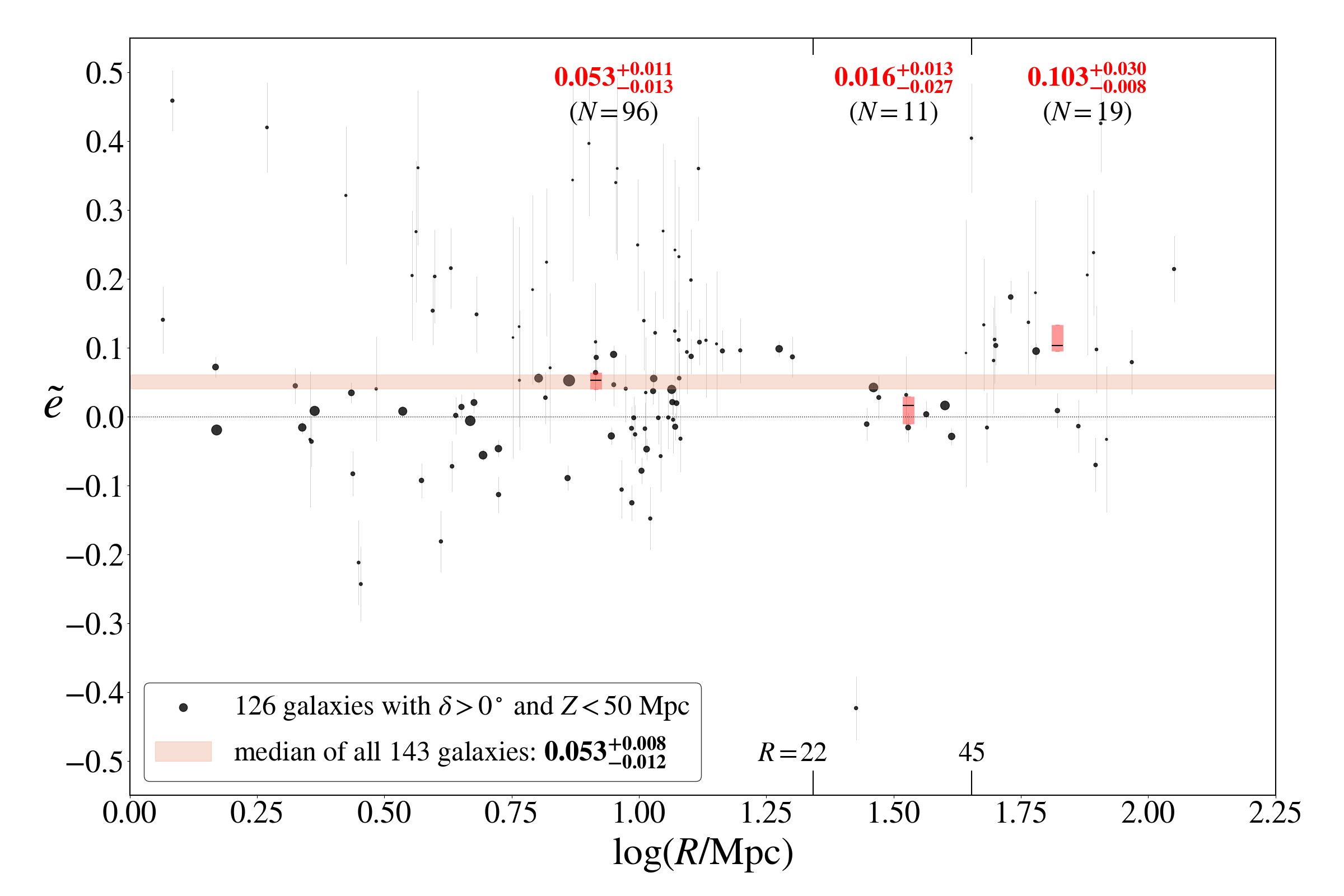}
    \caption{\small RC-fitted values of $\tilde{e}$ versus the cylindrical radius defined in Figure~\ref{map} for 126 SPARC galaxies in the Northern hemisphere. The size of a dot is inversely proportional to the uncertainty on $\tilde{e}$. The light-red band shows the median value and its uncertainty for all galaxies shown in Figure~\ref{eN2}. The red bars show median values (black line) and their uncertainties in three radial bins. The number of galaxies in each radial bin and their median values with bootstrap 1$\sigma$ uncertainties are indicated at the top. $\tilde{e}$ is systematically lower in the underdense second bin and higher in the third bin. 
    } 
  \label{eN2R}
\end{figure*}

We consider a cylindrical space of radius $R = 150$~Mpc and height $Z=50$~Mpc in the northern sky (declination $\delta >0^\circ$), where the vast majority of fitted galaxies (126) are located. In Figure~\ref{map}, the SPARC galaxies are projected on the equatorial plane of this cylindrical space together with galaxies from the 2M++ catalog \citep{LH2011}. At $22 \lesssim R/ {\rm Mpc} \lesssim 45$, there is an under-dense region in which SPARC galaxies tend to give lower $\tilde{e}$ values than the median. At $45 \lesssim R/{\rm Mpc} \lesssim 100$, instead, there are two massive structures: the CfA2 great wall \citep{GH1989} towards Right Ascension $\alpha\simeq180^{\circ}$ and the Perseus-Pisces supercluster \citep{JE1978} towards $\alpha\simeq45^{\circ}$. Several SPARC galaxies are located in these over-dense structures and tend to give higher $\tilde{e}$ value than the median. This seems qualitatively consistent with an EFE interpretation of the $\tilde{e}$ values from RC fits. The situation is more complex at $R<22$ Mpc due to a finger-like structure towards $\alpha=190^{\circ}$, which varies with height and disappears at $Z>11$ Mpc. The precise geometry of this structure may affect the $\tilde{e}$ values in a way that cannot be simply inferred by looking at projected maps. However, note that the two ``golden galaxies'' NGC5033 and NGC5055 highlighted in \pap\ are near the Virgo supercluster, while the control galaxy NGC6674 is in the underdense ring. The other control galaxy NGC1090 is in the southern sky, so it is not included in Figure~\ref{map}.

Figure~\ref{eN2R} shows the distribution of $\tilde{e}$ versus cylindrical radius $R$ for the 126 SPARC galaxies within the cylindrical space. It is evident that galaxies in the underdense region at $22\hspace{1ex}{\rm Mpc}<R<45\hspace{1ex}{\rm Mpc}$ have low values of $\tilde{e}$ consistent with zero environmental fields. The median value within this radial bin is $0.016_{-0.027}^{+0.015}$ (with bootstrap uncertainties), thus consistent with no-EFE detection as expected from the large-scale distribution of galaxies. On the other hand, galaxies in a radial bin at $R>45$ Mpc, in which the CfA2 great wall and the Persus-Pisces supercluster are located, display a median value of $\tilde{e} = 0.103_{-0.008}^{+0.030}$, significantly different from zero at $\sim 13\sigma$. This shows a striking contrast with the middle bin. The majority of the galaxies in the third bin, indeed, are associated with the CfA2 great wall or the Perseus-Pisces supercluster. 

\begin{figure*}
  \centering
  \includegraphics[width=0.49\textwidth]{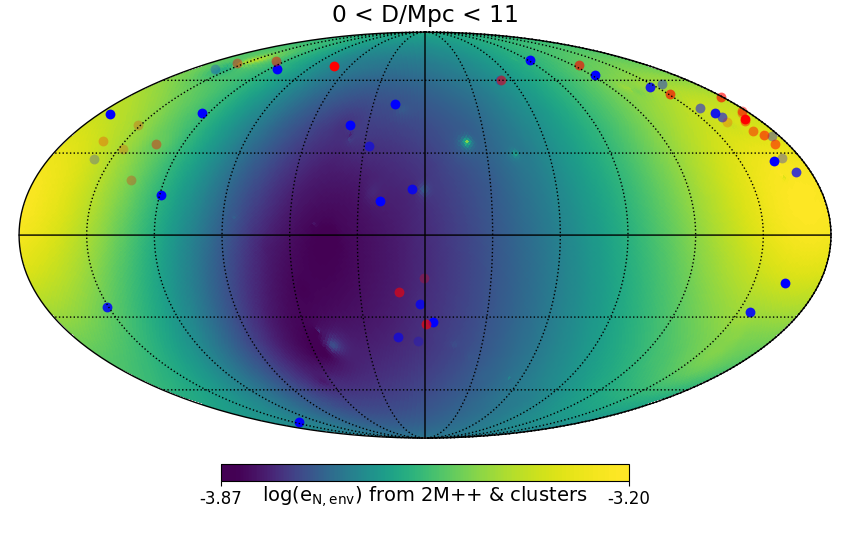}
  \includegraphics[width=0.49\textwidth]{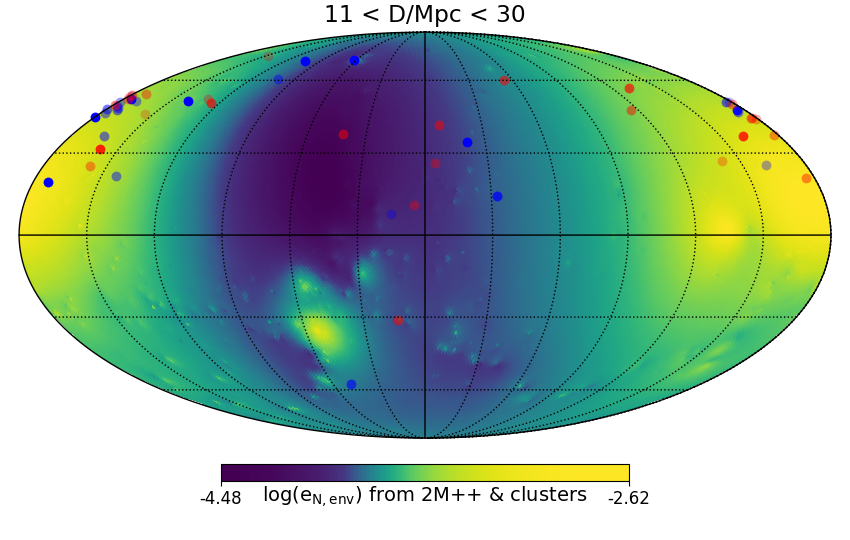}
  \includegraphics[width=0.49\textwidth]{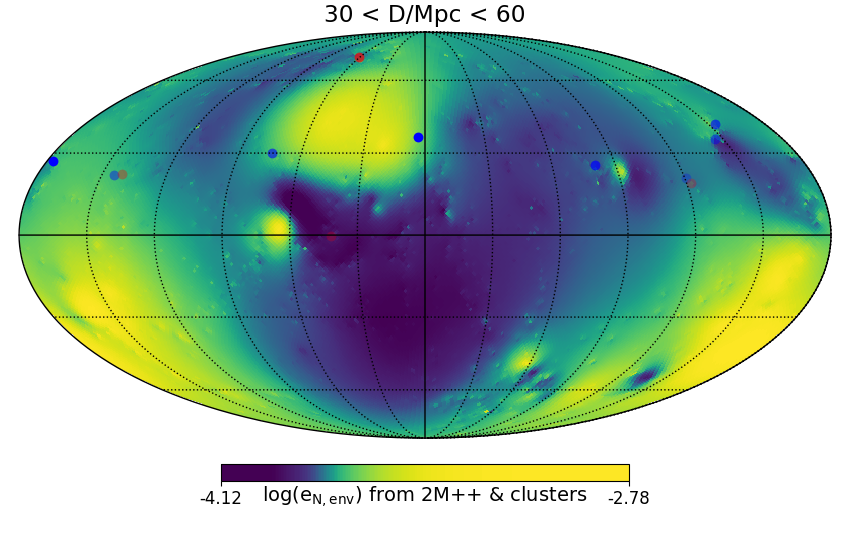}
  \includegraphics[width=0.49\textwidth]{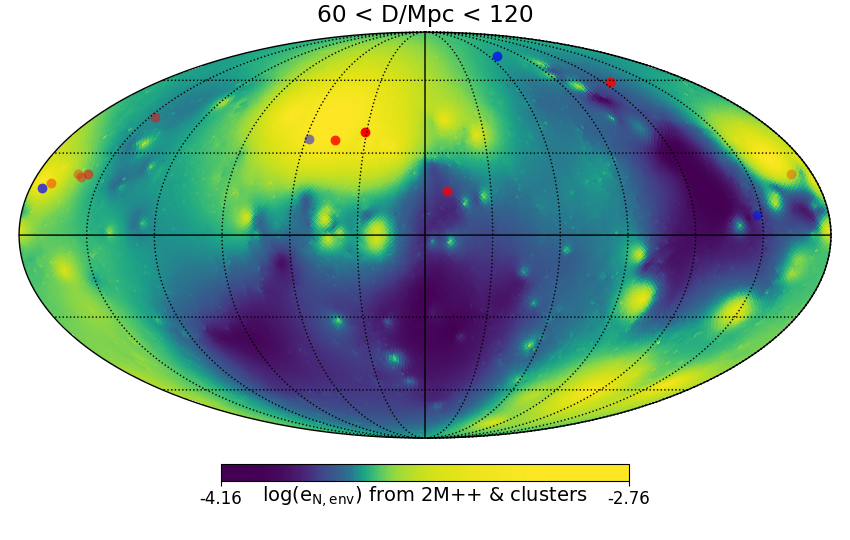}
  \caption{All-sky distributions of the environmental field \genv\ from 2M++ galaxies and MCXC clusters in Mollweide projection and equatorial coordinates averaged across various distance ranges. The location of SPARC galaxies with independent estimates of \gN\ from RC-fits are shown: red and blue points indicate $\tilde{e}_{\rm fit}\ge 0.053$ and $\tilde{e}_{\rm fit}<0.053$, respectively; the opacity of the point increases with $|\tilde{e}_{\rm fit}-0.053|$/(uncertainty in $\tilde{e}_{\rm fit}$) as in Figure~\ref{map}.
   }
  \label{fig:Moll}
\end{figure*}

\begin{figure}
  \centering
  \includegraphics[width=0.52\textwidth, trim=0mm 0mm 0mm 10mm, clip]{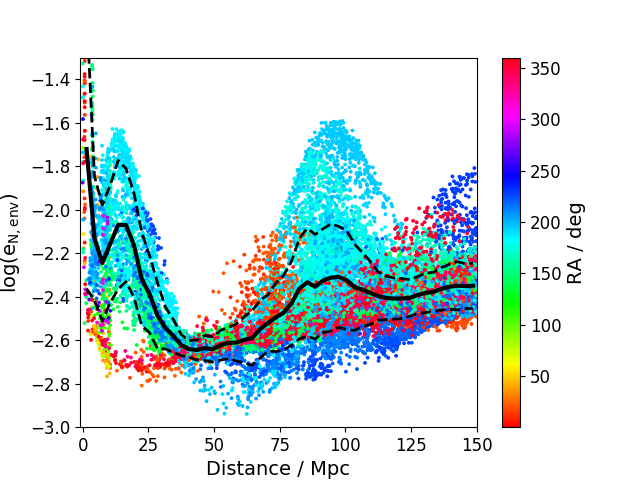}
  \caption{Variation of \eenv with distance for the galaxies in the NSA and Karachentsev catalogs. Individual galaxies are colour-coded by Right Ascension (RA). The black lines show the mean trend (solid) and standard deviation (dashed) in bins of distance. This plot assumes the max clustering model for the missing baryons (cf.\ Figure~\ref{fig:genv_SPARC}).}
  \label{fig:genv_NSA}
\end{figure}

\begin{figure}
  \centering
  \includegraphics[width=0.5\textwidth, trim=0mm 0mm 0mm 10mm, clip]{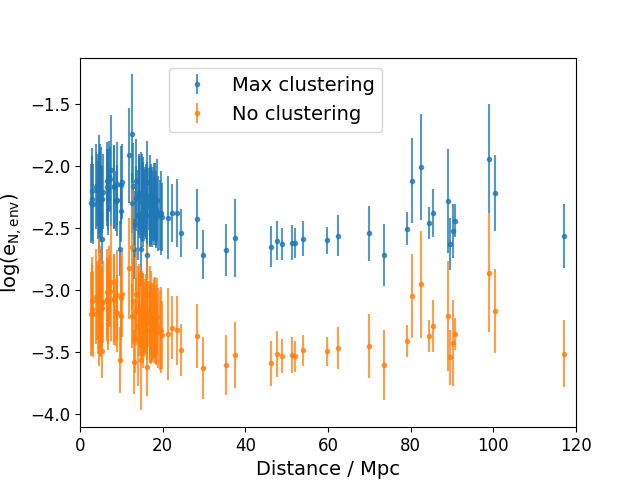}
  \caption{Variation of \eenv with distance for the SPARC galaxies within the NSA footprint. The ``max clustering'' model (blue) assumes that missing baryons are effectively coincident with observed structures, while the ``no clustering'' model (orange) distributes them uniformly in space. See Section~\ref{sec:mocks} for details.}
  \label{fig:genv_SPARC}
\end{figure}

These results demonstrate qualitative consistency between the Newtonian external field inferred from galactic RCs and the large-scale galaxy distribution. To make a more detailed comparison, Figure~\ref{fig:Moll} shows a full-sky Mollweide projection of the mean environmental field \genv\ from the 2M++ plus MCXC calculation (Section~\ref{sec:method_allsky}) over different distance bins: $0 < D/\text{Mpc} < 11$, $11 < D/\text{Mpc} < 30$, $30 < D/\text{Mpc} < 60$ and $60 < D/\text{Mpc} < 120$. These are chosen to represent (1) the Local Volume, (2) the remainder of the overdense region we identify in the SDSS footprint (see Section~\ref{sec:res_sdss}), (3) the underdense region further out, and (4) the CfA2 great wall and the Perseus-Pisces supercluster regions. Note that here we use three-dimensional distance $D$ from the Milky Way, rather than cylindrical radius $R$, to provide an alternative perspective. Again we find qualitative agreement between the RC-fitted field \gfit\ and the environmental field \genv\ in terms of their spatial variations. In particular, the Local Void \citep{Tully2019} is clearly visible as the blue regions with low values of \genv\ at $D<11$ Mpc; galaxies within the void tend to have values of \gfit\ below the average, as expected.

\subsection{Quantitative comparison in the SDSS footprint} \label{sec:res_sdss}

\begin{figure*}
  \centering
  \includegraphics[width=1.\linewidth, trim=0mm 8mm 0mm 0mm, clip]{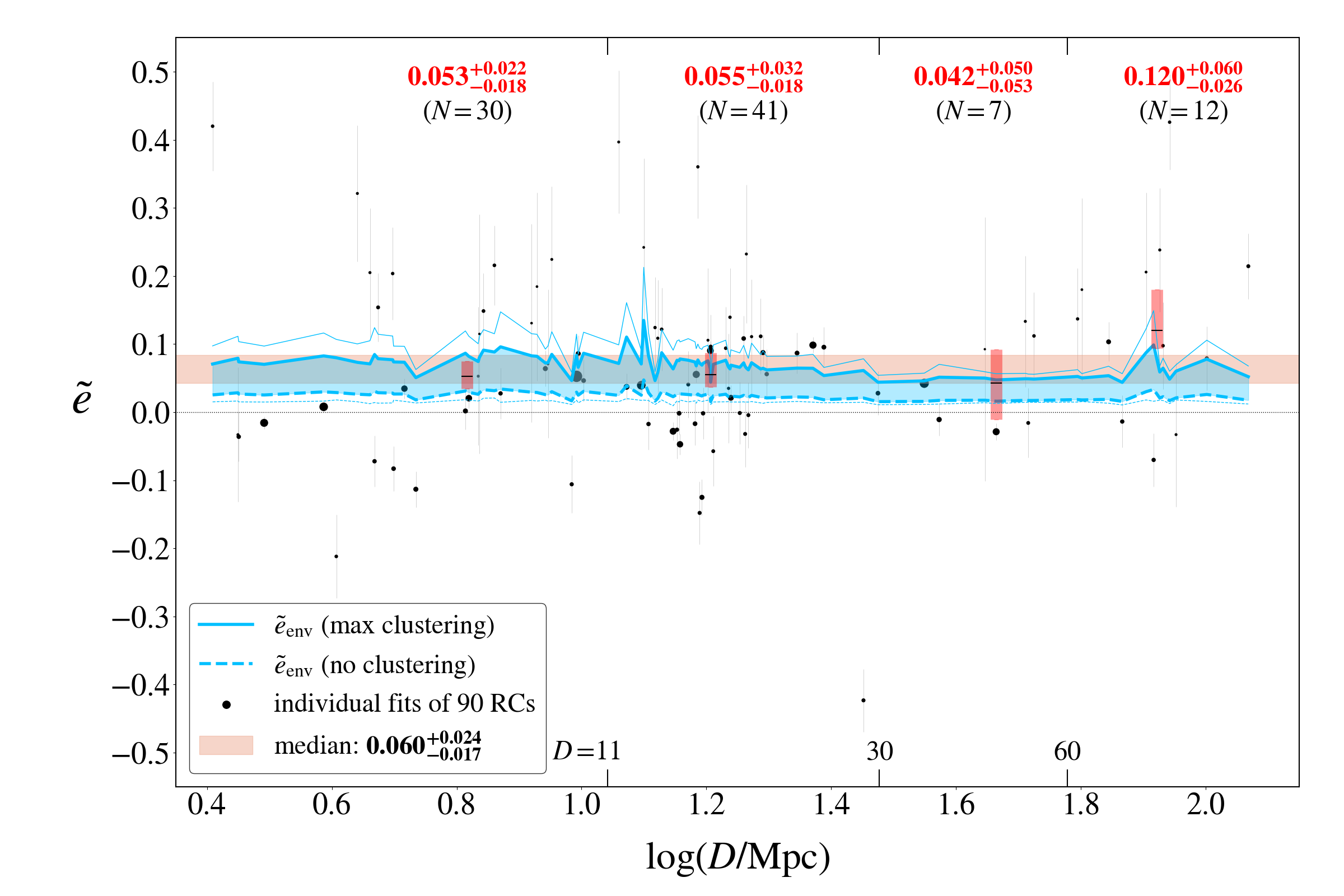}
    \vspace{0.0truecm}  
    \caption{\small The external Newtonian gravitational fields from RC-fits (dots with errorbars), parametrized by $\tilde{e}$, are compared with those from the large-scale distribution of baryonic matter (blue band) as a function of distance to 90 SPARC galaxies that probe the low acceleration regime and are within the SDSS footprint. The size of each black dot is inversely proportional to its statistical uncertainty. The median value of \etfit\ and the associated uncertainty is represented by the light-salmon band. The medians and associated bootstrap uncertainties in 4 distance bins are shown by red bars, with the values indicated at the top. The blue band is obtained considering the ``no clustering'' (dashed line) and ``max clustering'' (solid line) models for the missing baryons (see Section~\ref{sec:mocks}), interpolating between the \etenv\ values recorded at the location of each SPARC galaxy. The thin blue lines show the statistical 1$\sigma$ uncertainty for each clustering model. The agreement of $\tilde{e}_{\rm env}$ with the median values of $\tilde{e}_{\rm fit}$ is remarkable and not expected \textit{a priori} outside a MOND context.
    }
  \label{eN2env}
\end{figure*}

\begin{figure*}
  \centering
  \includegraphics[width=0.95\linewidth, trim=0mm 0mm 0mm 0mm, clip]{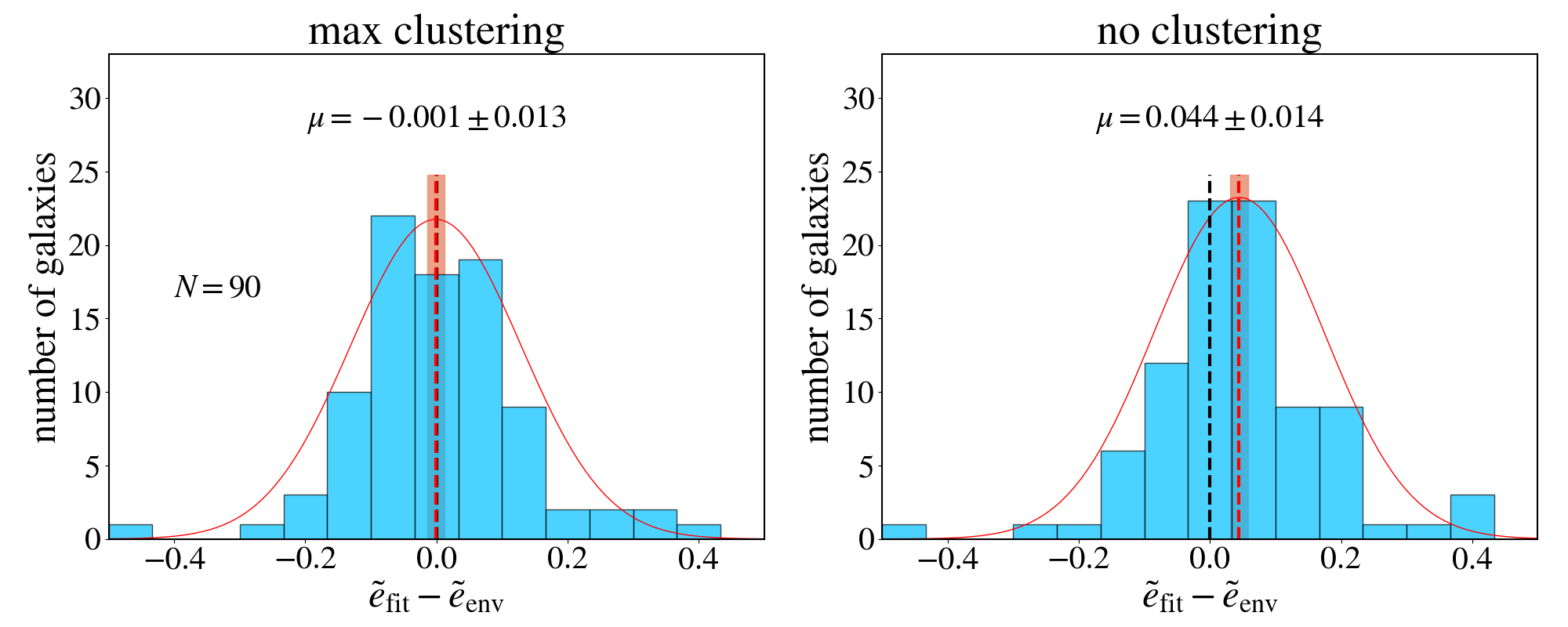}
    \vspace{0.0truecm}  
    \caption{\small Statistical comparison of $\tilde{e}_{\rm fit}$ and $\tilde{e}_{\rm env}$ for the subsample of 90 SPARC galaxies shown in Figure~\ref{eN2env}. The max clustering and no clustering cases (see Section~\ref{sec:env}) are considered. The excellent statistical match with the max clustering model indicates that ``missing'' cosmic baryons are strongly correlated with observed structures. 
    } 
  \label{deleN2}
\end{figure*}

We now consider a quantitative comparison of $\tilde{e}$ with the environmental field estimated in Section~\ref{sec:method_sdss}. To begin, we calculate \genv\ at the positions of the Karachentsev and NSA galaxies themselves out to 150 Mpc. This provides a fair tracing of the gravitational field over the nearby large scale structure. Figure~\ref{fig:genv_NSA} plots the mean \genv\ values as a function of distance. We use here the ``max clustering'' model, which we will find below to match the fitted values better than the ``no clustering'' case, although the qualitative trends are the same. We see that \genv\ falls from $D=0$ to $D\simeq50$ Mpc, then rises again out to $\sim$100 Mpc before levelling off. For both $D<40$ Mpc and $80 < D/\text{Mpc} < 120$, \genv\ is significantly larger towards $\alpha \simeq180^\circ$ than elsewhere, indicating excess structure in that direction. This is largely due to the Virgo cluster at small distance and the CfA2 great wall at $D\simeq100$ Mpc. The part of the SDSS footprint in the Southern Galactic Cap $(-25^\circ < \alpha < 25^\circ, -9^\circ < \delta < 30^\circ)$ has a weaker field on the whole, especially at $D\simeq20$ Mpc where it can be seen as a disjoint band of low-\genv\ points. Very similar trends are seen if the distance $D$ is replaced by the cylindrical distance $R$ adopted in Figures~\ref{map} and \ref{eN2R}.

We now evaluate \genv\ at the positions of the SPARC galaxies. Restricting to the SDSS footprint retains 109 SPARC galaxies, although only 90 of these have a measured $\tilde{e}$ satisfying the cut $x_{0,3}<-10.6$. Figure~\ref{fig:genv_SPARC} shows $\log(e_\text{N,env})$ for the SPARC galaxies as a function of distance, including their full uncertainties, for both the no clustering and max clustering models. Although the statistical uncertainties are considerable, the distance trend is still clearly visible, as is the factor $\sim$8 difference between the two clustering models driven by the influence of baryons not found in stars, cold gas, or the ICM.

Figure~\ref{eN2env} exhibits individual values of \etfit\ against $\tilde{e}_{\rm env}$ for the sample of 90 SPARC galaxies as a function of distance. Remarkably, the median value of \etfit\ from rotation-curve fits is well within the allowed values from the two clustering models. This is highly nontrivial because, in a general DM context, $\tilde{e}$ is merely a fitting parameter to describe the outer shapes of rotation curves. It could have taken a value across orders of magnitude, so there is no \textit{a priori} reason why its median value should lie within the boundaries of the $\tilde{e}_{\rm env}$ calculation from the large-scale distribution of baryonic matter. Moreover, both \etfit\ and \etenv\ are lower than average in the underdense region $30 < D/\text{Mpc} < 60$ while clearly higher than average in the overdense region $D>60$Mpc, consistent with the trend of \etfit\ for all 126 galaxies within $Z<50$ Mpc in the northern sky shown in Figure~\ref{eN2R}. 

To check the agreement between \etfit\ and $\tilde{e}_{\rm env}$ in more detail, Figure~\ref{deleN2} shows the distribution of their differences. Remarkably, there is an excellent statistical match between \etfit\ and $\tilde{e}_{\rm env}$ of the max clustering case while there is some tension with the no clustering case. This result suggests that accounting for missing baryons is important in MOND. Moreover, it agrees with previous indications that the missing baryons are likely significantly correlated with galaxies and groups (e.g. \citealt{WHIM_1, WHIM_2, WHIM_3, WHIM_4}). 

\begin{figure}
  \centering
  \includegraphics[width=1.\linewidth, trim=0mm 0mm 0mm 0mm, clip]{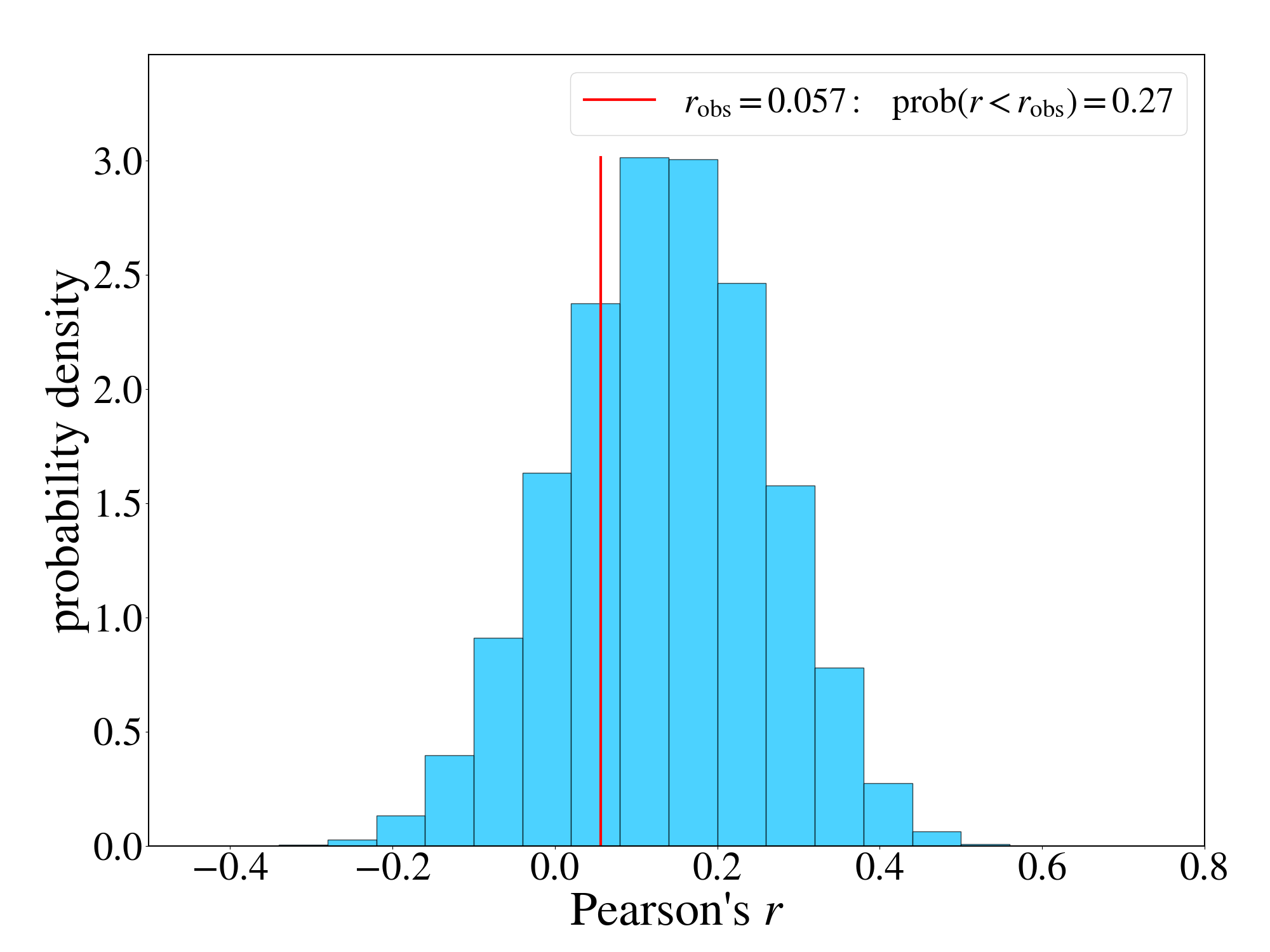}
    \vspace{-0.5truecm}  
        \caption{\small Observed value of Pearson's linear correlation coefficient $r$ between \etenv\ and \etfit\ for the sample shown in Figures~\ref{eN2env} and \ref{deleN2} (red line) and the expected probability distribution from many Monte Carlo realizations assuming \etfit\ = \etenv\ and scattering by the measured uncertainties of both quantities (histogram). For \etenv\ we use the average of the max clustering and no clustering cases.
    } 
  \label{pr}
\end{figure}

Finally, we check directly the correlation between \etenv\ and \etfit\ for the sample of 90 galaxies shown in Figures~\ref{eN2env} and \ref{deleN2} using Pearson's linear correlation coefficient $r$. Because \etfit\ has a large uncertainty ranging from 0.004 to 0.194 with a median value of 0.047 for these galaxies, we expect the correlation to be weak. In Figure~\ref{pr} we compare the measured correlation coefficient with the expected distribution from a Monte Carlo simulation. This simulation assumes that \etfit=\etenv\ fundamentally, with \etfit\ scattered by the error budgets of both quantities. For \etenv\ we use the mean of the max clustering and no clustering values. For the uncertainty of \etenv, we take the average of the statistical errors of the max and no clustering cases for the statistical error and one half of the difference between the max and no clustering cases for the systematic error. For the uncertainty of \etfit, we take the measured values (see Appendix~\ref{sec:app}) for the statistical error and 0.02 for the systematic error of our model (see Sec.~\ref{sec:efit_comp}). These errors are added in quadrature for each galaxy, with the result used to scatter \etfit\ in each Monte Carlo realization. 

We find the measured coefficient of $r=0.057$ to agree well with the distribution from the mock data, with a 27\% probability of a lower $r$ under the hypothesis \etfit=\etenv. The mock data have a RMS scatter of $0.077\pm 0.008$ for \etfit. This is about 3/5 of the measured scatter for this sample, indicating that the uncertainties of \etfit\ and/or \etenv\ may be somewhat underestimated. Regardless of the precise error model, it is clear that the theoretically expected correlation between \etfit\ and \etenv\ values of individual galaxies cannot be directly inferred given the current observational uncertainties. The existing data, however, are consistent with the existence of an intrinsic correlation between \etfit\ and \etenv.

\section{Discussion and Conclusion}  \label{sec:disc}

Using a simple, generic fitting function (Equation~\ref{eq:nueN}) we investigate the EFE in the RCs of 143 SPARC galaxies, using a dimensionless parameter $\tilde{e}$ related to the external Newtonian gravitational field. We find that $\tilde{e}$ is well correlated with the observed large scale structure of baryonic mass distribution (Figure~\ref{map}). The fitted values are consistent with zero in an underdense ring region at cylindrical radii $22\hspace{1ex}{\rm Mpc}<R<45\hspace{1ex}{\rm Mpc}$, while they are a factor of two higher than the average external field at or near the CfA2 great wall and the Perseus-Pisces supercluster (Figure~\ref{eN2R}). In this high density region the  EFE is statistically detected at $\sim$13$\sigma$. The correlation between \etfit\ and large-scale structure extends qualitatively across the whole sky (Figure~\ref{fig:Moll}).

Next, we compare the RC-fitted values of $\tilde{e}$ with fully independent estimates of the Newtonian fields from the baryonic mass distribution in the nearby Universe (using the 2M++, NSA, Karachentsev and MCXC catalogues). We include corrections for undetected galaxies below the magnitude limits of the surveys, for cold and hot gas in galaxies, for the intracluster medium, and for the possible presence of ``missing baryons'' in the warm-hot intergalactic medium, with full propagation of uncertainties in a Monte Carlo framework. We find good agreement between \etfit\ with \etenv\ values, both in their distance dependence (Figure~\ref{eN2env}) and the average value of the Newtonian field (Figure~\ref{deleN2}). In particular, \etfit\ is well matched with \etenv\ when the missing baryons needed to bring the local baryon density into agreement with Big Bang Nucleosynthesis \citep[$\Omega_b h^2 = 0.022$,][]{primordialDH} are strongly clustered with galaxies and galaxy clusters.

The qualitative correlation between \etfit\ and \etenv\ is not very sensitive to the functional form assumed for the EFE model (Equation~\ref{eq:nueN}) because the relative values of inferred field strengths are similar for different models. However, the absolute value of \etfit\ may vary, leading to systematically different mean and median values. For example, we fitted RCs using an alternative function based on numerical simulations \citep{Hagh2019} and found a systematic shift of $\sim$0.02 in $\tilde{e}$ on average (Figure~\ref{efemodel}). Until the fitted values of $\tilde{e}$ of Equation~(\ref{eq:nueN}) are thoroughly tested through realistic numerical simulations of disk galaxies, systematic shifts of order $\sim 0.02$ are possible. This could have implications for the abundance and clustering of cosmic baryons. Our current values of \etfit\ are within the limits set by the ``max clustering'' and ``no clustering'' models of missing baryons, although the max clustering case is preferred (Figures~\ref{eN2env} and \ref{deleN2}).

It may be possible for \genv\ to exceed the prediction of the max clustering model due to the gravitating scalar fields present in some relativistic MOND theories, such as that of~\citet{Skor2020}. Indeed, the EFE may manifest itself somewhat differently in this theory than in AQUAL and QUMOND because the free function in the Lagrangian (which plays the role of the interpolating function in AQUAL) has an oscillating piece related to the time-variation of the scalar field as well as a more traditional MOND term. It will therefore be important to determine the exact behaviour of the EFE in this and similar theories. 

The agreement between the external fields from RC fits and large-scale structure of baryonic matter that we have discovered is predicted by MOND modified gravity theories, supplementing the evidence for the EFE (and \textit{a fortiori} a violation of the SEP) presented in \pap. This correlation is nontrivial from the $\Lambda$CDM point of view. It is not naturally expected that the density of the cosmic web would affect the dynamics of the disk in the inner region of the DM halo, never mind in a manner mimicking the MOND EFE. We are not aware of any $\Lambda$CDM simulations predicting that RCs in stronger-field regions would tend to decline while those in low density regions do not. This adds to the ``small-scale problems'' with $\Lambda$CDM \citep{Bullock, Kroupa}. In this context, we note that challenges to $\Lambda$CDM cosmology have also been emphasized recently on the scales of galaxy clusters and voids \citep{KBCvoid2020,ElGordo2021}.

The external Newtonian fields encapsulated in $\tilde{e}$ will provide a benchmark for more detailed numerical studies of the MOND EFE, taking full account of the relative orientation of the external field and the three-dimensional morphology and kinematics of galactic disks. Various MOND theories may have different EFE predictions for RCs of disk galaxies under the same \genv. Although our fitted values of \et\ are model-dependent, they are empirical in nature and may distinguish existing theories when accurate numerical results become available. Here we used a fitting function that approximately agrees with existing AQUAL simulations (see Appendix~\ref{sec:form}), so the agreement between \gfit\ and \genv\ may indicate that AQUAL is close to the correct nonrelativistic theory.

Besides distinguishing between MOND formulations, our results may be used to infer the properties of ``missing baryons'' in a MOND context. Our preliminary result in this regard is that baryons in the warm-hot intergalactic medium needed for $\Omega_b \simeq 0.046$ \citep{baryoncensus} are likely strongly clustered with galaxies and clusters.

In summary, the agreement between our fit results and the observed baryonic large-scale structure matches well the MOND prediction and reinforces the detection of the EFE in \pap, pointing to a breakdown of the SEP. These results support the modified gravity hypothesis and reveal the possibility of using the internal dynamics of galaxies to study the large-scale distribution of cosmic baryons.

\vspace{0.3in}

\acknowledgements
We thank Indranil Banik, Andrey Kravtsov, Pengfei Li, Mordehai Milgrom, Ravi Sheth and Paolo Tozzi for useful discussions. We also thank the anonymous referee for insightful comments that helped us improve the presentation significantly. K-H.C. is supported by the National Research Foundation of Korea(NRF) grant funded by the Korea government(MSIT) (No.\ NRF-2019R1F1A1062477). H.D. is supported by St John's college, Oxford, and acknowledges financial support from ERC Grant No. 693024 and the Beecroft Trust. S.S.M. is supported in part by NASA ADAP 80NSSC19k0570 and NSF PHY-1911909.

\newpage

\appendix

\section{A simple  {one-dimensional} formalism  {for} the EFE}\label{sec:form}

In the MOND context, it is hard to quantify generally the effect of a constant external gravitational field on a galaxy disk because, for a test particle along a given orbit, the angle between the internal acceleration and the external gravitational field will vary continually, leading to a time-varying EFE. Broadly speaking, we expect that the outermost orbits will become slightly non-circular and possibly tilted with respect to the inner galaxy plane. Asymmetric, lopsided, and/or warped H\,{\small I} disks are frequently observed in nearby galaxies (e.g., \citealt{San2008}) but their relation to the EFE can only be studied with detailed numerical simulations. For example, \cite{BM2000} show that the asymmetric warp of the Milky Way may be induced by the EFE from the Magellanic Clouds, while \cite{Ban2020} reproduce the dynamical properties of the Local-Group spiral galaxy M33 as the result of the EFE from Andromeda. The main EFE, however, is a decrease in the circular velocity in the MOND regime. 

In this work we are particularly interested in how the EFE depends on the external Newtonian field. Here we describe a simple one-dimensional formalism to model the EFE-induced outer decline of RCs, neglecting higher order effects due to the full three-dimensional nature of the problem, such as the development of non-circular motions, warps, and asymmetries. Incidentally, this formalism turns out to be equivalent in AQUAL and QUMOND with a heuristic MOND relation for the constant external field.

\subsection{The EFE in AQUAL and QUMOND}
If a system is freely falling under a constant MOND external field $g_\text{e}$ and we approximate the EFE by a one-dimensional AQUAL equation (see Section~6.3 of \citealt{FM2012}), the gravitational acceleration $g$ of a test particle within the system can be determined by
\begin{equation}
\mu\left(\frac{g+g_{\rm e}}{a_0}\right) g + \left[\mu\left(\frac{g+g_{\rm e}}{a_0}\right) - \mu\left(\frac{g_{\rm e}}{a_0}\right)\right] g_{\rm e} = g_{\rm N},
  \label{eq:aqual1d}
\end{equation}
where $g_{\rm N}$ is the internal Newtonian field and $\mu(x)$ is the MOND interpolating function (IF). If we assume the simple IF $\mu(x)=x/(1+x)$, then the ratio $g/g_\text{N}$ is given by Equation~(6) of \pap\ as 
\begin{equation}
  \nu_e(y) = \frac{1}{2} + \sqrt{\left(\frac{1}{2}- e\frac{A_e}{y}\right)^2+\frac{B_e}{y}}- e\frac{A_e }{y},
  \label{eq:nue}
\end{equation}
where $A_e = (1+e/2)/(1+e)$, which is redefined here so that its limit tends to 1, and $B_e=1+e$. For typical values $e\lesssim 0.1$ (\pap) we have $A_e \approx B_e \approx 1$. It is then apparent that the EFE-dominating last term scales approximately linearly with $e$. For $e\rightarrow0$ (i.e., truly isolated galaxies), one recovers $\nu_0(y)=1/2 + \sqrt{1/4+1/y}$ which is the inverse function of the simple IF. 

If the external Newtonian field of the above system is $g_\text{Ne}$ and we use the QUMOND formalism, $g$ is given by
 \begin{equation}
g= \nu\left(\frac{g_{\rm N}+g_{\rm Ne}}{a_0}\right) g_{\rm N} + \left[\nu\left(\frac{g_{\rm N}+g_{\rm Ne}}{a_0}\right) - \nu\left(\frac{g_{\rm Ne}}{a_0}\right)\right] g_{\rm Ne},
  \label{eq:qumond1d}
\end{equation}
where $\nu(y)=\nu_0(y)$. Then, the ratio $g/g_\text{N}$ is given by Equation~(\ref{eq:nueN}). Although it is not obvious algebraically, one can show that Equation~(\ref{eq:nueN}) is equivalent to Equation~(\ref{eq:nue}) for positive (physical) external field with the transformation $e^2/(1+e)=e_\text{N}$. This means that although Equation~(\ref{eq:nueN}) is derived directly from Equation~(\ref{eq:qumond1d}), it is also consistent with Equation~(\ref{eq:aqual1d}). 
 
 \subsection{Comparison with the analytic point-mass weak-field limit and numerical simulations}
\label{sec:efit_comp}

Our fitting function (Equation~(\ref{eq:nueN})) is a heuristic first-order model of EFE that happens to be equivalent in AQUAL and QUMOND with the correspondence $\tilde{e}=e/\sqrt{1+e}$. For general dynamical systems, for the same Newtonian external field, these two Lagrangian theories predict different values of the ratio $g/g_\text{N}$, which can be calculated only through numerical methods. Here we compare Equation~(\ref{eq:nueN}) with analytic solutions of a point mass in the deep-MOND limit as well as previously published numerical results.

Far away from any finite mass distribution, the point-mass approximation will be valid regardless of the mass distribution. For the QUMOND EFE limit, we start from Equation~(60) by \cite{Mil2010} that gives the approximate analytic internal potential $\psi$ for a point mass $M$ under an external Newtonian field of strength $g_{\rm Ne}=e_{\rm N}a_0$. Taking the gradient of $\psi$ and considering an angle $\theta$ between $\vec{g}_{\rm Ne}$ and the rotation axis of a test particle on a quasi-circular orbit, the azimuthal average of the radial acceleration $g = |\hat{r}\cdot \vec{\nabla}\psi|$ is given by
\begin{equation}
  \langle g \rangle_\text{QUMOND} = \frac{GM}{r^2} \nu_0(y) \left(1+\frac{\hat{\nu}_0(y)}{2}\right) \left. \left(1-\frac{ \hat{\nu}_0(y)}{2+\hat{\nu}_0(y)} \frac{\sin^2\theta}{2}\right)\right|_{y=e_{\rm N}},
  \label{eq:point}
\end{equation}
where $\nu_0(y)=1/2 + \sqrt{1/4+1/y}$ and $\hat{\nu}_0(y)\equiv d \ln\nu_0(y)/d\ln y$ which tends to $-1/2$ at large $r$. Thus, at large $r$ the two extreme cases $\theta=0^\circ$ and $\theta=90^\circ$ show a difference of just $1/6$ of the value. In other words, the direction of the external field has a minor effect. Similarly, for the AQUAL limit, we start from Equation~(66) of \cite{Mil2010} considering the MOND external field $g_\text{e}=e \: a_0$ and obtain 
\begin{equation}
  \langle g \rangle_\text{AQUAL} = \frac{GM}{r^2} \frac{1}{\mu_0(x)\sqrt{1+L_0(x)}} \left. \left( 1 - \frac{L_0(x)}{1+L_0(x)} \frac{\sin^2\theta}{2} \right)^{-1/2} \right|_{x=e},
  \label{eq:pointa}
\end{equation}
where $\mu_0(x)=x/(1+x)$ and $L_0(x) \equiv d\ln\mu_0(x)/d\ln x$. 

\begin{figure}
  \centering
  \includegraphics[width=0.5\linewidth, trim=5mm 5mm 0mm 10mm, clip]{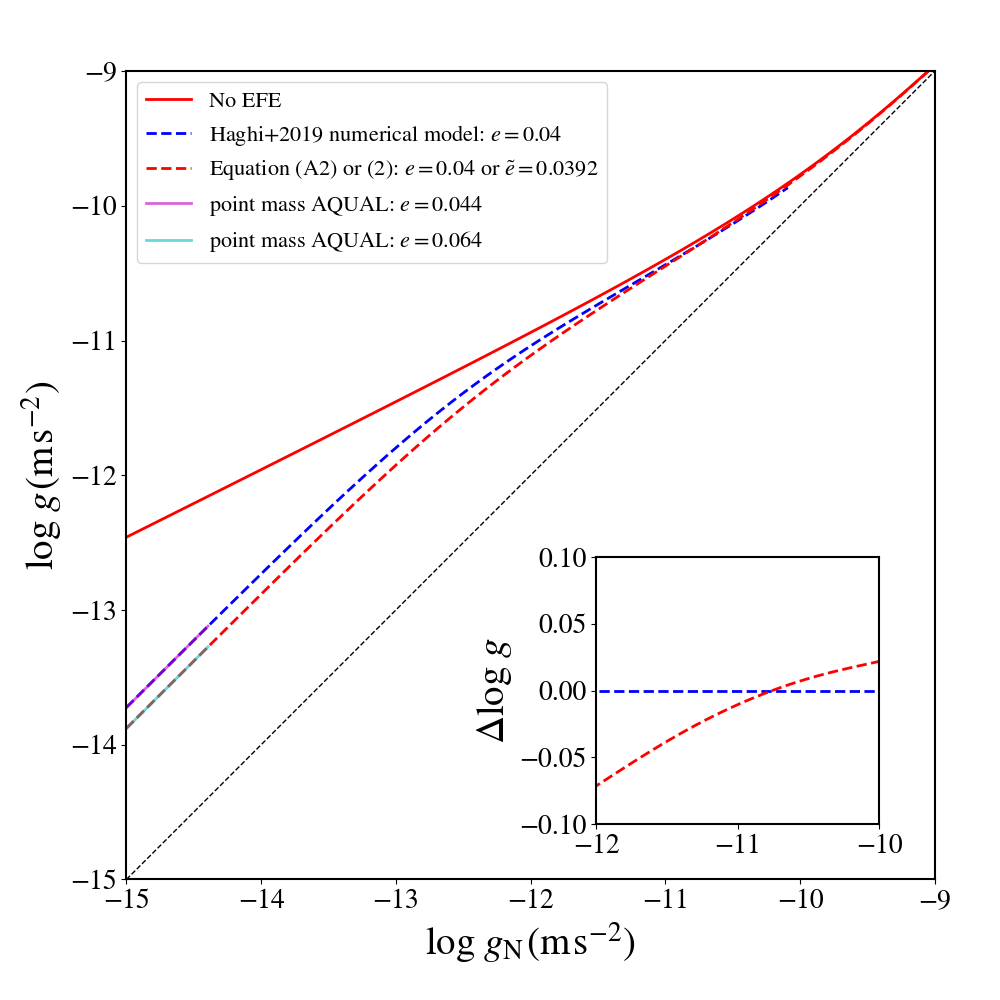}
    \caption{\small 
     The one-dimensional EFE model from Equation~(\ref{eq:nue}) (red dashed curve), or equivalently Equation~(\ref{eq:nueN}) with $\tilde{e}=e/\sqrt{1+e}$, is compared with the numerical simulation result by \cite{Hagh2019} (blue dashed curve). These are then compared with the analytic point-mass weak-field limit for different values of $e$, considering an average angle of 60$^{\circ}$ between the external field and the axis of a circular orbit. In the acceleration range probed by the SPARC RCs, the difference between our model and the numerical result is relatively small in terms of $e$ (or $\tilde{e}$). See the text for further details.}
   \label{efemodel}
\end{figure}

As for numerical results on the ratio $g/g_\text{N}$, we consider a functional relation derived by \cite{Hagh2019} from AQUAL-based numerical simulations of fully pressure-supported spherical systems. In a spherical shell of radius $r$ of a pressure-supported system, the measured velocity dispersion is a collective representation of various random orbits at different angles with respect to the external field. In this sense the velocity dispersion may capture a direction-averaged EFE. Then, for the case of an isotropic velocity distribution, the radial acceleration is given by $g(r)=3\sigma^2/r$, where $\sigma$ is the one-dimensional velocity dispersion. Thus, comparing $\sigma$ with and without an external field, one can obtain an EFE-dependent $g(r)/g_{\rm N}(r)$. \cite{Hagh2019} used the so-called ``standard'' IF (e.g., \citealt{FM2012}) for their simulations, so we transform their result to that for the simple IF by multiplying the EFE-dependent $g(r)/g_{\rm N}(r)$ by the ratio of the simple IF to the standard IF.

In Figure~\ref{efemodel} we compare our EFE model (Equation~(\ref{eq:nue}), or equivalently Equation~(\ref{eq:nueN})) with the numerical result by \cite{Hagh2019} and the weak-field analytic expectations of  QUMOND (Equation~(\ref{eq:point})) and AQUAL (Equation~(\ref{eq:pointa})). We note that QUMOND predicts 7 percent higher $g$ in the deep MOND limit. At $\log(g_{\rm N}/\mbox{m~s}^{-2})\approx -10.8$ our model matches the numerical result, but at higher (lower) accelerations our $g$ is higher (lower) than the numerical value. This means that for a declining RC covering the low acceleration range $\log(g_{\rm N}/\mbox{m~s}^{-2}) < -10.8$, our model will give a lower value of $\tilde{e}$ than the numerical model. For $\log(g_{\rm N}/\mbox{m~s}^{-2}) > -10.8$ the opposite may occur occasionally. In the weak-field limit, both our model and the numerical result give lower $g$ than the point-mass expectation. Our model function with  {$e=0.040$} matches the AQUAL point-mass limit with  {$e=0.064$}, exhibiting a difference of  {$0.024$}, whereas the numerical result exhibits a smaller difference of  {$0.004$}. Thus, the difference between our model and the numerical result for $\log(g_{\rm N}/\mbox{m~s}^{-2}) < -10.8$ is  {$0.020$} in terms of the AQUAL point-mass limit. A similar median difference of $\sim 0.02$ in the fitted value of $\tilde{e}$ is found from fitting the observed RCs. This is smaller than the typical individual uncertainties of $\sim 0.04$ -- $0.05$ in $\tilde{e}$.
 
\section{Fitted values of the parameters and the Newtonian environmental fields}\label{sec:app}

\begin{figure}[b]
  \centering
  \includegraphics[width=0.6\linewidth, trim=0mm 0mm 0mm 0mm, clip]{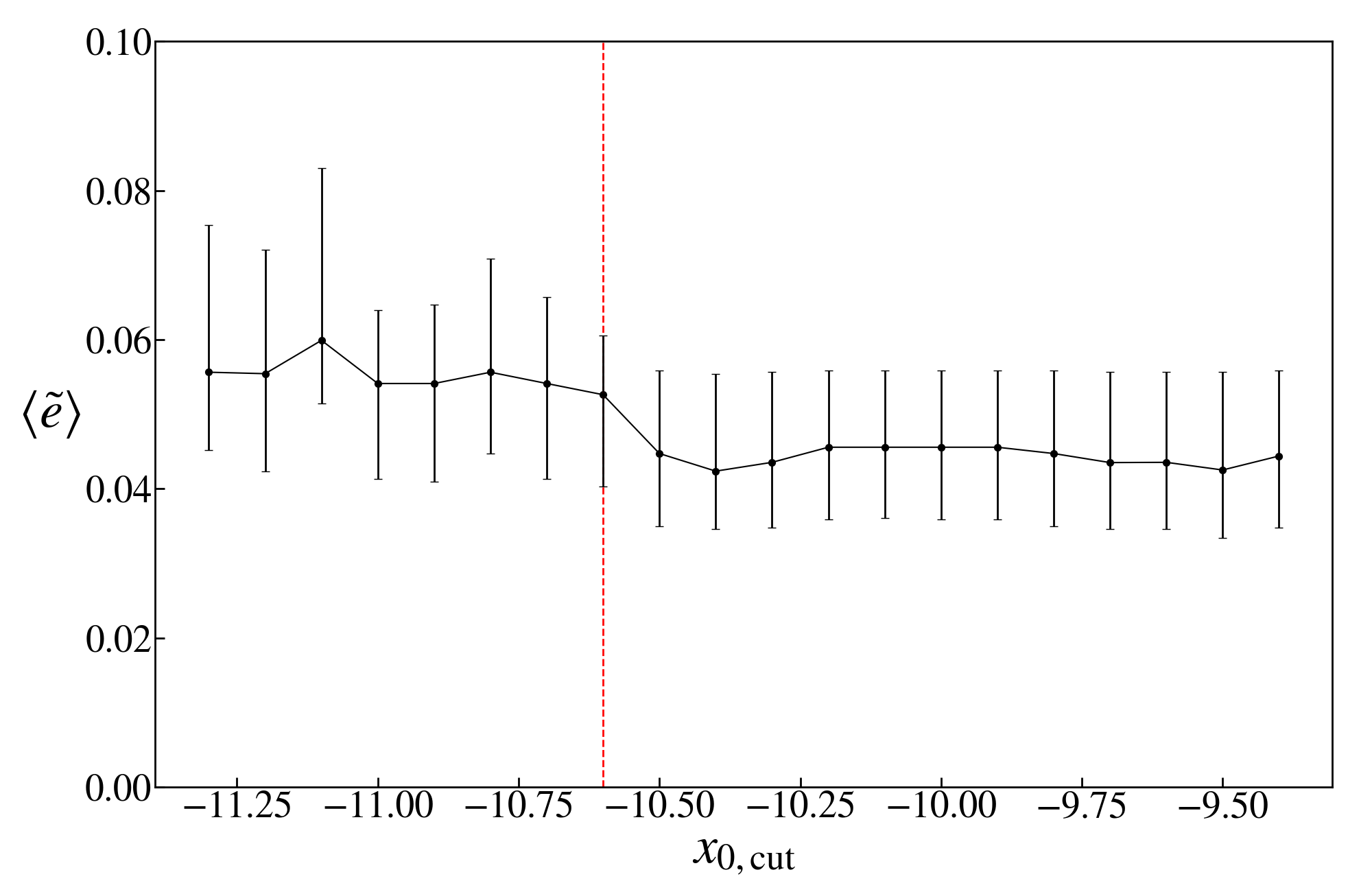}
    \vspace{-0.4truecm}  
    \caption{\small Dependence of the sample median of $\tilde{e}$ on the sample selection cut. Each sample is selected by $x_{0,3}<x_{0,{\rm cut}}$ where $x_{0,3}$ is the 3$^{\rm rd}$ lowest value of $x_0$ as defined in \pap\ (see the text for the details). The vertical dashed line indicates our selection. The errorbars indicate bootstrap uncertainties.  {As $x_{0,{\rm cut}}$ increases, galaxies without enough low-acceleration RC points start to be included in the sample so that $\langle\tilde{e}\rangle$ is biased to a lower value.} 
    } 
  \label{eN2_x0cut}
\end{figure}

 Here we provide the fitted values of \et\ and galactic parameters for all 162 galaxies considered in this work. We also describe how we select a statistical sample of 143 galaxies whose RCs reach low enough accelerations for the fitted \et\ values to be meaningful. As illustrated in Figure~4 of \pap\, only RCs that reach low enough accelerations can probe the EFE from the large-scale distribution of cosmic mass. In other words, if we used only RCs in a high-acceleration range, then the $g_{\rm N}$-$g$ relation would have little sensitivity to $\tilde{e}$ (or $e$), and consequently the median of fitted $\tilde{e}$ (or $e$) for such RCs will be biased towards  {zero}. We therefore select the RCs that can probe a low acceleration regime. We use the parameter $x_0$ introduced in \pap, which is obtained by projecting a point $(\log g_{\rm bar}, \log g_{\rm obs})$ (here $g_{\rm bar}$ means $g_{\rm N}$) to the curve corresponding to a flat RC (the red solid curve in Figure~\ref{efemodel}). An RC having low enough values of $x_0$ is expected to have sensitivity to $\tilde{e}$ (or $e$), so we require a RC to have at least 3 data points $(\log g_{\rm bar}, \log g_{\rm obs})$ satisfying $x_0 < x_{0,\rm{cut}}$. Here we do this by $x_{0,3}<x_{0,\rm{cut}}$ where $x_{0,3}$ is the 3$^{\rm rd}$ lowest value of $x_0$. For the threshold we choose $x_{0,\rm{cut}}=-10.6$, so that selected RCs have at least three values of $10^{x_0}$ lower than  $10^{-10.6}$~m~s$^{-2}$. Figure~\ref{eN2_x0cut} shows that, at a higher threshold than this value, the median of $\tilde{e}$ starts to decrease indicating that the selected sample starts to be affected by about 20 RCs in a high acceleration range with little sensitivity to $\tilde{e}$. 

Because $x_0$ is always on a flat RC by construction, when we select RCs by $x_0$, there is no selection bias to a more declining (or rising) RC. This can be seen clearly in Figure~\ref{eN2_x0cut}: as the threshold $x_{0,{\rm cut}}$ decreases from $-10.6$, the sample median of $\tilde{e}$ remains the same. However, if one selects RCs by requiring $\log g_{\rm obs}(R)$ to be lower than a certain threshold, then the selection is biased in favor of more declining RCs. This is because for a given $g_{\rm bar}(R)$, lower values of $g_{\rm obs}(R)$, thus declining RCs, are more likely to be selected by chance due to the unavoidable observational scatter in $g_{\rm obs}(R)$. Selection based on $\log g_{\rm bar}(R)$ is biased in the opposite sense. Namely, less declining RCs are more likely to be selected by requiring a certain number of values of $x$ to be lower than a certain threshold. This is because for a given  $g_{\rm obs}(R)$, a more declining Newtonian RC $g_{\rm bar}(R)$ is more likely to be selected by chance. Our selection based on $x_0$ is robust against possible bias.

\begin{figure*}
  \centering
  \includegraphics[width=1.\linewidth, trim=0mm 15mm 0mm 0mm, clip]{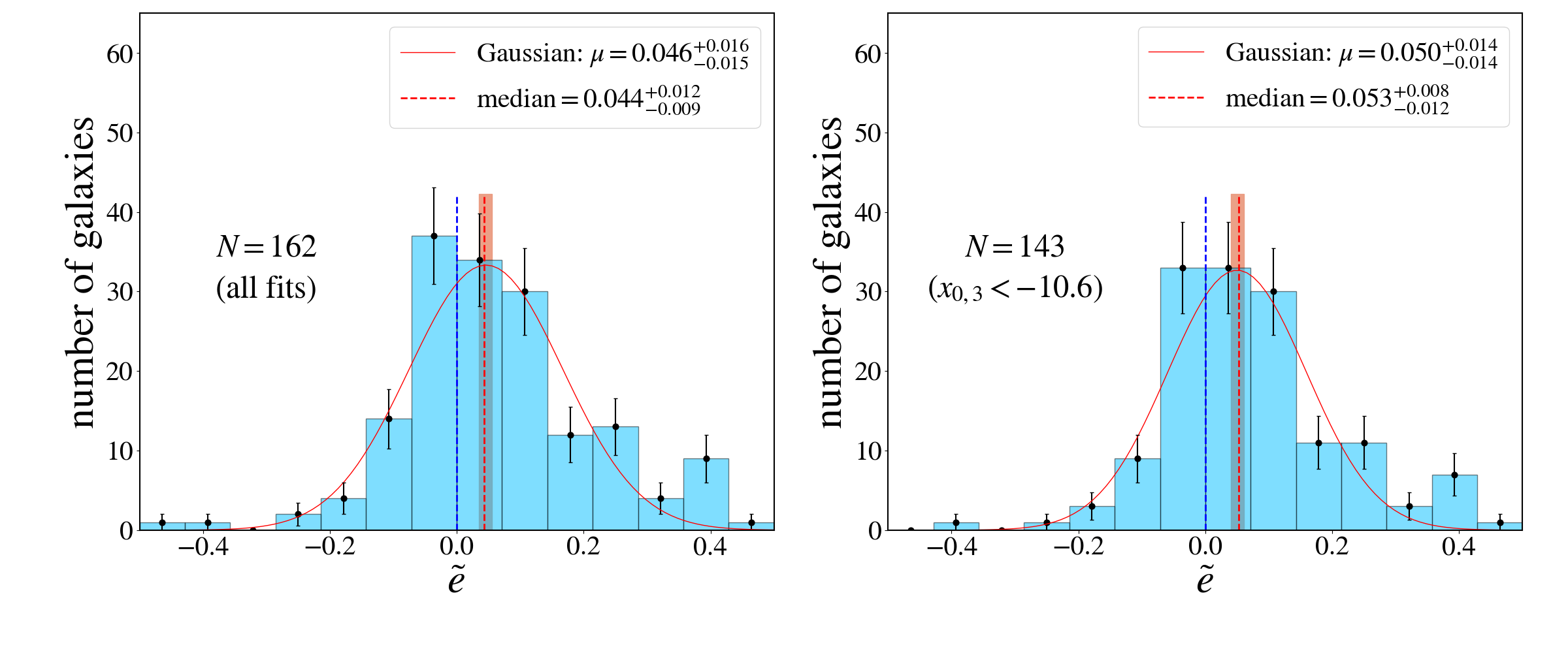}
    \vspace{-0.4truecm}  
    \caption{\small Distribution of the fitted values of $\tilde{e}$. 
        The left panel shows 162 SPARC galaxies with reliable RCs ($Q<3$, see \citealt{Lel2016}), while the right panel shows our selected sample of 143 galaxies with RCs sensitive to the EFE, as indicated in Figure~\ref{eN2_x0cut}. The errorbars on the histogram show Poissonian uncertainties $\sqrt{N_j}$, where $N_j$ is the number of galaxies in each bin. The red curve indicates a Gaussian fit whose rms width is $\sim$0.11. The uncertainty on the median or on the Gaussian mean is estimated through a bootstrap method.} 
  \label{eN2}
\end{figure*}

Figure~\ref{eN2} shows the distribution of the fitted $\tilde{e}$ values for the selected sample of 143 galaxies as well as for the full sample. The median values for the samples are $\langle \tilde{e} \rangle = 0.044_{-0.009}^{+0.012}$ (full sample) and $\langle \tilde{e} \rangle = 0.053_{-0.012}^{+0.008}$ (selected sample), the latter of which corresponds to $\langle g_{\rm Ne} \rangle \simeq 0.0028 a_0$. Here the uncertainty is estimated from the Monte Carlo distribution of the medians of many bootstrap resamples of the data (this method is simply referred to as ``bootstrap'' in this paper). This is a $\sim$4.4$\sigma$ detection of positive $\tilde{e}$. Also, from the histogram we have 97 cases of $\tilde{e}>0$ out of 143. The null hypothesis that $\tilde{e}$ is as likely to be $<$0 as $>$0 is ruled out at $4.1\sigma$ confidence based on binomial statistics. For the (contaminated) full sample the significance of $\tilde{e}>0$ is $3.5\sigma$ by binomial statistics. These results reinforce the results by \pap\ based on Equation~(\ref{eq:nue}) and a different sample selection.

Table~\ref{tab:fit} lists the RC-fitted values of the model parameters for all 162 SPARC galaxies modeled through MCMC simulations. Values of $x_{0,3}$ (the 3$^{\rm rd}$ lowest value of $x_0$ as defined above) are given for the purpose of selecting galaxies with $x_{0,3}<-10.6$ for statistical analyses. The fitted value and its uncertainty are derived from the 50 percentile and the 15.9 and 84.1 percentiles of the posterior PDF respectively.

\begin{figure*}
  \centering
  \includegraphics[width=1.\linewidth, trim=0mm 18mm 0mm 5mm, clip]{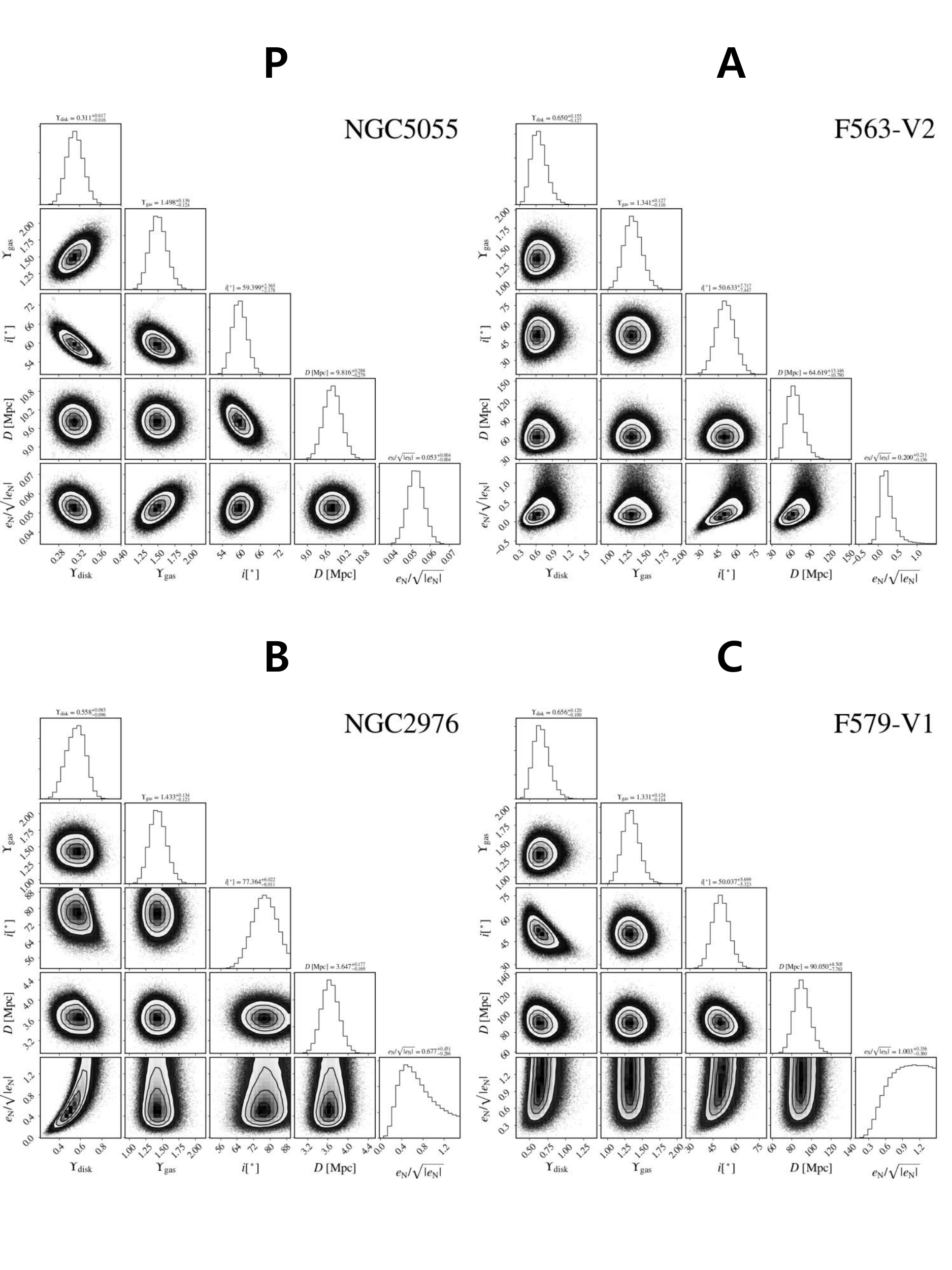}
    \vspace{0.0truecm}  
    \caption{\small Examples showing the posterior PDFs of our model parameters for four types of PDF-quality described in the text of the Appendix. For types A, B, and C, an extended prior range for $\tilde{e}\equiv e_{\rm N}/\sqrt{|e_{\rm N}|}$ is used to illustrate the behavior of its PDF. Thus, the values of $\tilde{e}$ for these cases are different from those given in Table~\ref{tab:fit}. 
    } 
  \label{corner}
\end{figure*}

We assign four quality flags based on the shape of the $\tilde{e}$ posterior: P, A, B, C. In most cases the PDF is well peaked and the 50 percentile agrees well with the best-fit. Those galaxies are assigned PDF-quality `P'. An example (NGC~5055) is exhibited in Figure~\ref{corner} (top-left panel). For the remaining 24 galaxies with quality flags A, B, or C, the PDF of $\tilde{e}$ extends beyond the prior range $-0.5< \tilde{e} < +0.5$, and is not well peaked in some cases. For 9 galaxies with PDF-quality `A', the PDF is well peaked within or a little outside the prior limits. An example (F563-V2) for this type is exhibited in Figure~\ref{corner} (top-right panel). For 8 galaxies with PDF-quality `B' the PDF has a peak with a long tail in the positive direction of $\tilde{e}$. An example (NGC~2976) for this type is exhibited in Figure~\ref{corner} (bottom-left panel). Finally, for 7 galaxies with PDF-quality `C', the PDF is not well peaked but stays nearly flat as $\tilde{e}$ increases. An example (F579-V1) for this type is exhibited in Figure~\ref{corner} (bottom-right panel). For the galaxies with PDF-quality of `A', `B', or `C', the lower bound is still within the prior range. For those galaxies, the given value of $\tilde{e}$ in Table~\ref{tab:fit} may be considered as a lower bound within the prior range.

Table~\ref{tab:env} lists the Newtonian environmental field strengths of 109 SPARC galaxies residing in the SDSS footprint for the ``max clustering'' and ``no clustering'' cases as described in Section~\ref{sec:method_sdss}.

Tables are also provided at \url{http://home.sejong.ac.kr/~chae} and \url{http://astroweb.cwru.edu/SPARC/.}

\setcounter{table}{1}
\begin{table}
\caption{Fitted model parameters}\label{tab:fit}
\begin{center}
  \begin{tabular}{lcccccccc} \hline
 galaxy & PDF-quality  & $x_{0,3}$  & $\tilde{e}$  &  $D$ [Mpc]  & $i$ [$^\circ$] & $\Upsilon_{\rm gas}$ & $\Upsilon_{\rm disk}$  & $\Upsilon_{\rm bulge}$  \\
 \hline
       CamB & C & $ -11.629 $ & $ 0.459 _{ -0.055 } ^{ + 0.030 } $ & $ 3.12 _{ -0.21 } ^{ + 0.22 } $ & $ 60.74 _{ -4.67 } ^{ + 4.80 } $ & $ 1.28 _{ -0.11 } ^{ + 0.12 } $ & $ 0.34 _{ -0.05 } ^{ + 0.06 } $ & --- \\  
     D512-2 & P & $ -11.421 $ & $ 0.106 _{ -0.081 } ^{ + 0.117 } $ & $ 15.94 _{ -3.73 } ^{ + 4.82 } $ & $ 61.25 _{ -9.10 } ^{ + 9.10 } $ & $ 1.36 _{ -0.12 } ^{ + 0.13 } $ & $ 0.52 _{ -0.11 } ^{ + 0.13 } $ & --- \\  
     D564-8 & P & $ -11.987 $ & $ 0.064 _{ -0.025 } ^{ + 0.026 } $ & $ 8.75 _{ -0.27 } ^{ + 0.28 } $ & $ 61.38 _{ -7.61 } ^{ + 7.48 } $ & $ 1.37 _{ -0.12 } ^{ + 0.13 } $ & $ 0.41 _{ -0.08 } ^{ + 0.10 } $ & --- \\  
     D631-7 & P & $ -11.656 $ & $ -0.089 _{ -0.019 } ^{ + 0.017 } $ & $ 7.48 _{ -0.17 } ^{ + 0.17 } $ & $ 38.72 _{ -2.27 } ^{ + 2.36 } $ & $ 1.12 _{ -0.09 } ^{ + 0.10 } $ & $ 0.26 _{ -0.04 } ^{ + 0.05 } $ & --- \\  
     DDO064 & P & $ -11.113 $ & $ 0.053 _{ -0.079 } ^{ + 0.110 } $ & $ 6.83 _{ -1.50 } ^{ + 2.12 } $ & $ 63.19 _{ -4.67 } ^{ + 4.68 } $ & $ 1.35 _{ -0.12 } ^{ + 0.13 } $ & $ 0.53 _{ -0.10 } ^{ + 0.13 } $ & --- \\  
     DDO154 & P & $ -11.686 $ & $ 0.008 _{ -0.008 } ^{ + 0.008 } $ & $ 3.86 _{ -0.16 } ^{ + 0.17 } $ & $ 61.88 _{ -2.55 } ^{ + 2.64 } $ & $ 1.42 _{ -0.11 } ^{ + 0.12 } $ & $ 0.20 _{ -0.03 } ^{ + 0.03 } $ & --- \\  
     DDO161 & P & $ -11.549 $ & $ -0.045 _{ -0.010 } ^{ + 0.010 } $ & $ 3.36 _{ -0.30 } ^{ + 0.40 } $ & $ 73.87 _{ -7.98 } ^{ + 8.01 } $ & $ 1.37 _{ -0.12 } ^{ + 0.13 } $ & $ 0.33 _{ -0.06 } ^{ + 0.07 } $ & --- \\  
     DDO168 & P & $ -11.057 $ & $ -0.212 _{ -0.065 } ^{ + 0.053 } $ & $ 4.05 _{ -0.19 } ^{ + 0.20 } $ & $ 35.60 _{ -3.58 } ^{ + 3.72 } $ & $ 1.24 _{ -0.11 } ^{ + 0.12 } $ & $ 0.44 _{ -0.09 } ^{ + 0.11 } $ & --- \\  
     DDO170 & P & $ -11.545 $ & $ 0.037 _{ -0.017 } ^{ + 0.020 } $ & $ 11.81 _{ -1.50 } ^{ + 1.86 } $ & $ 66.27 _{ -6.42 } ^{ + 6.65 } $ & $ 1.28 _{ -0.11 } ^{ + 0.12 } $ & $ 0.67 _{ -0.12 } ^{ + 0.14 } $ & --- \\  
ESO079-G014 & P & $ -10.556 $ & $ -0.031 _{ -0.058 } ^{ + 0.057 } $ & $ 28.02 _{ -3.88 } ^{ + 4.54 } $ & $ 79.98 _{ -4.63 } ^{ + 4.55 } $ & $ 1.38 _{ -0.12 } ^{ + 0.13 } $ & $ 0.56 _{ -0.09 } ^{ + 0.11 } $ & --- \\  
ESO116-G012 & P & $ -10.999 $ & $ -0.066 _{ -0.044 } ^{ + 0.043 } $ & $ 13.40 _{ -2.01 } ^{ + 2.45 } $ & $ 74.69 _{ -2.92 } ^{ + 2.91 } $ & $ 1.39 _{ -0.12 } ^{ + 0.13 } $ & $ 0.46 _{ -0.08 } ^{ + 0.10 } $ & --- \\  
ESO444-G084 & P & $ -11.107 $ & $ -0.095 _{ -0.035 } ^{ + 0.030 } $ & $ 4.68 _{ -0.41 } ^{ + 0.45 } $ & $ 33.05 _{ -2.90 } ^{ + 3.02 } $ & $ 1.34 _{ -0.12 } ^{ + 0.13 } $ & $ 0.47 _{ -0.09 } ^{ + 0.12 } $ & --- \\  
ESO563-G021 & P & $ -10.569 $ & $ -0.032 _{ -0.035 } ^{ + 0.033 } $ & $ 66.25 _{ -7.19 } ^{ + 8.04 } $ & $ 83.63 _{ -2.82 } ^{ + 2.77 } $ & $ 1.45 _{ -0.13 } ^{ + 0.14 } $ & $ 0.69 _{ -0.09 } ^{ + 0.10 } $ & --- \\  
     F563-1 & P & $ -11.698 $ & $ 0.112 _{ -0.052 } ^{ + 0.066 } $ & $ 53.07 _{ -8.42 } ^{ + 10.20 } $ & $ 37.60 _{ -3.86 } ^{ + 4.00 } $ & $ 1.35 _{ -0.12 } ^{ + 0.13 } $ & $ 0.61 _{ -0.10 } ^{ + 0.12 } $ & --- \\  
    F563-V2 & A & $ -11.145 $ & $ 0.179 _{ -0.124 } ^{ + 0.152 } $ & $ 63.34 _{ -10.21 } ^{ + 11.89 } $ & $ 49.74 _{ -7.16 } ^{ + 7.26 } $ & $ 1.34 _{ -0.12 } ^{ + 0.13 } $ & $ 0.64 _{ -0.12 } ^{ + 0.15 } $ & --- \\  
    F565-V2 & P & $ -11.433 $ & $ -0.016 _{ -0.050 } ^{ + 0.047 } $ & $ 51.95 _{ -8.86 } ^{ + 10.61 } $ & $ 65.71 _{ -9.00 } ^{ + 8.93 } $ & $ 1.36 _{ -0.12 } ^{ + 0.13 } $ & $ 0.50 _{ -0.10 } ^{ + 0.13 } $ & --- \\  
     F568-1 & P & $ -11.014 $ & $ -0.033 _{ -0.100 } ^{ + 0.100 } $ & $ 89.63 _{ -8.74 } ^{ + 9.69 } $ & $ 32.63 _{ -4.54 } ^{ + 4.65 } $ & $ 1.34 _{ -0.12 } ^{ + 0.13 } $ & $ 0.59 _{ -0.12 } ^{ + 0.15 } $ & --- \\  
     F568-3 & P & $ -11.365 $ & $ 0.238 _{ -0.081 } ^{ + 0.104 } $ & $ 84.40 _{ -7.59 } ^{ + 8.27 } $ & $ 62.04 _{ -6.94 } ^{ + 7.10 } $ & $ 1.42 _{ -0.12 } ^{ + 0.13 } $ & $ 0.47 _{ -0.07 } ^{ + 0.08 } $ & --- \\  
    F568-V1 & P & $ -11.227 $ & $ 0.098 _{ -0.055 } ^{ + 0.066 } $ & $ 85.42 _{ -7.47 } ^{ + 8.23 } $ & $ 64.08 _{ -6.36 } ^{ + 6.73 } $ & $ 1.32 _{ -0.11 } ^{ + 0.12 } $ & $ 0.81 _{ -0.13 } ^{ + 0.16 } $ & --- \\  
     F571-8 & A & $ -11.087 $ & $ -0.423 _{ -0.047 } ^{ + 0.052 } $ & $ 28.31 _{ -2.73 } ^{ + 3.66 } $ & $ 82.96 _{ -5.09 } ^{ + 4.27 } $ & $ 1.38 _{ -0.12 } ^{ + 0.13 } $ & $ 0.23 _{ -0.04 } ^{ + 0.04 } $ & --- \\  
    F571-V1 & P & $ -11.480 $ & $ 0.206 _{ -0.108 } ^{ + 0.134 } $ & $ 80.24 _{ -7.25 } ^{ + 7.94 } $ & $ 44.18 _{ -8.24 } ^{ + 8.08 } $ & $ 1.38 _{ -0.12 } ^{ + 0.13 } $ & $ 0.47 _{ -0.09 } ^{ + 0.11 } $ & --- \\  
     F574-1 & P & $ -11.170 $ & $ 0.079 _{ -0.042 } ^{ + 0.048 } $ & $ 100.40 _{ -8.66 } ^{ + 9.48 } $ & $ 75.12 _{ -7.12 } ^{ + 7.08 } $ & $ 1.32 _{ -0.11 } ^{ + 0.12 } $ & $ 0.78 _{ -0.12 } ^{ + 0.14 } $ & --- \\  
    F579-V1 & C & $ -11.130 $ & $ 0.425 _{ -0.089 } ^{ + 0.054 } $ & $ 87.49 _{ -7.64 } ^{ + 8.38 } $ & $ 44.32 _{ -4.59 } ^{ + 5.05 } $ & $ 1.32 _{ -0.11 } ^{ + 0.12 } $ & $ 0.58 _{ -0.09 } ^{ + 0.11 } $ & --- \\  
     F583-1 & P & $ -11.392 $ & $ 0.032 _{ -0.051 } ^{ + 0.055 } $ & $ 35.78 _{ -6.84 } ^{ + 8.42 } $ & $ 67.60 _{ -4.46 } ^{ + 4.53 } $ & $ 1.24 _{ -0.10 } ^{ + 0.11 } $ & $ 0.96 _{ -0.15 } ^{ + 0.16 } $ & --- \\  
     F583-4 & P & $ -11.393 $ & $ 0.081 _{ -0.062 } ^{ + 0.079 } $ & $ 52.51 _{ -8.74 } ^{ + 10.61 } $ & $ 64.47 _{ -8.26 } ^{ + 8.40 } $ & $ 1.37 _{ -0.12 } ^{ + 0.13 } $ & $ 0.49 _{ -0.10 } ^{ + 0.12 } $ & --- \\  
     IC2574 & P & $ -11.599 $ & $ 0.072 _{ -0.014 } ^{ + 0.016 } $ & $ 4.01 _{ -0.18 } ^{ + 0.19 } $ & $ 81.40 _{ -4.83 } ^{ + 4.56 } $ & $ 1.66 _{ -0.12 } ^{ + 0.13 } $ & $ 0.19 _{ -0.02 } ^{ + 0.03 } $ & --- \\  
     IC4202 & P & $ -10.370  $ & $ 0.169 _{ -0.048 } ^{ + 0.053 } $ & $ 99.01 _{ -7.19 } ^{ + 7.73 } $ & $ 89.33 _{ -0.74 } ^{ + 0.47 } $ & $ 1.26 _{ -0.11 } ^{ + 0.11 } $ & $ 0.90 _{ -0.09 } ^{ + 0.09 } $ & $ 0.44 _{ -0.04 } ^{ + 0.04 } $  \\
   KK98-251 & P & $ -11.418 $ & $ 0.269 _{ -0.099 } ^{ + 0.125 } $ & $ 7.38 _{ -1.52 } ^{ + 1.68 } $ & $ 62.87 _{ -4.52 } ^{ + 4.56 } $ & $ 1.37 _{ -0.12 } ^{ + 0.13 } $ & $ 0.47 _{ -0.09 } ^{ + 0.12 } $ & --- \\  
    NGC0024 & P & $ -10.977 $ & $ -0.004 _{ -0.016 } ^{ + 0.017 } $ & $ 7.47 _{ -0.34 } ^{ + 0.35 } $ & $ 67.25 _{ -2.70 } ^{ + 2.71 } $ & $ 1.34 _{ -0.12 } ^{ + 0.13 } $ & $ 0.99 _{ -0.10 } ^{ + 0.11 } $ & --- \\  
    NGC0055 & P & $ -11.385 $ & $ 0.050 _{ -0.019 } ^{ + 0.021 } $ & $ 1.94 _{ -0.09 } ^{ + 0.09 } $ & $ 75.20 _{ -3.11 } ^{ + 3.15 } $ & $ 1.33 _{ -0.11 } ^{ + 0.12 } $ & $ 0.21 _{ -0.03 } ^{ + 0.04 } $ & --- \\  
    NGC0100 & P & $ -11.304 $ & $ -0.106 _{ -0.044 } ^{ + 0.039 } $ & $ 9.64 _{ -1.56 } ^{ + 1.88 } $ & $ 88.80 _{ -0.91 } ^{ + 0.75 } $ & $ 1.39 _{ -0.12 } ^{ + 0.14 } $ & $ 0.39 _{ -0.07 } ^{ + 0.09 } $ & --- \\  
    NGC0247 & P & $ -11.096 $ & $ 0.182 _{ -0.046 } ^{ + 0.056 } $ & $ 3.76 _{ -0.18 } ^{ + 0.19 } $ & $ 75.58 _{ -2.82 } ^{ + 2.82 } $ & $ 1.29 _{ -0.11 } ^{ + 0.12 } $ & $ 1.03 _{ -0.11 } ^{ + 0.12 } $ & --- \\  
    NGC0289 & P & $ -11.673 $ & $ 0.118 _{ -0.026 } ^{ + 0.031 } $ & $ 19.90 _{ -2.65 } ^{ + 3.16 } $ & $ 54.33 _{ -4.00 } ^{ + 4.10 } $ & $ 1.43 _{ -0.12 } ^{ + 0.13 } $ & $ 0.44 _{ -0.06 } ^{ + 0.07 } $ & --- \\  
    NGC0300 & P & $ -11.496 $ & $ -0.009 _{ -0.026 } ^{ + 0.026 } $ & $ 2.03 _{ -0.09 } ^{ + 0.10 } $ & $ 47.22 _{ -4.57 } ^{ + 5.34 } $ & $ 1.34 _{ -0.12 } ^{ + 0.13 } $ & $ 0.40 _{ -0.06 } ^{ + 0.08 } $ & --- \\  
    NGC0801 & P & $ -10.840 $ & $ 0.174 _{ -0.023 } ^{ + 0.024 } $ & $ 68.41 _{ -5.84 } ^{ + 6.50 } $ & $ 79.93 _{ -1.00 } ^{ + 1.01 } $ & $ 1.44 _{ -0.13 } ^{ + 0.14 } $ & $ 0.60 _{ -0.06 } ^{ + 0.07 } $ & --- \\  
    NGC0891 & P & $ -10.407  $ & $ -0.118 _{ -0.022 } ^{ + 0.022 } $ & $ 9.83 _{ -0.44 } ^{ + 0.46 } $ & $ 89.32 _{ -0.73 } ^{ + 0.48 } $ & $ 1.34 _{ -0.11 } ^{ + 0.13 } $ & $ 0.33 _{ -0.02 } ^{ + 0.02 } $ & $ 0.52 _{ -0.06 } ^{ + 0.06 } $  \\
    NGC1003 & P & $ -11.640 $ & $ -0.056 _{ -0.009 } ^{ + 0.008 } $ & $ 6.53 _{ -0.58 } ^{ + 0.65 } $ & $ 70.19 _{ -4.49 } ^{ + 4.53 } $ & $ 1.22 _{ -0.10 } ^{ + 0.11 } $ & $ 0.77 _{ -0.10 } ^{ + 0.11 } $ & --- \\  
    NGC1090 & P & $ -11.006 $ & $ 0.059 _{ -0.023 } ^{ + 0.024 } $ & $ 31.86 _{ -4.02 } ^{ + 4.65 } $ & $ 65.31 _{ -2.87 } ^{ + 2.89 } $ & $ 1.36 _{ -0.12 } ^{ + 0.13 } $ & $ 0.52 _{ -0.08 } ^{ + 0.09 } $ & --- \\  
    NGC2403 & P & $ -11.512 $ & $ -0.019 _{ -0.005 } ^{ + 0.005 } $ & $ 3.59 _{ -0.13 } ^{ + 0.13 } $ & $ 72.06 _{ -2.22 } ^{ + 2.27 } $ & $ 0.76 _{ -0.05 } ^{ + 0.06 } $ & $ 0.39 _{ -0.02 } ^{ + 0.02 } $ & --- \\  
    NGC2683 & P & $ -11.126  $ & $ 0.086 _{ -0.025 } ^{ + 0.028 } $ & $ 9.88 _{ -0.45 } ^{ + 0.47 } $ & $ 81.02 _{ -4.53 } ^{ + 4.38 } $ & $ 1.41 _{ -0.12 } ^{ + 0.14 } $ & $ 0.56 _{ -0.05 } ^{ + 0.05 } $ & $ 0.69 _{ -0.14 } ^{ + 0.17 } $  \\  
    NGC2841 & P & $ -11.008  $ & $ -0.028 _{ -0.013 } ^{ + 0.013 } $ & $ 14.02 _{ -0.91 } ^{ + 0.98 } $ & $ 82.96 _{ -5.40 } ^{ + 4.46 } $ & $ 1.31 _{ -0.11 } ^{ + 0.12 } $ & $ 0.91 _{ -0.09 } ^{ + 0.10 } $ & $ 0.96 _{ -0.07 } ^{ + 0.08 } $  \\  
    NGC2903 & P & $ -11.325 $ & $ 0.040 _{ -0.008 } ^{ + 0.008 } $ & $ 12.46 _{ -0.88 } ^{ + 0.97 } $ & $ 69.16 _{ -2.76 } ^{ + 2.76 } $ & $ 1.26 _{ -0.10 } ^{ + 0.11 } $ & $ 0.18 _{ -0.02 } ^{ + 0.02 } $ & --- \\  
    NGC2915 & P & $ -11.655 $ & $ -0.053 _{ -0.014 } ^{ + 0.013 } $ & $ 4.12 _{ -0.19 } ^{ + 0.20 } $ & $ 62.27 _{ -3.37 } ^{ + 3.41 } $ & $ 1.35 _{ -0.12 } ^{ + 0.13 } $ & $ 0.58 _{ -0.09 } ^{ + 0.11 } $ & --- \\  
 \hline
\end{tabular}
\end{center}
\end{table}
\newpage

\setcounter{table}{1}
\begin{table}
\caption{(continued) Fitted model parameters}
\begin{center}
  \begin{tabular}{lcccccccc} \hline
    galaxy & PDF-quality  & $x_{0,3}$  & $\tilde{e}$ &  $D$ [Mpc] & $i$ [$^\circ$] & $\Upsilon_{\rm gas}$ & $\Upsilon_{\rm disk}$  & $\Upsilon_{\rm bulge}$  \\
    \hline
    NGC2955 & P & $ -10.505  $ & $ 0.030 _{ -0.050 } ^{ + 0.051 } $ & $ 90.28 _{ -7.34 } ^{ + 8.10 } $ & $ 59.69 _{ -4.86 } ^{ + 5.30 } $ & $ 1.43 _{ -0.13 } ^{ + 0.14 } $ & $ 0.32 _{ -0.04 } ^{ + 0.05 } $ & $ 0.72 _{ -0.08 } ^{ + 0.09 } $  \\
    NGC2976 & B & $ -10.253 $ & $ 0.380 _{ -0.112 } ^{ + 0.083 } $ & $ 3.62 _{ -0.17 } ^{ + 0.18 } $ & $ 76.57 _{ -6.12 } ^{ + 6.22 } $ & $ 1.44 _{ -0.12 } ^{ + 0.13 } $ & $ 0.46 _{ -0.06 } ^{ + 0.06 } $ & --- \\  
    NGC2998 & P & $ -10.813 $ & $ 0.103 _{ -0.027 } ^{ + 0.029 } $ & $ 69.88 _{ -7.57 } ^{ + 8.60 } $ & $ 58.67 _{ -1.95 } ^{ + 1.94 } $ & $ 1.44 _{ -0.13 } ^{ + 0.14 } $ & $ 0.54 _{ -0.07 } ^{ + 0.09 } $ & --- \\  
    NGC3109 & P & $ -11.535 $ & $ 0.012 _{ -0.010 } ^{ + 0.010 } $ & $ 1.40 _{ -0.06 } ^{ + 0.07 } $ & $ 76.85 _{ -3.81 } ^{ + 3.90 } $ & $ 1.68 _{ -0.13 } ^{ + 0.14 } $ & $ 0.24 _{ -0.04 } ^{ + 0.05 } $ & --- \\  
    NGC3198 & P & $ -11.532 $ & $ 0.055 _{ -0.011 } ^{ + 0.012 } $ & $ 15.27 _{ -1.08 } ^{ + 1.17 } $ & $ 75.62 _{ -2.71 } ^{ + 2.73 } $ & $ 1.36 _{ -0.12 } ^{ + 0.13 } $ & $ 0.43 _{ -0.04 } ^{ + 0.04 } $ & --- \\  
    NGC3521 & P & $ -10.304 $ & $ -0.116 _{ -0.067 } ^{ + 0.061 } $ & $ 6.59 _{ -0.96 } ^{ + 1.11 } $ & $ 78.29 _{ -4.36 } ^{ + 4.35 } $ & $ 1.42 _{ -0.12 } ^{ + 0.14 } $ & $ 0.57 _{ -0.09 } ^{ + 0.11 } $ & --- \\  
    NGC3726 & P & $ -11.048 $ & $ -0.002 _{ -0.031 } ^{ + 0.031 } $ & $ 14.33 _{ -1.39 } ^{ + 1.54 } $ & $ 52.19 _{ -1.95 } ^{ + 1.97 } $ & $ 1.34 _{ -0.12 } ^{ + 0.13 } $ & $ 0.44 _{ -0.06 } ^{ + 0.07 } $ & --- \\  
    NGC3741 & P & $ -11.943 $ & $ -0.016 _{ -0.009 } ^{ + 0.009 } $ & $ 3.10 _{ -0.15 } ^{ + 0.16 } $ & $ 69.51 _{ -3.86 } ^{ + 3.92 } $ & $ 1.35 _{ -0.12 } ^{ + 0.13 } $ & $ 0.34 _{ -0.06 } ^{ + 0.07 } $ & --- \\  
    NGC3769 & P & $ -11.675 $ & $ 0.021 _{ -0.018 } ^{ + 0.019 } $ & $ 17.35 _{ -1.59 } ^{ + 1.75 } $ & $ 70.20 _{ -1.97 } ^{ + 1.98 } $ & $ 1.41 _{ -0.12 } ^{ + 0.14 } $ & $ 0.38 _{ -0.06 } ^{ + 0.07 } $ & --- \\  
    NGC3877 & A & $ -10.200 $ & $ 0.238 _{ -0.155 } ^{ + 0.157 } $ & $ 17.45 _{ -1.85 } ^{ + 2.06 } $ & $ 76.04 _{ -1.00 } ^{ + 0.99 } $ & $ 1.39 _{ -0.12 } ^{ + 0.13 } $ & $ 0.50 _{ -0.07 } ^{ + 0.08 } $ & --- \\  
    NGC3893 & P & $ -10.658 $ & $ -0.032 _{ -0.049 } ^{ + 0.048 } $ & $ 18.28 _{ -1.86 } ^{ + 2.06 } $ & $ 49.70 _{ -1.90 } ^{ + 1.91 } $ & $ 1.41 _{ -0.12 } ^{ + 0.14 } $ & $ 0.45 _{ -0.06 } ^{ + 0.07 } $ & --- \\  
    NGC3917 & P & $ -10.822 $ & $ 0.111 _{ -0.050 } ^{ + 0.057 } $ & $ 19.37 _{ -2.05 } ^{ + 2.30 } $ & $ 79.33 _{ -1.96 } ^{ + 1.97 } $ & $ 1.39 _{ -0.12 } ^{ + 0.13 } $ & $ 0.61 _{ -0.08 } ^{ + 0.10 } $ & --- \\  
    NGC3949 & P & $ -10.017 $ & $ -0.024 _{ -0.180 } ^{ + 0.216 } $ & $ 16.90 _{ -1.85 } ^{ + 2.09 } $ & $ 54.98 _{ -1.97 } ^{ + 1.98 } $ & $ 1.40 _{ -0.12 } ^{ + 0.13 } $ & $ 0.43 _{ -0.06 } ^{ + 0.07 } $ & --- \\  
    NGC3953 & B & $ -10.070 $ & $ 0.390 _{ -0.115 } ^{ + 0.077 } $ & $ 18.83 _{ -1.97 } ^{ + 2.20 } $ & $ 62.12 _{ -0.99 } ^{ + 1.00 } $ & $ 1.41 _{ -0.12 } ^{ + 0.14 } $ & $ 0.60 _{ -0.07 } ^{ + 0.08 } $ & --- \\  
    NGC3972 & P & $ -10.550 $ & $ -0.096 _{ -0.071 } ^{ + 0.075 } $ & $ 17.03 _{ -1.92 } ^{ + 2.17 } $ & $ 77.01 _{ -1.00 } ^{ + 1.00 } $ & $ 1.38 _{ -0.12 } ^{ + 0.13 } $ & $ 0.46 _{ -0.08 } ^{ + 0.09 } $ & --- \\  
    NGC3992 & P & $ -10.731 $ & $ 0.095 _{ -0.028 } ^{ + 0.030 } $ & $ 24.45 _{ -1.94 } ^{ + 2.10 } $ & $ 56.74 _{ -1.92 } ^{ + 1.92 } $ & $ 1.42 _{ -0.12 } ^{ + 0.14 } $ & $ 0.68 _{ -0.08 } ^{ + 0.09 } $ & --- \\  
    NGC4010 & P & $ -10.715 $ & $ -0.057 _{ -0.050 } ^{ + 0.051 } $ & $ 16.26 _{ -1.80 } ^{ + 2.01 } $ & $ 88.80 _{ -0.91 } ^{ + 0.75 } $ & $ 1.41 _{ -0.13 } ^{ + 0.14 } $ & $ 0.36 _{ -0.06 } ^{ + 0.07 } $ & --- \\  
    NGC4013 & P & $ -11.065  $ & $ -0.047 _{ -0.016 } ^{ + 0.015 } $ & $ 14.38 _{ -1.24 } ^{ + 1.35 } $ & $ 88.80 _{ -0.91 } ^{ + 0.75 } $ & $ 1.38 _{ -0.12 } ^{ + 0.13 } $ & $ 0.48 _{ -0.07 } ^{ + 0.08 } $ & $ 0.82 _{ -0.16 } ^{ + 0.20 } $  \\  
    NGC4051 & B & $ -10.205 $ & $ 0.335 _{ -0.150 } ^{ + 0.114 } $ & $ 17.29 _{ -1.87 } ^{ + 2.08 } $ & $ 49.32 _{ -2.80 } ^{ + 2.82 } $ & $ 1.39 _{ -0.12 } ^{ + 0.13 } $ & $ 0.48 _{ -0.07 } ^{ + 0.08 } $ & --- \\  
    NGC4068 & A & $ -11.195 $ & $ 0.321 _{ -0.109 } ^{ + 0.108 } $ & $ 4.37 _{ -0.21 } ^{ + 0.22 } $ & $ 48.19 _{ -4.97 } ^{ + 4.79 } $ & $ 1.39 _{ -0.12 } ^{ + 0.13 } $ & $ 0.43 _{ -0.08 } ^{ + 0.10 } $ & --- \\  
    NGC4085 & P & $ -10.175 $ & $ -0.170 _{ -0.106 } ^{ + 0.114 } $ & $ 15.12 _{ -1.67 } ^{ + 1.89 } $ & $ 81.87 _{ -2.02 } ^{ + 2.01 } $ & $ 1.39 _{ -0.12 } ^{ + 0.13 } $ & $ 0.33 _{ -0.05 } ^{ + 0.06 } $ & --- \\  
    NGC4088 & P & $ -10.621 $ & $ 0.041 _{ -0.047 } ^{ + 0.049 } $ & $ 14.82 _{ -1.50 } ^{ + 1.68 } $ & $ 68.71 _{ -2.00 } ^{ + 1.99 } $ & $ 1.40 _{ -0.12 } ^{ + 0.13 } $ & $ 0.35 _{ -0.05 } ^{ + 0.05 } $ & --- \\  
    NGC4100 & P & $ -10.925 $ & $ 0.087 _{ -0.025 } ^{ + 0.026 } $ & $ 19.53 _{ -1.92 } ^{ + 2.13 } $ & $ 73.69 _{ -1.94 } ^{ + 1.95 } $ & $ 1.40 _{ -0.12 } ^{ + 0.14 } $ & $ 0.57 _{ -0.07 } ^{ + 0.08 } $ & --- \\  
    NGC4138 & P & $ -10.717  $ & $ 0.111 _{ -0.073 } ^{ + 0.087 } $ & $ 18.72 _{ -1.90 } ^{ + 2.11 } $ & $ 54.34 _{ -2.77 } ^{ + 2.78 } $ & $ 1.40 _{ -0.12 } ^{ + 0.13 } $ & $ 0.56 _{ -0.09 } ^{ + 0.11 } $ & $ 0.68 _{ -0.14 } ^{ + 0.17 } $  \\
    NGC4157 & P & $ -10.880  $ & $ -0.017 _{ -0.031 } ^{ + 0.031 } $ & $ 15.20 _{ -1.46 } ^{ + 1.61 } $ & $ 81.93 _{ -3.02 } ^{ + 2.98 } $ & $ 1.39 _{ -0.12 } ^{ + 0.13 } $ & $ 0.42 _{ -0.06 } ^{ + 0.06 } $ & $ 0.65 _{ -0.13 } ^{ + 0.16 } $  \\  
    NGC4183 & P & $ -11.314 $ & $ 0.108 _{ -0.030 } ^{ + 0.034 } $ & $ 18.19 _{ -1.80 } ^{ + 2.03 } $ & $ 82.23 _{ -1.97 } ^{ + 1.96 } $ & $ 1.33 _{ -0.11 } ^{ + 0.13 } $ & $ 0.69 _{ -0.09 } ^{ + 0.11 } $ & --- \\  
    NGC4214 & P & $ -11.329 $ & $ -0.036 _{ -0.036 } ^{ + 0.034 } $ & $ 2.82 _{ -0.13 } ^{ + 0.14 } $ & $ 18.24 _{ -1.52 } ^{ + 1.63 } $ & $ 1.35 _{ -0.12 } ^{ + 0.13 } $ & $ 0.43 _{ -0.08 } ^{ + 0.11 } $ & --- \\  
    NGC4217 & P & $ -10.604  $ & $ -0.148 _{ -0.046 } ^{ + 0.044 } $ & $ 15.46 _{ -1.40 } ^{ + 1.53 } $ & $ 85.96 _{ -1.97 } ^{ + 1.90 } $ & $ 1.41 _{ -0.12 } ^{ + 0.14 } $ & $ 0.86 _{ -0.15 } ^{ + 0.18 } $ & $ 0.23 _{ -0.03 } ^{ + 0.03 } $  \\  
    NGC4559 & P & $ -11.130 $ & $ 0.028 _{ -0.036 } ^{ + 0.038 } $ & $ 7.42 _{ -1.03 } ^{ + 1.25 } $ & $ 67.19 _{ -0.99 } ^{ + 0.99 } $ & $ 1.38 _{ -0.12 } ^{ + 0.13 } $ & $ 0.46 _{ -0.07 } ^{ + 0.09 } $ & --- \\  
    NGC5005 & P & $ -9.819  $ & $ -0.121 _{ -0.191 } ^{ + 0.243 } $ & $ 16.16 _{ -1.15 } ^{ + 1.23 } $ & $ 68.15 _{ -1.96 } ^{ + 1.98 } $ & $ 1.42 _{ -0.12 } ^{ + 0.14 } $ & $ 0.49 _{ -0.08 } ^{ + 0.08 } $ & $ 0.54 _{ -0.07 } ^{ + 0.08 } $  \\
    NGC5033 & P & $ -11.306  $ & $ 0.098 _{ -0.011 } ^{ + 0.012 } $ & $ 23.48 _{ -1.82 } ^{ + 1.96 } $ & $ 66.26 _{ -0.99 } ^{ + 0.99 } $ & $ 1.45 _{ -0.12 } ^{ + 0.13 } $ & $ 0.43 _{ -0.05 } ^{ + 0.06 } $ & $ 0.28 _{ -0.04 } ^{ + 0.04 } $  \\
    NGC5055 & P & $ -11.416 $ & $ 0.053 _{ -0.004 } ^{ + 0.004 } $ & $ 9.82 _{ -0.28 } ^{ + 0.29 } $ & $ 59.40 _{ -2.19 } ^{ + 2.38 } $ & $ 1.50 _{ -0.12 } ^{ + 0.14 } $ & $ 0.31 _{ -0.02 } ^{ + 0.02 } $ & --- \\  
    NGC5371 & P & $ -10.367 $ & $ 0.249 _{ -0.043 } ^{ + 0.042 } $ & $ 16.44 _{ -2.15 } ^{ + 2.49 } $ & $ 52.10 _{ -1.99 } ^{ + 2.01 } $ & $ 1.36 _{ -0.12 } ^{ + 0.13 } $ & $ 1.38 _{ -0.20 } ^{ + 0.23 } $ & --- \\  
    NGC5585 & P & $ -11.208 $ & $ -0.083 _{ -0.033 } ^{ + 0.032 } $ & $ 5.00 _{ -0.69 } ^{ + 0.81 } $ & $ 51.78 _{ -1.95 } ^{ + 1.95 } $ & $ 1.38 _{ -0.12 } ^{ + 0.13 } $ & $ 0.36 _{ -0.06 } ^{ + 0.07 } $ & --- \\  
    NGC5907 & P & $ -10.991 $ & $ 0.090 _{ -0.013 } ^{ + 0.013 } $ & $ 16.08 _{ -0.76 } ^{ + 0.80 } $ & $ 87.51 _{ -1.88 } ^{ + 1.55 } $ & $ 1.31 _{ -0.11 } ^{ + 0.12 } $ & $ 0.65 _{ -0.04 } ^{ + 0.04 } $ & --- \\  
    NGC5985 & P & $ -10.685  $ & $ 0.174 _{ -0.030 } ^{ + 0.032 } $ & $ 72.59 _{ -8.01 } ^{ + 8.94 } $ & $ 62.13 _{ -1.90 } ^{ + 1.89 } $ & $ 1.34 _{ -0.11 } ^{ + 0.13 } $ & $ 0.43 _{ -0.06 } ^{ + 0.07 } $ & $ 1.86 _{ -0.23 } ^{ + 0.26 } $  \\
    NGC6015 & P & $ -10.872 $ & $ -0.093 _{ -0.026 } ^{ + 0.024 } $ & $ 8.05 _{ -0.78 } ^{ + 0.86 } $ & $ 60.87 _{ -1.93 } ^{ + 1.95 } $ & $ 1.37 _{ -0.12 } ^{ + 0.13 } $ & $ 1.68 _{ -0.18 } ^{ + 0.21 } $ & --- \\  
    NGC6195 & P & $ -10.479  $ & $ -0.012 _{ -0.036 } ^{ + 0.036 } $ & $ 110.16 _{ -8.67 } ^{ + 9.46 } $ & $ 59.77 _{ -4.21 } ^{ + 4.43 } $ & $ 1.46 _{ -0.13 } ^{ + 0.14 } $ & $ 0.29 _{ -0.04 } ^{ + 0.05 } $ & $ 0.81 _{ -0.08 } ^{ + 0.09 } $  \\  
    NGC6503 & P & $ -11.601 $ & $ 0.008 _{ -0.006 } ^{ + 0.006 } $ & $ 6.79 _{ -0.28 } ^{ + 0.29 } $ & $ 75.79 _{ -1.84 } ^{ + 1.86 } $ & $ 1.37 _{ -0.12 } ^{ + 0.13 } $ & $ 0.41 _{ -0.03 } ^{ + 0.03 } $ & --- \\  
    NGC6674 & P & $ -10.969  $ & $ -0.016 _{ -0.023 } ^{ + 0.020 } $ & $ 37.38 _{ -5.29 } ^{ + 6.18 } $ & $ 52.20 _{ -5.25 } ^{ + 5.50 } $ & $ 1.37 _{ -0.12 } ^{ + 0.13 } $ & $ 1.08 _{ -0.28 } ^{ + 0.33 } $ & $ 1.39 _{ -0.41 } ^{ + 0.59 } $  \\  
    NGC6789 & A & $ -10.592 $ & $ -0.268 _{ -0.127 } ^{ + 0.124 } $ & $ 3.52 _{ -0.17 } ^{ + 0.18 } $ & $ 46.93 _{ -5.60 } ^{ + 6.32 } $ & $ 1.35 _{ -0.12 } ^{ + 0.13 } $ & $ 0.51 _{ -0.10 } ^{ + 0.13 } $ & --- \\  
    NGC6946 & P & $ -10.854  $ & $ 0.045 _{ -0.027 } ^{ + 0.025 } $ & $ 4.25 _{ -0.49 } ^{ + 0.55 } $ & $ 41.92 _{ -1.81 } ^{ + 1.82 } $ & $ 1.40 _{ -0.12 } ^{ + 0.13 } $ & $ 0.49 _{ -0.06 } ^{ + 0.07 } $ & $ 0.56 _{ -0.06 } ^{ + 0.07 } $  \\ 
    NGC7331 & P & $ -10.758  $ & $ -0.079 _{ -0.020 } ^{ + 0.018 } $ & $ 12.26 _{ -0.83 } ^{ + 0.89 } $ & $ 74.96 _{ -1.98 } ^{ + 1.99 } $ & $ 1.34 _{ -0.11 } ^{ + 0.12 } $ & $ 0.42 _{ -0.04 } ^{ + 0.04 } $ & $ 0.63 _{ -0.12 } ^{ + 0.14 } $  \\  
  \hline
\end{tabular}
\end{center}
\end{table}

\newpage

\setcounter{table}{1}
\begin{table}
\caption{(continued) Fitted model parameters}
\begin{center}
  \begin{tabular}{lcccccccc} \hline
 galaxy & PDF-quality  & $x_{0,3}$  & $\tilde{e}$ &  $D$ [Mpc]  & $i$ [$^\circ$] & $\Upsilon_{\rm gas}$ & $\Upsilon_{\rm disk}$  & $\Upsilon_{\rm bulge}$  \\
 \hline
    NGC7793 & P & $ -10.944 $ & $ 0.233 _{ -0.057 } ^{ + 0.069 } $ & $ 3.59 _{ -0.17 } ^{ + 0.18 } $ & $ 69.18 _{ -5.78 } ^{ + 6.00 } $ & $ 1.44 _{ -0.13 } ^{ + 0.14 } $ & $ 0.33 _{ -0.03 } ^{ + 0.04 } $ & --- \\  
    NGC7814 & P & $ -10.563  $ & $ -0.110 _{ -0.020 } ^{ + 0.020 } $ & $ 14.75 _{ -0.61 } ^{ + 0.63 } $ & $ 89.33 _{ -0.73 } ^{ + 0.47 } $ & $ 1.40 _{ -0.12 } ^{ + 0.14 } $ & $ 0.83 _{ -0.12 } ^{ + 0.13 } $ & $ 0.58 _{ -0.05 } ^{ + 0.05 } $  \\  
   UGC00128 & P & $ -11.681 $ & $ 0.016 _{ -0.007 } ^{ + 0.007 } $ & $ 49.32 _{ -5.77 } ^{ + 6.79 } $ & $ 52.54 _{ -4.88 } ^{ + 5.71 } $ & $ 1.12 _{ -0.09 } ^{ + 0.10 } $ & $ 1.78 _{ -0.19 } ^{ + 0.21 } $ & --- \\  
   UGC00191 & P & $ -10.944 $ & $ 0.096 _{ -0.038 } ^{ + 0.048 } $ & $ 16.10 _{ -2.60 } ^{ + 3.27 } $ & $ 47.98 _{ -4.24 } ^{ + 4.39 } $ & $ 1.28 _{ -0.11 } ^{ + 0.12 } $ & $ 0.79 _{ -0.11 } ^{ + 0.13 } $ & --- \\  
   UGC00634 & P & $ -11.179 $ & $ 0.028 _{ -0.029 } ^{ + 0.032 } $ & $ 29.84 _{ -5.01 } ^{ + 6.22 } $ & $ 41.35 _{ -4.85 } ^{ + 5.35 } $ & $ 1.39 _{ -0.12 } ^{ + 0.13 } $ & $ 0.45 _{ -0.08 } ^{ + 0.10 } $ & --- \\  
   UGC00731 & P & $ -11.203 $ & $ -0.243 _{ -0.056 } ^{ + 0.049 } $ & $ 4.39 _{ -0.71 } ^{ + 0.85 } $ & $ 54.98 _{ -3.04 } ^{ + 3.05 } $ & $ 1.22 _{ -0.11 } ^{ + 0.12 } $ & $ 0.61 _{ -0.14 } ^{ + 0.19 } $ & --- \\  
   UGC00891 & P & $ -11.216 $ & $ -0.113 _{ -0.028 } ^{ + 0.024 } $ & $ 5.43 _{ -0.77 } ^{ + 0.90 } $ & $ 56.28 _{ -4.91 } ^{ + 5.04 } $ & $ 1.34 _{ -0.12 } ^{ + 0.13 } $ & $ 0.42 _{ -0.08 } ^{ + 0.10 } $ & --- \\  
   UGC01230 & C & $ -11.686 $ & $ 0.404 _{ -0.100 } ^{ + 0.069 } $ & $ 49.83 _{ -7.50 } ^{ + 9.02 } $ & $ 38.61 _{ -4.63 } ^{ + 5.00 } $ & $ 1.32 _{ -0.11 } ^{ + 0.12 } $ & $ 0.59 _{ -0.10 } ^{ + 0.12 } $ & --- \\  
   UGC01281 & P & $ -11.411 $ & $ 0.014 _{ -0.017 } ^{ + 0.019 } $ & $ 5.32 _{ -0.23 } ^{ + 0.24 } $ & $ 89.33 _{ -0.73 } ^{ + 0.47 } $ & $ 1.42 _{ -0.12 } ^{ + 0.13 } $ & $ 0.45 _{ -0.08 } ^{ + 0.09 } $ & --- \\  
   UGC02023 & C & $ -11.154 $ & $ 0.344 _{ -0.178 } ^{ + 0.112 } $ & $ 8.88 _{ -1.93 } ^{ + 2.51 } $ & $ 23.95 _{ -4.72 } ^{ + 5.01 } $ & $ 1.36 _{ -0.12 } ^{ + 0.13 } $ & $ 0.45 _{ -0.09 } ^{ + 0.11 } $ & --- \\  
   UGC02259 & P & $ -11.108 $ & $ 0.198 _{ -0.060 } ^{ + 0.081 } $ & $ 15.96 _{ -2.65 } ^{ + 3.34 } $ & $ 44.35 _{ -2.72 } ^{ + 2.74 } $ & $ 1.30 _{ -0.11 } ^{ + 0.12 } $ & $ 0.89 _{ -0.13 } ^{ + 0.16 } $ & --- \\  
   UGC02487 & P & $ -11.058  $ & $ 0.095 _{ -0.011 } ^{ + 0.011 } $ & $ 73.74 _{ -8.02 } ^{ + 9.10 } $ & $ 46.16 _{ -3.44 } ^{ + 3.58 } $ & $ 1.49 _{ -0.13 } ^{ + 0.15 } $ & $ 0.58 _{ -0.10 } ^{ + 0.12 } $ & $ 0.59 _{ -0.09 } ^{ + 0.10 } $  \\
   UGC02885 & P & $ -10.968  $ & $ 0.009 _{ -0.025 } ^{ + 0.025 } $ & $ 81.59 _{ -6.33 } ^{ + 6.83 } $ & $ 66.66 _{ -3.54 } ^{ + 3.63 } $ & $ 1.44 _{ -0.13 } ^{ + 0.14 } $ & $ 0.44 _{ -0.06 } ^{ + 0.06 } $ & $ 0.92 _{ -0.10 } ^{ + 0.11 } $  \\
   UGC02916 & P & $ -10.824  $ & $ 0.258 _{ -0.060 } ^{ + 0.074 } $ & $ 58.40 _{ -5.72 } ^{ + 6.44 } $ & $ 58.57 _{ -3.81 } ^{ + 3.95 } $ & $ 1.39 _{ -0.12 } ^{ + 0.13 } $ & $ 1.10 _{ -0.13 } ^{ + 0.15 } $ & $ 0.42 _{ -0.04 } ^{ + 0.05 } $  \\
   UGC02953 & P & $ -11.147  $ & $ -0.006 _{ -0.006 } ^{ + 0.006 } $ & $ 13.51 _{ -0.76 } ^{ + 0.87 } $ & $ 64.55 _{ -3.02 } ^{ + 3.06 } $ & $ 1.51 _{ -0.13 } ^{ + 0.15 } $ & $ 0.57 _{ -0.02 } ^{ + 0.03 } $ & $ 0.58 _{ -0.02 } ^{ + 0.02 } $  \\  
   UGC03205 & P & $ -10.823  $ & $ 0.004 _{ -0.019 } ^{ + 0.018 } $ & $ 42.35 _{ -4.07 } ^{ + 4.52 } $ & $ 70.74 _{ -3.55 } ^{ + 3.58 } $ & $ 1.33 _{ -0.11 } ^{ + 0.12 } $ & $ 0.63 _{ -0.08 } ^{ + 0.09 } $ & $ 1.30 _{ -0.13 } ^{ + 0.14 } $  \\
   UGC03546 & P & $ -10.772  $ & $ 0.020 _{ -0.023 } ^{ + 0.022 } $ & $ 24.33 _{ -2.87 } ^{ + 3.31 } $ & $ 60.83 _{ -4.21 } ^{ + 4.33 } $ & $ 1.39 _{ -0.12 } ^{ + 0.13 } $ & $ 0.58 _{ -0.08 } ^{ + 0.10 } $ & $ 0.43 _{ -0.05 } ^{ + 0.06 } $  \\
   UGC03580 & P & $ -11.295  $ & $ -0.046 _{ -0.012 } ^{ + 0.012 } $ & $ 15.17 _{ -1.24 } ^{ + 1.38 } $ & $ 67.11 _{ -3.55 } ^{ + 3.58 } $ & $ 1.51 _{ -0.13 } ^{ + 0.14 } $ & $ 0.47 _{ -0.05 } ^{ + 0.06 } $ & $ 0.15 _{ -0.02 } ^{ + 0.02 } $  \\  
   UGC04278 & P & $ -10.918 $ & $ -0.181 _{ -0.047 } ^{ + 0.039 } $ & $ 5.85 _{ -0.89 } ^{ + 0.97 } $ & $ 88.01 _{ -2.17 } ^{ + 1.40 } $ & $ 1.38 _{ -0.12 } ^{ + 0.13 } $ & $ 0.37 _{ -0.07 } ^{ + 0.08 } $ & --- \\  
   UGC04325 & B & $ -10.700 $ & $ 0.340 _{ -0.119 } ^{ + 0.105 } $ & $ 14.00 _{ -2.13 } ^{ + 2.46 } $ & $ 44.10 _{ -2.70 } ^{ + 2.74 } $ & $ 1.26 _{ -0.11 } ^{ + 0.12 } $ & $ 1.09 _{ -0.17 } ^{ + 0.20 } $ & --- \\  
   UGC04483 & P & $ -11.307 $ & $ 0.141 _{ -0.043 } ^{ + 0.051 } $ & $ 3.36 _{ -0.29 } ^{ + 0.31 } $ & $ 58.91 _{ -2.94 } ^{ + 2.95 } $ & $ 1.37 _{ -0.12 } ^{ + 0.13 } $ & $ 0.48 _{ -0.09 } ^{ + 0.11 } $ & --- \\  
   UGC04499 & P & $ -11.197 $ & $ 0.109 _{ -0.069 } ^{ + 0.090 } $ & $ 13.25 _{ -2.64 } ^{ + 3.54 } $ & $ 51.37 _{ -2.89 } ^{ + 2.90 } $ & $ 1.38 _{ -0.12 } ^{ + 0.13 } $ & $ 0.49 _{ -0.08 } ^{ + 0.10 } $ & --- \\  
   UGC05005 & P & $ -11.663 $ & $ 0.133 _{ -0.078 } ^{ + 0.106 } $ & $ 51.40 _{ -8.80 } ^{ + 10.78 } $ & $ 52.27 _{ -8.81 } ^{ + 8.95 } $ & $ 1.42 _{ -0.13 } ^{ + 0.14 } $ & $ 0.39 _{ -0.07 } ^{ + 0.09 } $ & --- \\  
   UGC05253 & P & $ -11.332  $ & $ 0.056 _{ -0.008 } ^{ + 0.008 } $ & $ 20.87 _{ -1.95 } ^{ + 2.22 } $ & $ 52.13 _{ -2.99 } ^{ + 3.07 } $ & $ 1.40 _{ -0.12 } ^{ + 0.13 } $ & $ 0.30 _{ -0.03 } ^{ + 0.04 } $ & $ 0.39 _{ -0.03 } ^{ + 0.03 } $  \\
   UGC05414 & P & $ -11.015 $ & $ 0.071 _{ -0.090 } ^{ + 0.114 } $ & $ 8.84 _{ -2.02 } ^{ + 2.72 } $ & $ 55.45 _{ -2.96 } ^{ + 2.98 } $ & $ 1.40 _{ -0.12 } ^{ + 0.13 } $ & $ 0.42 _{ -0.08 } ^{ + 0.10 } $ & --- \\  
   UGC05716 & P & $ -11.585 $ & $ 0.087 _{ -0.026 } ^{ + 0.031 } $ & $ 22.14 _{ -3.07 } ^{ + 3.76 } $ & $ 65.53 _{ -7.04 } ^{ + 7.60 } $ & $ 1.25 _{ -0.10 } ^{ + 0.11 } $ & $ 0.86 _{ -0.09 } ^{ + 0.11 } $ & --- \\  
   UGC05721 & P & $ -11.391 $ & $ 0.046 _{ -0.029 } ^{ + 0.032 } $ & $ 10.08 _{ -1.36 } ^{ + 1.59 } $ & $ 67.73 _{ -4.23 } ^{ + 4.26 } $ & $ 1.39 _{ -0.12 } ^{ + 0.14 } $ & $ 0.51 _{ -0.09 } ^{ + 0.11 } $ & --- \\  
   UGC05750 & P & $ -11.405 $ & $ 0.137 _{ -0.059 } ^{ + 0.080 } $ & $ 62.33 _{ -10.19 } ^{ + 12.27 } $ & $ 72.17 _{ -7.99 } ^{ + 7.93 } $ & $ 1.37 _{ -0.12 } ^{ + 0.13 } $ & $ 0.55 _{ -0.11 } ^{ + 0.13 } $ & --- \\  
   UGC05764 & A & $ -11.286 $ & $ 0.360 _{ -0.082 } ^{ + 0.085 } $ & $ 15.34 _{ -2.07 } ^{ + 2.27 } $ & $ 74.71 _{ -6.66 } ^{ + 6.73 } $ & $ 1.11 _{ -0.09 } ^{ + 0.10 } $ & $ 2.67 _{ -0.29 } ^{ + 0.33 } $ & --- \\  
   UGC05829 & B & $ -11.254 $ & $ 0.115 _{ -0.157 } ^{ + 0.237 } $ & $ 6.86 _{ -1.70 } ^{ + 2.09 } $ & $ 41.85 _{ -9.62 } ^{ + 9.75 } $ & $ 1.27 _{ -0.11 } ^{ + 0.12 } $ & $ 0.69 _{ -0.17 } ^{ + 0.18 } $ & --- \\  
   UGC05918 & P & $ -11.569 $ & $ 0.040 _{ -0.062 } ^{ + 0.077 } $ & $ 7.36 _{ -1.78 } ^{ + 2.40 } $ & $ 48.64 _{ -4.87 } ^{ + 4.88 } $ & $ 1.32 _{ -0.11 } ^{ + 0.13 } $ & $ 0.58 _{ -0.12 } ^{ + 0.15 } $ & --- \\  
   UGC05986 & P & $ -11.159 $ & $ -0.017 _{ -0.037 } ^{ + 0.037 } $ & $ 12.80 _{ -1.96 } ^{ + 2.32 } $ & $ 88.05 _{ -2.12 } ^{ + 1.37 } $ & $ 1.49 _{ -0.13 } ^{ + 0.15 } $ & $ 0.37 _{ -0.06 } ^{ + 0.08 } $ & --- \\  
   UGC05999 & B & $ -11.309 $ & $ 0.093 _{ -0.171 } ^{ + 0.241 } $ & $ 44.25 _{ -7.24 } ^{ + 8.66 } $ & $ 24.99 _{ -7.61 } ^{ + 8.98 } $ & $ 1.37 _{ -0.12 } ^{ + 0.13 } $ & $ 0.48 _{ -0.09 } ^{ + 0.11 } $ & --- \\  
   UGC06399 & P & $ -11.074 $ & $ -0.004 _{ -0.045 } ^{ + 0.049 } $ & $ 18.50 _{ -2.20 } ^{ + 2.51 } $ & $ 75.22 _{ -1.98 } ^{ + 1.98 } $ & $ 1.37 _{ -0.12 } ^{ + 0.13 } $ & $ 0.54 _{ -0.10 } ^{ + 0.12 } $ & --- \\  
   UGC06446 & P & $ -11.364 $ & $ 0.139 _{ -0.059 } ^{ + 0.076 } $ & $ 17.29 _{ -2.98 } ^{ + 3.79 } $ & $ 54.16 _{ -2.77 } ^{ + 2.81 } $ & $ 1.28 _{ -0.11 } ^{ + 0.12 } $ & $ 0.92 _{ -0.13 } ^{ + 0.15 } $ & --- \\  
   UGC06614 & P & $ -11.009  $ & $ -0.070 _{ -0.040 } ^{ + 0.037 } $ & $ 82.52 _{ -7.45 } ^{ + 8.17 } $ & $ 30.92 _{ -2.83 } ^{ + 3.13 } $ & $ 1.42 _{ -0.13 } ^{ + 0.14 } $ & $ 0.47 _{ -0.09 } ^{ + 0.11 } $ & $ 0.57 _{ -0.11 } ^{ + 0.13 } $  \\  
   UGC06628 & C & $ -11.250 $ & $ 0.397 _{ -0.129 } ^{ + 0.075 } $ & $ 11.46 _{ -2.57 } ^{ + 3.33 } $ & $ 19.93 _{ -3.43 } ^{ + 3.87 } $ & $ 1.34 _{ -0.12 } ^{ + 0.13 } $ & $ 0.48 _{ -0.09 } ^{ + 0.11 } $ & --- \\  
   UGC06667 & P & $ -11.279 $ & $ -0.125 _{ -0.027 } ^{ + 0.025 } $ & $ 15.59 _{ -1.66 } ^{ + 1.82 } $ & $ 88.80 _{ -0.91 } ^{ + 0.75 } $ & $ 1.31 _{ -0.11 } ^{ + 0.12 } $ & $ 0.52 _{ -0.11 } ^{ + 0.14 } $ & --- \\  
   UGC06786 & P & $ -11.242  $ & $ -0.029 _{ -0.013 } ^{ + 0.012 } $ & $ 46.16 _{ -4.00 } ^{ + 4.34 } $ & $ 68.02 _{ -2.69 } ^{ + 2.73 } $ & $ 1.49 _{ -0.13 } ^{ + 0.14 } $ & $ 0.36 _{ -0.05 } ^{ + 0.05 } $ & $ 0.42 _{ -0.04 } ^{ + 0.04 } $  \\  
   UGC06818 & P & $ -11.254 $ & $ -0.002 _{ -0.035 } ^{ + 0.038 } $ & $ 15.66 _{ -1.93 } ^{ + 2.21 } $ & $ 74.64 _{ -3.04 } ^{ + 3.04 } $ & $ 1.42 _{ -0.13 } ^{ + 0.14 } $ & $ 0.31 _{ -0.06 } ^{ + 0.07 } $ & --- \\  
   UGC06917 & P & $ -10.966 $ & $ -0.001 _{ -0.044 } ^{ + 0.046 } $ & $ 17.93 _{ -1.96 } ^{ + 2.22 } $ & $ 56.49 _{ -1.95 } ^{ + 1.96 } $ & $ 1.36 _{ -0.12 } ^{ + 0.13 } $ & $ 0.55 _{ -0.08 } ^{ + 0.09 } $ & --- \\  
   UGC06923 & P & $ -10.751 $ & $ 0.035 _{ -0.071 } ^{ + 0.083 } $ & $ 17.20 _{ -2.05 } ^{ + 2.35 } $ & $ 65.02 _{ -2.00 } ^{ + 1.99 } $ & $ 1.38 _{ -0.12 } ^{ + 0.13 } $ & $ 0.44 _{ -0.08 } ^{ + 0.10 } $ & --- \\  
   UGC06930 & P & $ -11.169 $ & $ 0.232 _{ -0.092 } ^{ + 0.120 } $ & $ 18.37 _{ -2.14 } ^{ + 2.41 } $ & $ 38.86 _{ -3.91 } ^{ + 4.06 } $ & $ 1.36 _{ -0.12 } ^{ + 0.13 } $ & $ 0.58 _{ -0.09 } ^{ + 0.11 } $ & --- \\  
 \hline
\end{tabular}
\end{center}
\end{table}

\newpage

\setcounter{table}{1}
\begin{table}
\caption{(continued) Fitted model parameters}
\begin{center}
  \begin{tabular}{lcccccccc} \hline
 galaxy & PDF-quality & $x_{0,3}$  & $\tilde{e}$ &  $D$ [Mpc]  & $i$ [$^\circ$] & $\Upsilon_{\rm gas}$ & $\Upsilon_{\rm disk}$  & $\Upsilon_{\rm bulge}$  \\
 \hline
   UGC06983 & P & $ -11.198 $ & $ 0.056 _{ -0.035 } ^{ + 0.037 } $ & $ 19.80 _{ -2.06 } ^{ + 2.31 } $ & $ 49.43 _{ -0.99 } ^{ + 0.98 } $ & $ 1.33 _{ -0.11 } ^{ + 0.13 } $ & $ 0.77 _{ -0.10 } ^{ + 0.12 } $ & --- \\  
   UGC07089 & P & $ -11.240 $ & $ 0.094 _{ -0.051 } ^{ + 0.063 } $ & $ 17.02 _{ -2.11 } ^{ + 2.43 } $ & $ 80.15 _{ -2.98 } ^{ + 2.98 } $ & $ 1.40 _{ -0.12 } ^{ + 0.14 } $ & $ 0.40 _{ -0.07 } ^{ + 0.09 } $ & --- \\  
   UGC07125 & P & $ -11.509 $ & $ 0.122 _{ -0.045 } ^{ + 0.063 } $ & $ 13.44 _{ -2.28 } ^{ + 3.16 } $ & $ 87.98 _{ -2.21 } ^{ + 1.42 } $ & $ 1.27 _{ -0.11 } ^{ + 0.12 } $ & $ 0.69 _{ -0.09 } ^{ + 0.11 } $ & --- \\  
   UGC07151 & P & $ -10.861 $ & $ 0.149 _{ -0.049 } ^{ + 0.059 } $ & $ 6.96 _{ -0.32 } ^{ + 0.34 } $ & $ 88.03 _{ -2.15 } ^{ + 1.38 } $ & $ 1.35 _{ -0.12 } ^{ + 0.13 } $ & $ 0.71 _{ -0.08 } ^{ + 0.09 } $ & --- \\  
   UGC07232 & P & $ -10.628 $ & $ -0.033 _{ -0.087 } ^{ + 0.099 } $ & $ 2.82 _{ -0.16 } ^{ + 0.17 } $ & $ 59.23 _{ -4.98 } ^{ + 5.01 } $ & $ 1.36 _{ -0.12 } ^{ + 0.13 } $ & $ 0.46 _{ -0.09 } ^{ + 0.11 } $ & --- \\  
   UGC07261 & B & $ -11.157 $ & $ 0.242 _{ -0.134 } ^{ + 0.153 } $ & $ 12.56 _{ -2.60 } ^{ + 3.23 } $ & $ 42.29 _{ -6.67 } ^{ + 7.11 } $ & $ 1.36 _{ -0.12 } ^{ + 0.13 } $ & $ 0.49 _{ -0.08 } ^{ + 0.10 } $ & --- \\  
   UGC07323 & P & $ -10.849 $ & $ 0.131 _{ -0.132 } ^{ + 0.167 } $ & $ 8.32 _{ -1.96 } ^{ + 2.37 } $ & $ 48.51 _{ -2.89 } ^{ + 2.91 } $ & $ 1.41 _{ -0.12 } ^{ + 0.13 } $ & $ 0.43 _{ -0.08 } ^{ + 0.10 } $ & --- \\  
   UGC07399 & P & $ -11.089 $ & $ -0.026 _{ -0.042 } ^{ + 0.041 } $ & $ 14.22 _{ -2.14 } ^{ + 2.53 } $ & $ 57.54 _{ -2.80 } ^{ + 2.81 } $ & $ 1.38 _{ -0.12 } ^{ + 0.13 } $ & $ 0.61 _{ -0.11 } ^{ + 0.13 } $ & --- \\  
   UGC07524 & P & $ -11.216 $ & $ 0.154 _{ -0.045 } ^{ + 0.052 } $ & $ 4.72 _{ -0.22 } ^{ + 0.24 } $ & $ 49.72 _{ -2.78 } ^{ + 2.78 } $ & $ 1.24 _{ -0.10 } ^{ + 0.11 } $ & $ 0.90 _{ -0.12 } ^{ + 0.13 } $ & --- \\  
   UGC07559 & P & $ -11.403 $ & $ 0.204 _{ -0.055 } ^{ + 0.073 } $ & $ 4.98 _{ -0.24 } ^{ + 0.25 } $ & $ 61.77 _{ -2.97 } ^{ + 2.95 } $ & $ 1.37 _{ -0.12 } ^{ + 0.13 } $ & $ 0.48 _{ -0.10 } ^{ + 0.12 } $ & --- \\  
   UGC07577 & B & $ -11.735 $ & $ 0.420 _{ -0.082 } ^{ + 0.056 } $ & $ 2.56 _{ -0.12 } ^{ + 0.13 } $ & $ 63.05 _{ -2.92 } ^{ + 2.94 } $ & $ 1.33 _{ -0.11 } ^{ + 0.12 } $ & $ 0.44 _{ -0.08 } ^{ + 0.09 } $ & --- \\  
   UGC07603 & P & $ -11.259 $ & $ -0.072 _{ -0.038 } ^{ + 0.035 } $ & $ 4.66 _{ -0.69 } ^{ + 0.80 } $ & $ 78.38 _{ -2.93 } ^{ + 2.94 } $ & $ 1.39 _{ -0.12 } ^{ + 0.13 } $ & $ 0.44 _{ -0.08 } ^{ + 0.10 } $ & --- \\  
   UGC07608 & A & $ -11.489 $ & $ 0.185 _{ -0.128 } ^{ + 0.160 } $ & $ 8.47 _{ -1.94 } ^{ + 2.34 } $ & $ 42.61 _{ -7.85 } ^{ + 7.61 } $ & $ 1.36 _{ -0.12 } ^{ + 0.13 } $ & $ 0.51 _{ -0.10 } ^{ + 0.13 } $ & --- \\  
   UGC07690 & P & $ -10.968 $ & $ 0.224 _{ -0.098 } ^{ + 0.126 } $ & $ 8.96 _{ -1.65 } ^{ + 2.02 } $ & $ 45.36 _{ -4.21 } ^{ + 4.35 } $ & $ 1.36 _{ -0.12 } ^{ + 0.13 } $ & $ 0.53 _{ -0.09 } ^{ + 0.12 } $ & --- \\  
   UGC07866 & P & $ -11.399 $ & $ 0.205 _{ -0.080 } ^{ + 0.107 } $ & $ 4.58 _{ -0.22 } ^{ + 0.23 } $ & $ 47.67 _{ -4.75 } ^{ + 4.69 } $ & $ 1.35 _{ -0.12 } ^{ + 0.13 } $ & $ 0.53 _{ -0.11 } ^{ + 0.13 } $ & --- \\  
   UGC08286 & P & $ -11.135 $ & $ 0.021 _{ -0.014 } ^{ + 0.015 } $ & $ 6.60 _{ -0.20 } ^{ + 0.21 } $ & $ 88.10 _{ -2.07 } ^{ + 1.34 } $ & $ 1.32 _{ -0.11 } ^{ + 0.12 } $ & $ 1.14 _{ -0.08 } ^{ + 0.09 } $ & --- \\  
   UGC08490 & P & $ -11.588 $ & $ 0.035 _{ -0.015 } ^{ + 0.015 } $ & $ 5.20 _{ -0.43 } ^{ + 0.47 } $ & $ 55.39 _{ -2.52 } ^{ + 2.58 } $ & $ 1.37 _{ -0.12 } ^{ + 0.13 } $ & $ 0.67 _{ -0.09 } ^{ + 0.11 } $ & --- \\  
   UGC08550 & P & $ -11.388 $ & $ 0.002 _{ -0.026 } ^{ + 0.027 } $ & $ 6.52 _{ -0.86 } ^{ + 1.00 } $ & $ 88.00 _{ -2.17 } ^{ + 1.40 } $ & $ 1.28 _{ -0.11 } ^{ + 0.12 } $ & $ 0.72 _{ -0.11 } ^{ + 0.13 } $ & --- \\  
   UGC08699 & P & $ -10.863  $ & $ -0.011 _{ -0.024 } ^{ + 0.023 } $ & $ 37.40 _{ -4.04 } ^{ + 4.53 } $ & $ 80.73 _{ -5.93 } ^{ + 5.31 } $ & $ 1.38 _{ -0.12 } ^{ + 0.13 } $ & $ 0.63 _{ -0.10 } ^{ + 0.12 } $ & $ 0.68 _{ -0.07 } ^{ + 0.08 } $  \\  
   UGC08837 & P & $ -11.286 $ & $ 0.216 _{ -0.049 } ^{ + 0.063 } $ & $ 7.25 _{ -0.34 } ^{ + 0.36 } $ & $ 80.74 _{ -4.70 } ^{ + 4.47 } $ & $ 1.43 _{ -0.12 } ^{ + 0.13 } $ & $ 0.40 _{ -0.07 } ^{ + 0.08 } $ & --- \\  
   UGC09037 & P & $ -11.097 $ & $ -0.014 _{ -0.037 } ^{ + 0.038 } $ & $ 73.49 _{ -6.18 } ^{ + 6.79 } $ & $ 63.78 _{ -4.74 } ^{ + 4.87 } $ & $ 1.45 _{ -0.13 } ^{ + 0.14 } $ & $ 0.22 _{ -0.03 } ^{ + 0.03 } $ & --- \\  
   UGC09133 & P & $ -11.361  $ & $ 0.042 _{ -0.007 } ^{ + 0.007 } $ & $ 35.41 _{ -3.04 } ^{ + 3.58 } $ & $ 64.78 _{ -4.44 } ^{ + 4.59 } $ & $ 1.50 _{ -0.13 } ^{ + 0.15 } $ & $ 0.83 _{ -0.09 } ^{ + 0.09 } $ & $ 0.72 _{ -0.04 } ^{ + 0.04 } $  \\
   UGC09992 & C & $ -11.354 $ & $ 0.361 _{ -0.138 } ^{ + 0.098 } $ & $ 9.52 _{ -1.97 } ^{ + 2.53 } $ & $ 35.87 _{ -5.88 } ^{ + 6.37 } $ & $ 1.34 _{ -0.12 } ^{ + 0.13 } $ & $ 0.50 _{ -0.10 } ^{ + 0.12 } $ & --- \\  
   UGC10310 & A & $ -11.125 $ & $ 0.270 _{ -0.136 } ^{ + 0.142 } $ & $ 16.38 _{ -3.31 } ^{ + 3.93 } $ & $ 40.81 _{ -4.78 } ^{ + 4.88 } $ & $ 1.32 _{ -0.11 } ^{ + 0.12 } $ & $ 0.66 _{ -0.12 } ^{ + 0.14 } $ & --- \\  
   UGC11455 & P & $ -10.651 $ & $ -0.035 _{ -0.028 } ^{ + 0.025 } $ & $ 72.22 _{ -7.36 } ^{ + 8.17 } $ & $ 89.33 _{ -0.73 } ^{ + 0.47 } $ & $ 1.42 _{ -0.12 } ^{ + 0.14 } $ & $ 0.46 _{ -0.06 } ^{ + 0.07 } $ & --- \\  
   UGC11557 & C & $ -10.882 $ & $ 0.360 _{ -0.168 } ^{ + 0.101 } $ & $ 18.22 _{ -3.39 } ^{ + 4.20 } $ & $ 34.01 _{ -5.55 } ^{ + 5.91 } $ & $ 1.40 _{ -0.12 } ^{ + 0.13 } $ & $ 0.35 _{ -0.07 } ^{ + 0.09 } $ & --- \\  
   UGC11820 & P & $ -11.301 $ & $ -0.015 _{ -0.019 } ^{ + 0.019 } $ & $ 12.15 _{ -2.30 } ^{ + 3.03 } $ & $ 44.21 _{ -5.90 } ^{ + 6.75 } $ & $ 1.20 _{ -0.10 } ^{ + 0.11 } $ & $ 0.98 _{ -0.13 } ^{ + 0.15 } $ & --- \\  
   UGC11914 & A & $ -9.760  $ & $ -0.448 _{ -0.037 } ^{ + 0.057 } $ & $ 9.50 _{ -1.09 } ^{ + 1.32 } $ & $ 49.36 _{ -3.52 } ^{ + 3.64 } $ & $ 1.42 _{ -0.13 } ^{ + 0.14 } $ & $ 0.27 _{ -0.04 } ^{ + 0.04 } $ & $ 0.81 _{ -0.07 } ^{ + 0.06 } $  \\ 
   UGC12506 & P & $ -10.921 $ & $ 0.214 _{ -0.044 } ^{ + 0.050 } $ & $ 117.01 _{ -9.62 } ^{ + 10.55 } $ & $ 86.11 _{ -3.10 } ^{ + 2.48 } $ & $ 1.43 _{ -0.13 } ^{ + 0.14 } $ & $ 1.04 _{ -0.11 } ^{ + 0.13 } $ & --- \\  
   UGC12632 & P & $ -11.405 $ & $ 0.249 _{ -0.085 } ^{ + 0.114 } $ & $ 13.17 _{ -2.49 } ^{ + 2.93 } $ & $ 49.21 _{ -2.77 } ^{ + 2.79 } $ & $ 1.25 _{ -0.11 } ^{ + 0.11 } $ & $ 1.05 _{ -0.14 } ^{ + 0.16 } $ & --- \\  
   UGC12732 & P & $ -11.443 $ & $ 0.124 _{ -0.055 } ^{ + 0.077 } $ & $ 13.11 _{ -2.52 } ^{ + 3.32 } $ & $ 48.12 _{ -4.79 } ^{ + 4.98 } $ & $ 1.26 _{ -0.11 } ^{ + 0.12 } $ & $ 0.86 _{ -0.10 } ^{ + 0.12 } $ & --- \\  
    UGCA442 & P & $ -11.393 $ & $ -0.051 _{ -0.013 } ^{ + 0.012 } $ & $ 4.20 _{ -0.20 } ^{ + 0.21 } $ & $ 51.12 _{ -3.59 } ^{ + 3.93 } $ & $ 1.29 _{ -0.11 } ^{ + 0.12 } $ & $ 0.45 _{ -0.09 } ^{ + 0.11 } $ & --- \\  
    UGCA444 & P & $ -11.550 $ & $ 0.061 _{ -0.022 } ^{ + 0.024 } $ & $ 0.95 _{ -0.05 } ^{ + 0.05 } $ & $ 78.76 _{ -3.88 } ^{ + 3.83 } $ & $ 1.25 _{ -0.11 } ^{ + 0.12 } $ & $ 0.56 _{ -0.12 } ^{ + 0.15 } $ & --- \\  
 \hline
\end{tabular}
\end{center}

\end{table}

\newpage
\setcounter{table}{2}
\begin{table}
\caption{Newtonian environmental field strength $\log(e_{\rm N,env})$}\label{tab:env}
\begin{center}
  \begin{tabular}{lcc}
  \hline
 galaxy & max clustering & no clustering  \\
 \hline
   D512-2 & $-2.245 \pm 0.261 $  & $-3.155 \pm 0.285 $ \\ 
   D564-8 & $-2.283 \pm 0.242 $  & $-3.178 \pm 0.271 $ \\ 
   DDO064 & $-2.257 \pm 0.299 $  & $-3.162 \pm 0.317 $ \\ 
   DDO154 & $-2.166 \pm 0.344 $  & $-3.052 \pm 0.383 $ \\ 
   DDO168 & $-2.194 \pm 0.287 $  & $-3.081 \pm 0.319 $ \\ 
   DDO170 & $-1.915 \pm 0.385 $  & $-2.825 \pm 0.407 $ \\ 
   F563-1 & $-2.628 \pm 0.127 $  & $-3.530 \pm 0.138 $ \\ 
  F563-V1 & $-2.585 \pm 0.139 $  & $-3.483 \pm 0.126 $ \\ 
  F563-V2 & $-2.598 \pm 0.110 $  & $-3.492 \pm 0.115 $ \\ 
  F565-V2 & $-2.620 \pm 0.124 $  & $-3.533 \pm 0.126 $ \\ 
   F567-2 & $-2.505 \pm 0.131 $  & $-3.412 \pm 0.129 $ \\ 
   F568-1 & $-2.440 \pm 0.134 $  & $-3.355 \pm 0.128 $ \\ 
   F568-3 & $-2.459 \pm 0.128 $  & $-3.368 \pm 0.129 $ \\ 
  F568-V1 & $-2.380 \pm 0.196 $  & $-3.289 \pm 0.211 $ \\ 
   F571-8 & $-2.427 \pm 0.240 $  & $-3.366 \pm 0.259 $ \\ 
  F571-V1 & $-2.118 \pm 0.340 $  & $-3.045 \pm 0.340 $ \\ 
   F574-1 & $-2.217 \pm 0.304 $  & $-3.169 \pm 0.336 $ \\ 
   F574-2 & $-2.282 \pm 0.417 $  & $-3.209 \pm 0.441 $ \\ 
  F579-V1 & $-2.630 \pm 0.207 $  & $-3.542 \pm 0.218 $ \\ 
   IC4202 & $-1.942 \pm 0.438 $  & $-2.858 \pm 0.479 $ \\ 
  NGC0100 & $-2.667 \pm 0.221 $  & $-3.561 \pm 0.268 $ \\ 
  NGC2683 & $-2.362 \pm 0.252 $  & $-3.207 \pm 0.381 $ \\ 
  NGC2841 & $-2.395 \pm 0.366 $  & $-3.288 \pm 0.483 $ \\ 
  NGC2903 & $-2.300 \pm 0.243 $  & $-3.208 \pm 0.257 $ \\ 
  NGC2955 & $-2.520 \pm 0.217 $  & $-3.425 \pm 0.344 $ \\ 
  NGC2998 & $-2.542 \pm 0.222 $  & $-3.451 \pm 0.258 $ \\ 
  NGC3198 & $-2.335 \pm 0.317 $  & $-3.248 \pm 0.376 $ \\ 
  NGC3726 & $-2.242 \pm 0.340 $  & $-3.159 \pm 0.369 $ \\ 
  NGC3741 & $-2.307 \pm 0.324 $  & $-3.195 \pm 0.347 $ \\ 
  NGC3769 & $-2.322 \pm 0.279 $  & $-3.252 \pm 0.311 $ \\ 
  NGC3877 & $-2.331 \pm 0.319 $  & $-3.266 \pm 0.309 $ \\ 
  NGC3893 & $-2.344 \pm 0.304 $  & $-3.272 \pm 0.316 $ \\ 
  NGC3917 & $-2.396 \pm 0.281 $  & $-3.325 \pm 0.318 $ \\ 
  NGC3949 & $-2.320 \pm 0.324 $  & $-3.256 \pm 0.331 $ \\ 
  NGC3953 & $-2.382 \pm 0.304 $  & $-3.306 \pm 0.323 $ \\ 
  NGC3972 & $-2.333 \pm 0.321 $  & $-3.268 \pm 0.316 $ \\ 
  NGC3992 & $-2.540 \pm 0.192 $  & $-3.479 \pm 0.209 $ \\ 
  NGC4010 & $-2.307 \pm 0.342 $  & $-3.242 \pm 0.318 $ \\ 
  NGC4013 & $-2.215 \pm 0.274 $  & $-3.142 \pm 0.290 $ \\ 
  NGC4051 & $-2.271 \pm 0.288 $  & $-3.196 \pm 0.346 $ \\ 
  NGC4068 & $-2.272 \pm 0.315 $  & $-3.157 \pm 0.356 $ \\ 
  NGC4085 & $-2.253 \pm 0.353 $  & $-3.196 \pm 0.348 $ \\ 
  NGC4088 & $-2.235 \pm 0.351 $  & $-3.175 \pm 0.353 $ \\ 
 \hline
\end{tabular}
\end{center}
\end{table}

\setcounter{table}{2}
\begin{table}
\caption{(continued) Newtonian environmental field strength $\log(e_{\rm N,env})$}
\begin{center}
  \begin{tabular}{lcc}
  \hline
 galaxy & max clustering & no clustering  \\
 \hline
  NGC4100 & $-2.382 \pm 0.337 $  & $-3.322 \pm 0.344 $ \\ 
  NGC4138 & $-2.278 \pm 0.336 $  & $-3.218 \pm 0.313 $ \\ 
  NGC4157 & $-2.266 \pm 0.341 $  & $-3.207 \pm 0.286 $ \\ 
  NGC4183 & $-2.280 \pm 0.302 $  & $-3.217 \pm 0.324 $ \\ 
  NGC4214 & $-2.263 \pm 0.340 $  & $-3.149 \pm 0.365 $ \\ 
  NGC4217 & $-2.285 \pm 0.262 $  & $-3.213 \pm 0.278 $ \\ 
  NGC4389 & $-2.316 \pm 0.288 $  & $-3.243 \pm 0.306 $ \\ 
  NGC4559 & $-2.036 \pm 0.445 $  & $-2.938 \pm 0.474 $ \\ 
  NGC5005 & $-2.132 \pm 0.334 $  & $-3.032 \pm 0.404 $ \\ 
  NGC5033 & $-2.381 \pm 0.272 $  & $-3.324 \pm 0.277 $ \\ 
  NGC5055 & $-2.155 \pm 0.314 $  & $-3.064 \pm 0.339 $ \\ 
  NGC5371 & $-2.268 \pm 0.249 $  & $-3.191 \pm 0.263 $ \\ 
  NGC5585 & $-2.264 \pm 0.269 $  & $-3.149 \pm 0.290 $ \\ 
  NGC5907 & $-2.395 \pm 0.193 $  & $-3.322 \pm 0.203 $ \\ 
  NGC7814 & $-2.670 \pm 0.307 $  & $-3.560 \pm 0.408 $ \\ 
 PGC51017 & $-2.261 \pm 0.259 $  & $-3.172 \pm 0.270 $ \\ 
 UGC00191 & $-2.713 \pm 0.201 $  & $-3.620 \pm 0.233 $ \\ 
 UGC00634 & $-2.714 \pm 0.199 $  & $-3.623 \pm 0.249 $ \\ 
 UGC00891 & $-2.590 \pm 0.201 $  & $-3.487 \pm 0.217 $ \\ 
 UGC04499 & $-2.470 \pm 0.265 $  & $-3.391 \pm 0.280 $ \\ 
 UGC05005 & $-2.619 \pm 0.141 $  & $-3.522 \pm 0.142 $ \\ 
 UGC05414 & $-2.277 \pm 0.287 $  & $-3.187 \pm 0.307 $ \\ 
 UGC05716 & $-2.378 \pm 0.235 $  & $-3.303 \pm 0.252 $ \\ 
 UGC05721 & $-2.126 \pm 0.304 $  & $-3.030 \pm 0.347 $ \\ 
 UGC05750 & $-2.561 \pm 0.163 $  & $-3.468 \pm 0.169 $ \\ 
 UGC05764 & $-2.232 \pm 0.261 $  & $-3.149 \pm 0.286 $ \\ 
 UGC05829 & $-2.189 \pm 0.324 $  & $-3.091 \pm 0.344 $ \\ 
 UGC05986 & $-2.163 \pm 0.348 $  & $-3.087 \pm 0.342 $ \\ 
 UGC05999 & $-2.601 \pm 0.160 $  & $-3.512 \pm 0.184 $ \\ 
 UGC06399 & $-2.412 \pm 0.233 $  & $-3.337 \pm 0.263 $ \\ 
 UGC06446 & $-2.394 \pm 0.240 $  & $-3.327 \pm 0.265 $ \\ 
 UGC06614 & $-2.011 \pm 0.426 $  & $-2.954 \pm 0.433 $ \\ 
 UGC06628 & $-2.291 \pm 0.324 $  & $-3.217 \pm 0.313 $ \\ 
 UGC06667 & $-2.322 \pm 0.286 $  & $-3.248 \pm 0.290 $ \\ 
 UGC06786 & $-2.649 \pm 0.165 $  & $-3.587 \pm 0.181 $ \\ 
 UGC06787 & $-2.416 \pm 0.337 $  & $-3.353 \pm 0.374 $ \\ 
 UGC06818 & $-2.289 \pm 0.282 $  & $-3.209 \pm 0.299 $ \\ 
 UGC06917 & $-2.364 \pm 0.276 $  & $-3.300 \pm 0.288 $ \\ 
 UGC06923 & $-2.359 \pm 0.252 $  & $-3.294 \pm 0.271 $ \\ 
 UGC06930 & $-2.377 \pm 0.282 $  & $-3.314 \pm 0.277 $ \\ 
 UGC06973 & $-2.286 \pm 0.312 $  & $-3.215 \pm 0.315 $ \\ 
 UGC06983 & $-2.414 \pm 0.273 $  & $-3.361 \pm 0.271 $ \\ 
 UGC07089 & $-2.234 \pm 0.326 $  & $-3.160 \pm 0.356 $ \\ 
 \hline
\end{tabular}
\end{center}
\end{table}

\setcounter{table}{2}
\begin{table}
\caption{(continued) Newtonian environmental field strength $\log(e_{\rm N,env})$}
\begin{center}
  \begin{tabular}{lcc}
  \hline
 galaxy & max clustering & no clustering  \\
 \hline
 UGC07125 & $-2.117 \pm 0.330 $  & $-3.031 \pm 0.348 $ \\ 
 UGC07151 & $-2.081 \pm 0.282 $  & $-2.965 \pm 0.321 $ \\ 
 UGC07232 & $-2.202 \pm 0.344 $  & $-3.090 \pm 0.360 $ \\ 
 UGC07261 & $-1.741 \pm 0.478 $  & $-2.650 \pm 0.513 $ \\ 
 UGC07323 & $-2.164 \pm 0.333 $  & $-3.065 \pm 0.349 $ \\ 
 UGC07399 & $-2.236 \pm 0.317 $  & $-3.165 \pm 0.304 $ \\ 
 UGC07524 & $-2.210 \pm 0.384 $  & $-3.088 \pm 0.445 $ \\ 
 UGC07559 & $-2.228 \pm 0.381 $  & $-3.107 \pm 0.430 $ \\ 
 UGC07577 & $-2.301 \pm 0.319 $  & $-3.192 \pm 0.341 $ \\ 
 UGC07603 & $-2.145 \pm 0.392 $  & $-3.030 \pm 0.446 $ \\ 
 UGC07608 & $-2.172 \pm 0.334 $  & $-3.078 \pm 0.366 $ \\ 
 UGC07690 & $-2.143 \pm 0.332 $  & $-3.043 \pm 0.368 $ \\ 
 UGC07866 & $-2.299 \pm 0.402 $  & $-3.187 \pm 0.427 $ \\ 
 UGC08286 & $-2.173 \pm 0.307 $  & $-3.071 \pm 0.337 $ \\ 
 UGC08490 & $-2.270 \pm 0.272 $  & $-3.147 \pm 0.343 $ \\ 
 UGC08550 & $-2.125 \pm 0.321 $  & $-3.014 \pm 0.349 $ \\ 
 UGC08699 & $-2.579 \pm 0.313 $  & $-3.524 \pm 0.265 $ \\ 
 UGC08837 & $-2.108 \pm 0.260 $  & $-3.004 \pm 0.289 $ \\ 
 UGC09037 & $-2.719 \pm 0.251 $  & $-3.603 \pm 0.282 $ \\ 
 UGC09133 & $-2.676 \pm 0.212 $  & $-3.603 \pm 0.241 $ \\ 
 UGC12506 & $-2.567 \pm 0.259 $  & $-3.511 \pm 0.272 $ \\ 
 UGC12732 & $-2.667 \pm 0.243 $  & $-3.577 \pm 0.258 $ \\ 
  UGCA281 & $-2.206 \pm 0.303 $  & $-3.097 \pm 0.329 $ \\ 
 \hline
\end{tabular}
\end{center}
\end{table}


\begin{thebibliography}{}

\bibitem[Angus et al.(2008)]{Angus} Angus, G.~W., Famaey, B., \& Buote, D.~A.\ 2008, \mnras, 387, 1470

\bibitem[Asencio et al.(2021)]{ElGordo2021} {{Asencio}, E.}, {{Banik}, I.}, {{Kroupa}, P.} 2021, \mnras, 500, 5249

\bibitem[Babyk et al.(2018)]{Babyk} Babyk, I.~V., McNamara, B.~R., Nulsen, P.~E.~J., et al.\ 2018, \apj, 857, 32

\bibitem[Banik et al.(2020)]{Ban2020} {{Banik}, I.}, {{Thies}, I.}, {{Famaey}, B.}, {{Candish}, G.}, {{Kroupa}, P.}, {{Ibata}, R.} 2020, \apj, 905, 135

 
\bibitem[Bekenstein \& Milgrom(1984)]{BM1984} {{Bekenstein}, J.}, {{Milgrom}, M.} 1984, {\apj}, {286}, 7 

\bibitem[Bernardi et al.(2013)]{Bernardi_SMF} Bernardi, M., Meert, A., Sheth, R.~K., et al.\ 2013, \mnras, 436, 697. doi:10.1093/mnras/stt1607

\bibitem[Brada \& Milgrom(2000)]{BM2000} {{Brada}, R.}, {{Milgrom}, M.} 2000, {\apj}, {531}, {21L}

\bibitem[Bullock \& Boylan-Kolchin(2017)]{Bullock} Bullock, J.~S. \& Boylan-Kolchin, M.\ 2017, \araa, 55, 343

\bibitem[Chae et al.(2019)]{Chae2019} {{Chae}, K.-H.} , {{Bernardi}, M.}, {{Sheth}, R.~K.}, {{Gong}, I.-T.} 2019, {\apj}, {877}, {18}

\bibitem[Chae et al.(2020a)]{Chae2020a} {{Chae}, K.-H.} , {{Bernardi}, M.}, {{Dom\'{i}nguez S\'{a}nchez}, H.}, {{Sheth}, R.~K.} 2020a, {\apj}, {903}, {L31}

\bibitem[Chae et al.(2020b)]{Chae2020b} {{Chae}, K.-H.} , {{Lelli}, F.}, {{Desmond}, H.}, {{McGaugh}, S. S.}, {{Li}, P.}, {{Schombert} J. M.} 2020b, {\apj}, {904}, {51} (Paper~I)

\bibitem[Chae et al.(2021)]{Chae2021} {{Chae}, K.-H.} , {{Lelli}, F.}, {{Desmond}, H.}, {{McGaugh}, S. S.}, {{Li}, P.}, {{Schombert} J. M.} 2021, {\apj}, {910}, {81}

\bibitem[Cooke et al.(2018)]{primordialDH} Cooke, R.~J., Pettini, M., \& Steidel, C.~C.\ 2018, \apj, 855, 102

\bibitem[Das et al.(2019)]{WHIM_3} Das, S., Mathur, S., Gupta, A., et al.\ 2019, \apj, 887, 257

\bibitem[Das et al.(2020)]{WHIM_4} Das, S., Mathur, S., \& Gupta, A.\ 2020, \apj, 897, 63

\bibitem[Deng et al.(2007)]{SGW1} Deng, X.-F., He, J.-Z., He, C.-G., et al.\ 2007, Acta Physica Polonica B, 38, 219

\bibitem[Einasto et al.(2010)]{SGW2} Einasto, M., Tago, E., Saar, E., et al.\ 2010, \aap, 522, A92

\bibitem[Einasto et al.(2011)]{SGW3} Einasto, M., Liivam{\"a}gi, L.~J., Tempel, E., et al.\ 2011, \apj, 736, 51

\bibitem[Ettori et al.(2019)]{Ettori2019} Ettori, S., Ghirardini, V., Eckert, D., et al.\ 2019, \aap, 621, A39

\bibitem[Famaey \& Binney(2005)]{FB2005} Famaey, B., Binney, J. 2005, \mnras, 363, 603

\bibitem[Famaey \& McGaugh(2012)]{FM2012} {{Famaey}, B.}, {{McGaugh}, S.~S.} 2012, {LRR} {15}, {10}

\bibitem[Famaey et al.(2018)]{Famaey_DF2} Famaey, B., McGaugh, S., \& Milgrom, M.\ 2018, \mnras, 480, 473

\bibitem[Foreman-Mackey et al.(2013)]{emcee} Foreman-Mackey, D., Hogg, D. W., Lang, D., Goodman, J. 2013, \pasp, 125, 306 

\bibitem[Geller \& Huchra(1989)]{GH1989} {{Geller}, M. J. }, {{Huchra}, J. P.} 1989, {Science}, {246}, {879}  

\bibitem[G{\'o}rski et al.(2005)]{hp} G{\'o}rski, K.~M., Hivon, E., Banday, A.~J., et al.\ 2005, \apj, 622, 759

\bibitem[Haghi et al.(2016)]{Haghi2016} Haghi, H., Bazkiaei, A.~E., Zonoozi, A.~H., Kroupa, P., 2016, \mnras, 458, 4172.

\bibitem[Haghi et al.(2019)]{Hagh2019} {{Haghi}, H.}, {{Kroupa}, P.}, {{Banik}, I.}, {{Wu}, X.}, {{Zonoozi}, A. H.}, {{Javanmardi}, B.}, {{Ghari}, A.}, {{M\"{u}ller}, O.}, {{Dabringhausen}, J.}, {{Zhao}, H.} 2019, \mnras, 487, 2441

\bibitem[Haslbauer et al.(2020)]{KBCvoid2020} {{Haslbauer}, M.}, {{Banik}, I.}, {{Kroupa}, P.} 2020, \mnras, 499, 2845  

\bibitem[Hees et al.(2016)]{Hees} Hees, A., Famaey, B., Angus, G.~W., Gentile, G., 2016, \mnras, 455, 449.

\bibitem[Henriques et al.(2015)]{Henriques} Henriques, B.~M.~B., White, S.~D.~M., Thomas, P.~A., et al.\ 2015, \mnras, 451, 2663

\bibitem[Joeveer \& Einasto(1978)]{JE1978} {{Joeveer}, M.}, {{Einasto}, J.} 1978, in The Large Scale Structure of the Universe (IAU Symposium, No.\ 79), p.\ 241
  
\bibitem[Karachentsev et al.(2018)]{Kar1} Karachentsev, I.~D., Kaisina, E.~I., \& Makarov, D.~I.\ 2018, \mnras, 479, 4136

\bibitem[Karachentsev \& Kaisina(2019)]{Kar2} Karachentsev, I.~D. \& Kaisina, E.~I.\ 2019, Astrophysical Bulletin, 74, 111

\bibitem[Kroupa(2012)]{Kroupa} Kroupa, P.\ 2012, \pasa, 29, 395

\bibitem[Lavaux \& Hudson(2011)]{LH2011} {{Lavaux}, G.}, {{Hudson}, M. J.} 2011, {\mnras}, {416}, {2840} 
  
\bibitem[Lelli, McGaugh \& Schombert(2016)]{Lel2016} {{Lelli}, F.} , {{McGaugh}, S.~S.}, {{Schombert}, J.~M.} 2016, {\aj}, {152}, {157}

\bibitem[Lelli et al.(2017)]{Lel2017} {{Lelli}, F.}, {{McGaugh}, S.~S.}, {{Schombert}, J.~M.}, {{Pawlowski}, M.~S.} 2017, {\apj}, {836}, {152}

\bibitem[Li et al.(2018)]{Li2018} Li, P., Lelli, F., McGaugh, S., Schombert, J. 2018, \aap, 615, 70  

\bibitem[Li et al.(2020)]{Li2020} Li, P., Lelli, F., McGaugh, S., Schombert, J. 2020, \apjs, 247, 31 

\bibitem[Li et al.(2021)]{Li2021} Li, P., Lelli, F., McGaugh, S., Schombert, J., Chae, K.-H. 2021, \aap, 646, L13  

\bibitem[Li \& White(2009)]{Li_White} Li, C. \& White, S.~D.~M.\ 2009, \mnras, 398, 2177

\bibitem[Lim et al.(2020)]{WHIM_2} Lim, S.~H., Mo, H.~J., Wang, H., et al.\ 2020, \apj, 889, 48

\bibitem[Mantz et al.(2016)]{WtG} Mantz, A.~B., Allen, S.~W., Morris, R.~G., et al.\ 2016, \mnras, 463, 3582

\bibitem[Macquart et al.(2020)]{baryoncensus} Macquart, J.-P., Prochaska, J.~X., McQuinn, M., et al.\ 2020, \nat, 581, 391

\bibitem[McGaugh(2015)]{McGaugh2015} McGaugh, S.~S. 2015, CaJPH, 93, 250

\bibitem[McGaugh \& Milgrom(2013a)]{McGaugh_Milgrom} McGaugh, S. \& Milgrom, M.\ 2013, \apj, 766, 22.

\bibitem[McGaugh \& Milgrom(2013b)]{McGaugh_Milgrom_2} McGaugh, S. \& Milgrom, M.\ 2013, \apj, 775, 139.

\bibitem[McGaugh \& Schombert(2014)]{McGaugh_Schombert} McGaugh, S.~S. \& Schombert, J.~M.\ 2014, \aj, 148, 77

\bibitem[McGaugh, Lelli \& Schombert(2016)]{MLS2016} {{McGaugh}, S.~S.}, {{Lelli}, F.}, {{Schombert}, J.~M.} 2016, {\prl}, {117}, {201101}

\bibitem[McGaugh et al.(2020)]{McGaugh_MolecularGas} McGaugh, S.~S., Lelli, F., \& Schombert, J.~M.\ 2020, Research Notes of the American Astronomical Society, 4, 45

\bibitem[Milgrom(1983)]{Mil1983} {{Milgrom}, M.} 1983, {\apj}, {270}, 365

\bibitem[Milgrom(2008)]{Mil2008} {{Milgrom}, M.} 2008, NewAR, {51}, 906

\bibitem[Milgrom(2010)]{Mil2010} {{Milgrom}, M.} 2010, {\mnras}, {403}, 886 

\bibitem[Milgrom(2014)]{Mil2014} {{Milgrom}, M.} 2014, Scholarpedia, 9(6):31410 

\bibitem[Moffett et al.(2016)]{GAMA_SMF} Moffett, A.~J., Ingarfield, S.~A., Driver, S.~P., et al.\ 2016, \mnras, 457, 1308

\bibitem[Nicastro et al.(2018)]{WHIM_1} Nicastro, F., Kaastra, J., Krongold, Y., et al.\ 2018, \nat, 558, 406

\bibitem[Piffaretti et al.(2011)]{MCXC} Piffaretti, R., Arnaud, M., Pratt, G.~W., et al.\ 2011, \aap, 534, A109

\bibitem[Noordermeer et al.(2007)]{Noordermeer2007} {{Noordermeer} E.}, {{van der Hulst}, J. M. }, {{Sancisi}, R.}, {{Swaters}, R. S.}, {{van Albada}, T. S.} 2007, {\mnras}, {376}, {1513}

\bibitem[Ramella, Geller \& Huchra(1992)]{Ram1992} {{Ramella} M.}, {{Geller}, M. J. }, {{Huchra}, J. P.} 1992, {\apj}, {384}, {396}

\bibitem[Sanders(1999)]{Sanders1999} Sanders, R. H. 1999, \apj, 512, L23 

\bibitem[Sancisi et al.(2008)]{San2008} {{Sancisi}, R.}, {{Fraternali}, F.}, {{Oosterloo}, T.}, {{van der Hulst}, T.} 2008, {A{\&}ARv}, {15}, {189}

\bibitem[Schombert et al.(2019)]{Schom2019} {{Schombert}, J.}, {{McGaugh}, S.}, {{Lelli}, F.} 2019, \mnras, 483, 1496

\bibitem[Shull et al.(2012)]{Shull2012} {{Shull}, J. M.}, {{Smith}, B. D.}, {{Danforth}, C. W.} 2018, {\apj}, {759}, {23}
  
\bibitem[Skordis \& Zlo\'{s}nik(2020)]{Skor2020} {{Skordis}, C.}, {{Zlo\'{s}nik}, T.} 2020, arXiv:2007.00082

\bibitem[Stiskalek et al.(2021)]{Stiskalek} Stiskalek, R., Desmond, H., Holvey, T., et al.\ 2021, arXiv:2101.02765

\bibitem[Su et al.(2015)]{Su} Su, Y., Irwin, J.~A., White, R.~E., et al.\ 2015, \apj, 806, 156

\bibitem[Tully et al.(2019)]{Tully2019} {Tully} R.~B., Pomar{\`e}de, D., Graziani, R., Courtois, H.~M., Hoffman, H., Shaya, E.~J., 2019 \apj, 880, 24

\bibitem[Voges et al.(1999)]{ROSAT} Voges, W., Aschenbach, B., Boller, T., et al.\ 1999, \aap, 349, 389

\bibitem[Will(2014)]{Will2014} {{Will}, C. M.} 2014, {LRR}, {{17}}, {4}

\bibitem[Wu \& Kroupa(2015)]{Wu_Kroupa} Wu, X. \& Kroupa, P.\ 2015, \mnras, 446, 330.

\bibitem[Zaninetti(2018)]{SGW4} Zaninetti, L.\ 2018, International Journal of Astronomy and Astrophysics, 8, 258

\bibitem[Zhang et al.(2011)]{HIFLUGCS} Zhang, Y.-Y., Andernach, H., Caretta, C.~A., et al.\ 2011, \aap, 526, A105

\end{thebibliography}
\end{document}